\documentclass[aps,prd,preprint,preprintnumbers,showpacs,nofootinbib]{revtex4-1}
\usepackage{graphicx}
\usepackage{multirow}
\usepackage{amsmath}
\usepackage{slashed}
\usepackage{soul}
\usepackage{array}
\begin{document}
\title{Theoretical study of the $\gamma d \to \pi^0\eta d$ reaction}
\author{A.~Mart\'inez~Torres$^{1,3}$\footnote{amartine@if.usp.br}}
\author{K.~P.~Khemchandani$^{2,3}$\footnote{kanchan.khemchandani@unifesp.br}}
\author{E. Oset$^{3}$\footnote{eulogio.oset@ific.uv.es}}
\preprint{}

 \affiliation{
 $^1$ Universidade de Sao Paulo, Instituto de Fisica, C.P. 05389-970, Sao Paulo,  Brazil.\\
$^2$ Universidade Federal de S\~ao Paulo, C.P. 01302-907, S\~ao Paulo, Brazil.\\
$^3$ Centro Mixto Universidad de Valencia-CSIC Institutos de Investigaci\'on de Paterna, Aptdo. 22085, 46071 Valencia, Spain}

\date{\today}

\begin{abstract}
We have done a theoretical study of the $\gamma d \to \pi^0 \eta d$ reaction starting with a realistic model for the $\gamma N \to \pi^0 \eta N$ reaction that reproduces cross sections and polarization observables at low energies and involves the $\gamma N \to \Delta(1700)\to \eta \Delta(1232) \to  \eta \pi^0 N$ process. For the coherent reaction in the deuteron we considered the impulse approximation together with the rescattering of the pions and the $\eta$ on a different nucleon than the one where they are produced. We found this second mechanism very important since it helps share between two nucleons the otherwise large momentum transfer of the reaction. Other contributions to the $\gamma d\to\pi^0\eta d$ reaction, involving the $\gamma N\to \pi^\pm\pi^0 N^\prime$ process, followed by the rescattering of the $\pi^\pm$ with another nucleon to give $\eta$ and a nucleon, have also been included. We find a natural explanation, tied to the dynamics of our model, for the shift of the $\eta-d$ mass distribution to lower invariant masses, and of the  $\pi^0-d$ mass distribution to larger invariant masses, compared to a phase space calculation. We also study theoretical uncertainties related to the large momenta of the deuteron wave function involved in the process as well as to the couplings present in the model. Striking differences are found with the experimental angular distribution and further theoretical investigations might be necessary.
\end{abstract}

\maketitle
\section{Introduction}
\begin{figure}[h!]
\includegraphics[width=0.6\textwidth]{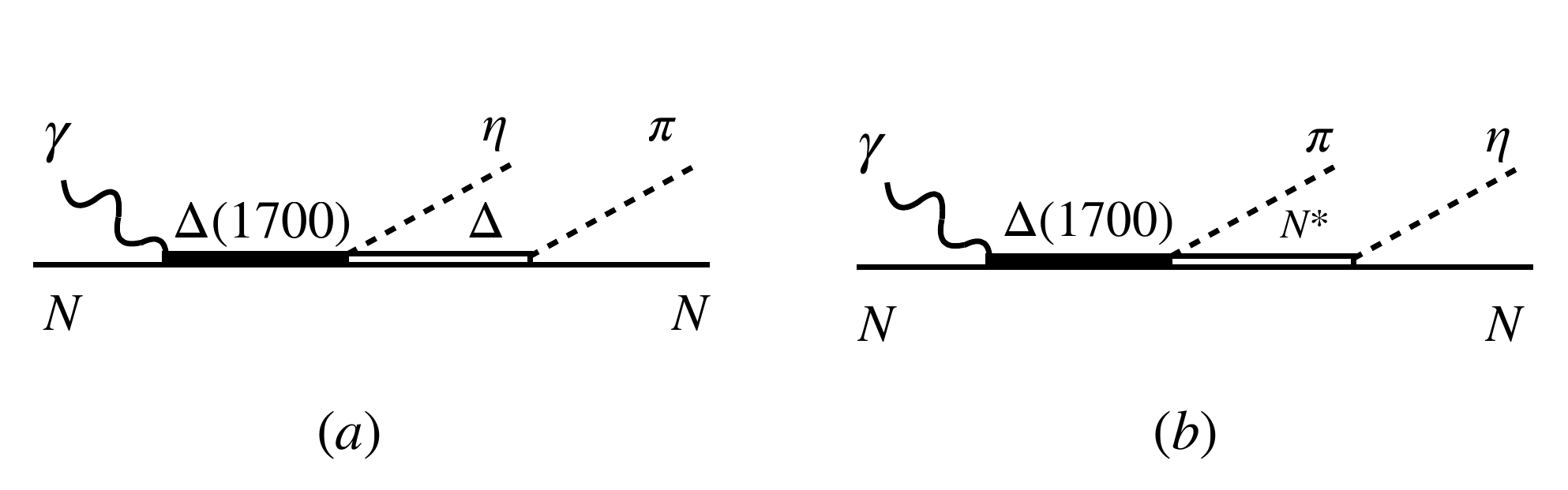}
\caption{Two of the diagrams present in the model of Ref.~\cite{Fix:2010bd}.}\label{Fig1}
\end{figure}

The $\gamma p \to \pi^0 \eta p$ reaction has shown a great potential to address relevant issues on hadron physics, such as the nature of some resonances and effects of triangle singularities. Prior to its measurement, predictions on the cross section at low energies were made in Ref.~\cite{Doring:2005bx}. Many different mechanisms were investigated in the former work, concluding that the reaction at low energies was largely dominated by $\gamma N \to \Delta(1700) \to \Delta(1232) \eta \to \pi^0 N \eta$. The predictions for this process were done by considering $\Delta(1700)$ (spin-parity $J^P=3/2^-$) to be a resonance which appears dynamically generated from the interaction of pseudoscalar mesons with the baryons of the $\Delta(1232)$ decuplet~\cite{Kolomeitsev:2003kt,Sarkar:2004jh}. The chiral unitary approach applied to the coupled channels $\Delta \pi$, $\Sigma^*K$, $\Delta \eta$ gives rise to two resonances, the one at higher energies being associated with the $\Delta(1700)$ of the Particle Data Group (PDG)~\cite{ParticleDataGroup:2020ssz}. It was found in Ref.~\cite{Sarkar:2004jh} that this resonance has a strong coupling to $\Delta \eta$ in $s$-wave, and hence, the  $\gamma N \to \Delta(1700) \to \Delta \eta$ followed be $\Delta \to \pi N$ provided  a natural mechanism to produce the $\gamma N \to  \pi^0 N \eta$ process. The predictions were soon corroborated by  measurements in Refs.~\cite{Nakabayashi:2006ut,Ajaka:2008zz,CB-ELSA:2007xbv,CB-ELSA:2008zkd,CrystalBallatMAMI:2009lze,CBELSATAPS:2014wvh,A2:2018vbv,CBELSATAPS:2014wvh,CBELSA:2009irj}. Subsequent models share the $\Delta(1700)$ mediated mechanism and add new terms, accounting for higher mass resonances that play a role at higher photon energies~\cite{Fix:2010bd,Egorov:2013ppa,Egorov:2020xdt}. The mechanism of Ref.~\cite{Doring:2005bx} was also found to lead to good results in the description of beam asymmetry in $\gamma p \to \pi^0\eta p$~\cite{Ajaka:2008zz} and of the $I^S$, $I^C$ polarization observables~\cite{Doring:2010fw} measured in Ref.~\cite{CBELSA:2009irj}. The same idea regarding the formation of $\Delta(1700)$ and its $\Delta \pi$ and $\Delta \eta$ decay is taken in Ref.~\cite{Doring:2006pt} to describe the data on $\pi^- p \to K^0 \pi^0 \Lambda$ and related reactions. In the model of Ref.~\cite{Fix:2010bd} an extra term is included that deserves some discussion. This is depicted in Fig.~\ref{Fig1}b. 
The diagrams in Fig.~\ref{Fig1} are evaluated phenomenologically in Ref.~\cite{Fix:2010bd} while in Ref.~\cite{Doring:2005bx} the coupling $\Delta(1700)\to \eta \Delta$ is taken from the chiral unitary approach of Ref.~\cite{Sarkar:2004jh}. The latter model does not provide information on the $\Delta(1700)\to \pi N^*(1535)$ transition, with $N^*(1535)$ being a dynamically generated resonance itself from the interaction of pseudoscalar-mesons and baryons~\cite{Inoue:2001ip}. Yet, as shown in the work of Ref.~\cite{Debastiani:2017dlz}, the $\gamma p \to \pi N^*\to \pi N \eta$ mechanism is 
possible within the approach of Ref.~\cite{Doring:2005bx} by considering $\eta N \to \eta N$ rescattering in the final state of the process depicted in Fig.~\ref{Fig1}a. 
\begin{figure}[h!]
\includegraphics[width=0.25\textwidth]{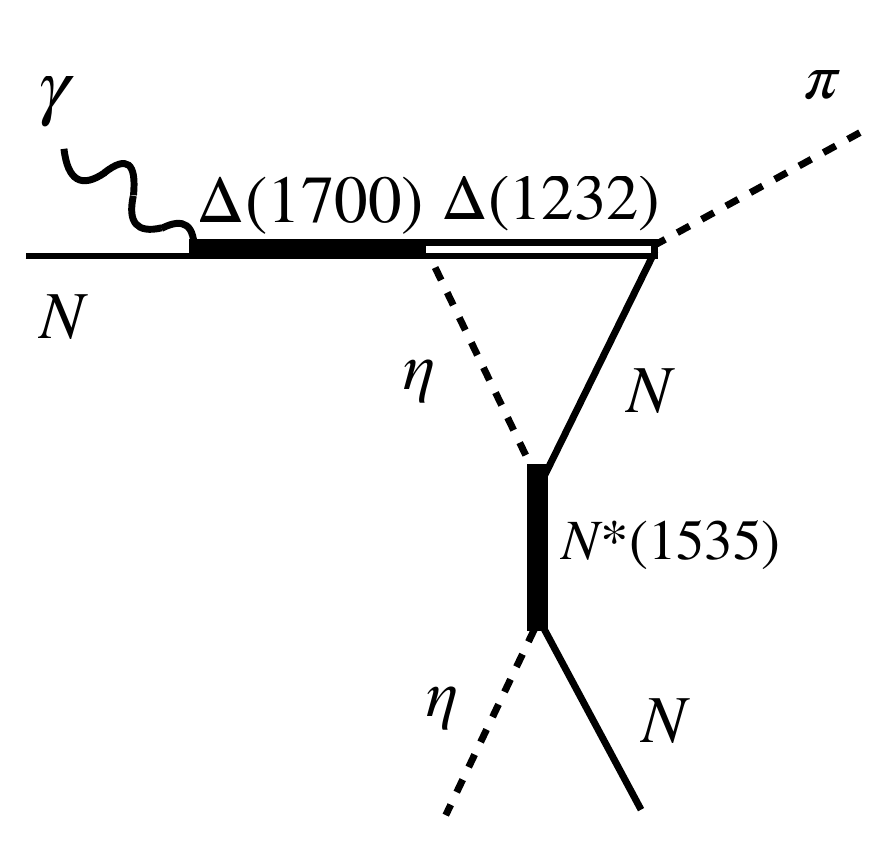}
\caption{Mechanism of Ref.~\cite{Debastiani:2017dlz} to  generate $\gamma p \to \pi N^*(1535)\to \pi \eta N$.}\label{Fig2}
\end{figure}
This is shown in Fig.~\ref{Fig2}, which depicts a triangle mechanism for $\pi N^*(1535)$ production. This mechanism was shown in Ref.~\cite{Debastiani:2017dlz} to produce a triangle singularity~\cite{Landau:1959fi} when the $\Delta(1232)$, $\eta$, and $N$ in the loop are placed on shell and $\Delta(1232)$ and $\pi$ go in the same direction. Yet, there is a subtlety in this  mechanism because there is a theorem in action, Schmid theorem, which states that the triangle singularity stemming from the diagram of Fig.~\ref{Fig2} due to the rescattering of the $\eta N$, when added to the tree-level of Fig.~\ref{Fig1}a does not change the cross section provided by the tree-level~\cite{Schmid:1967ojm}. This, as proven in Ref.~\cite{Debastiani:2018xoi}, is because the sum of the contributions of the $s$-wave tree level amplitude of Fig.~\ref{Fig1}(a) [$t^{(0)}_\text{tree}$] and that of Fig.~\ref{Fig2} (which we denote as $t_L$), where elastic scattering of $\eta N$ involves a triangular loop, is such that $t^{(0)}_\text{tree}+t_L=t^{(0)}_\text{tree}e^{2i\delta}$, with $\delta$ being here the $s$-wave $\eta N\to \eta N$ phase shift. Hence, one finds that $|t^{(0)}_\text{tree}+t_L|^2=|t^{(0)}_\text{tree}|^2$. The theorem holds exactly in the limit of $\Gamma_{\Delta(1232)} \to 0$~\cite{Debastiani:2018xoi}, but it was shown to hold to a good extent for finite widths of the intermediate state in the loop. In this sense, it was found in Ref.~\cite{Debastiani:2017dlz} that the mechanism of Fig.~\ref{Fig2} by itself produced a signal for $\gamma p \to \pi N^*(1535)\to\pi\eta N$ compatible with the experimental data extracted in Ref.~\cite{CBELSATAPS:2014wvh}, but when added coherently to the mechanism of Fig.~\ref{Fig1}a it did not change the integrated cross section appreciably (see Fig.~1 of Ref.~\cite{Debastiani:2017dlz}). Based on this fact we rely upon the mechanism of Fig.~\ref{Fig1}a for $\gamma p \to \pi^0 \eta p$  at low energies which was found to provide a reasonable cross section for the reaction~\cite{Doring:2005bx,Ajaka:2008zz}. In this context it is appropriate to mention that another triangle singularity in the $\gamma p\to\pi^0\eta p$ reaction has been explored at higher energies from a loop containing $a_0(980)$, $p$ and $\pi^0$~\cite{CBELSATAPS:2021osa}.

Having a good model for the $\gamma p \to \pi^0 \eta p$ reaction is a necessary step to make good  predictions for the $\gamma d \to \pi^0 \eta d$ reaction which is the purpose of our study. 

The  $\gamma d \to \pi^0 \eta d$ reaction cross section and the $\pi^0d$, $\eta d$ invariant mass distributions were reported in Ref.~\cite{Ishikawa:2021yyz} and further mass and angular distributions have been recently reported in Ref.~\cite{Ishikawa:2022mgt}. The experimental data are compared with theoretical results from Refs.~\cite{Egorov:2013ppa,Egorov:2020xdt}. The cross sections from Ref.~\cite{Egorov:2013ppa} are based on the impulse approximation, which means that one is summing coherently the $\gamma N \to \pi^0 \eta N$ amplitudes, with $N$ being the proton or the neutron of the deuteron,  and deuteron form factors are used in the evaluation. Effects on final state interaction of the eta with the final deuteron are also considered but they are found to be small. In Ref.~\cite{Egorov:2020xdt} some $\pi N\to \pi N$ and $\pi N \to \eta N$ rescattering mechanisms with the spectator nucleon are also considered, producing an increase of about 20-30$\%$ in the cross section. These mechanisms can be classified as exchange currents, where the exchanged particles are off-shell and some contain $\pi$ rescattering from subdominant mechanisms (Fig. 3 of Ref.~\cite{Egorov:2020xdt}). Our mechanism takes into account the rescattering of the produced $\pi$ and $\eta$ of the mechanism of the impulse approximation. These mesons can be on-shell in the rescattering process, hence enhancing the effect of the mechanism.

\begin{figure}[h!]
\includegraphics[width=0.6\textwidth]{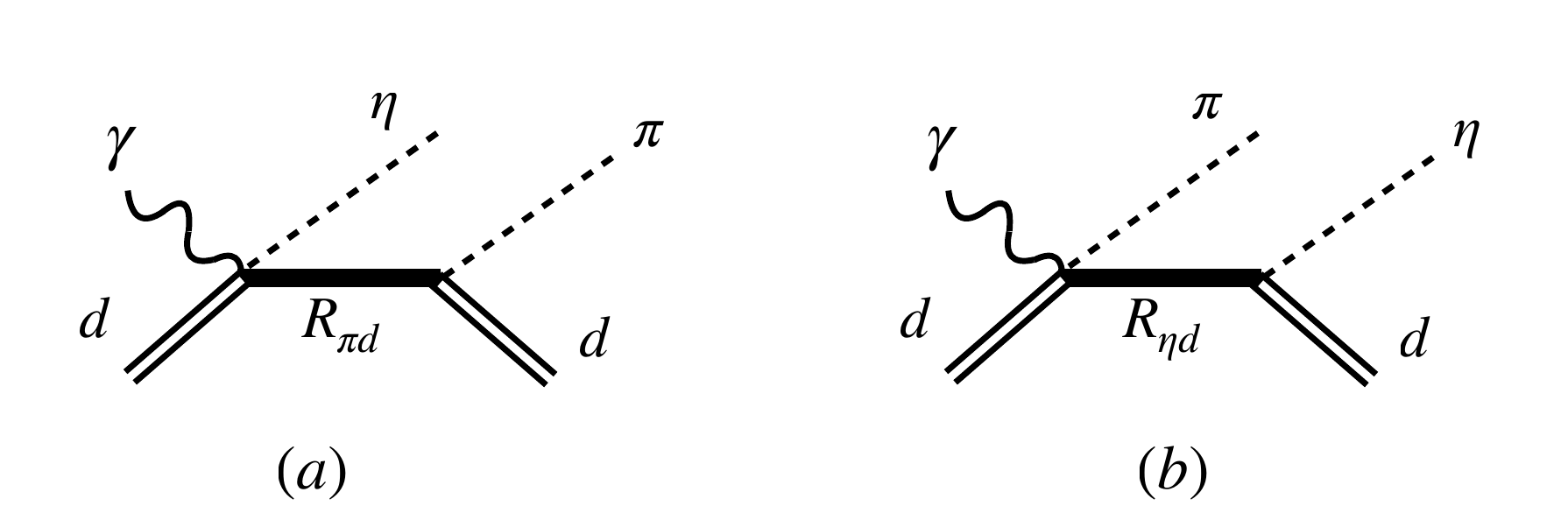}
\caption{Empirical amplitudes used in  Ref.~\cite{Ishikawa:2022mgt} to reproduce the $\gamma d \to \pi^0 \eta d$ observables.}\label{Fig3}
\end{figure}

While the works of Refs.~\cite{Egorov:2013ppa,Egorov:2020xdt} well reproduce the $\gamma d \to \pi^0 \eta d$ integrated cross sections, they show difficulties in reproducing invariant mass distributions and particularly the angular distributions. In view of this, a pure empirical model was suggested in Ref.~\cite{Ishikawa:2022mgt} based on the sum of the two amplitudes depicted in Fig.~\ref{Fig3}.
Two poles were assigned to the $\pi^0d$ and $\eta d$ subsystems and adjusting the parameters of the amplitudes a good reproduction of the observables was obtained. Given the complexity of the dynamics of the reaction, as shown in Refs.~\cite{Egorov:2013ppa,Egorov:2020xdt}, its substitution by the simple model of Fig.~\ref{Fig3} is not very satisfying from the theoretical point of view. This is the motivation for the present paper, complementary to those of Refs.~\cite{Egorov:2013ppa,Egorov:2020xdt}, where we shall investigate in detail the different mechanisms that contribute to the reaction and particularly the uncertainties tied to them.

One of the aims of the work of Refs.~\cite{Ishikawa:2021yyz,Ishikawa:2022mgt} was to show that the data demanded the presence of a bound $\eta d$ state or maybe a virtual state, and the shift of the peaks of $\pi^0 d$, $\eta d$ invariant mass distributions with respect to the phase space distributions were  considered in Ref.~\cite{Ishikawa:2021yyz} as a signal of the presence of this bound state. We shall see in the present work that these shifts are tied to the dynamics of the $\gamma p \to \pi^0 \eta p$ reaction, something that is also found using the model of Ref~\cite{Fix:2010bd} at the level of the impulse approximation in Ref.~\cite{Ishikawa:2022mgt}. 

The search for an $\eta$ bound state in nuclei has been persistent since its likely existence was suggested in Refs.~\cite{Bhalerao:1985cr,Haider:1986sa,Liu:1986rd}. The idea was pursued in Ref.~\cite{Chiang:1990ft} where a many body theory was developed considering the $N^*(1535)$ excitation and the modification of its properties in nuclei, concluding that while the binding was found to be certain in heavy nuclei, the width obtained was bigger than its binding, making thus difficult its experimental observation. The fact is that after many years of search no definite conclusion has been obtained about the existence of such bound state in nuclei~\cite{Machner:2014ona,Kelkar:2013lwa,Metag:2017yuh}. Modern calculations based on chiral unitary theory and many body theory~\cite{Inoue:2002xw,Garcia-Recio:2002xog} conclude  that while for medium and heavy nuclei bound states appear, their widths are fairly larger than the binding, in line with the findings of Ref.~\cite{Chiang:1990ft}.

Further, in case of light nuclei, since a likely $\eta d$ bound state is suggested in Refs.~\cite{Ishikawa:2021yyz,Ishikawa:2022mgt}, we would like to direct the attention of the reader to the discussion in section 7.3 of Ref.~\cite{Metag:2017yuh}. As discussed in the former work, in the case of $\eta\,{}^3\!He$, a pole with 0.3 MeV binding is suggested in Ref.~\cite{Wilkin:2007aa}, and an approximate Breit-Wigner structure with centroid at $-0.3$ MeV below the threshold is found in Ref.~\cite{Xie:2016zhs}, but the pole is in the continuum. Similarly in Ref.~\cite{Barnea:2017epo} it is concluded that deeper potentials than present ones  would be needed to bind
 the $\eta$ in  ${}^3\!He$ and ${}^4\!He$. A similar conclusion is reached in Ref.~\cite{Xie:2018aeg} for the $\eta\,{}^4\!He$ from the study of the $dd \to \eta\, ^4\!He$ reaction at threshold, where again, while some accumulation of strength in the $\eta\,{}^4\!He$ scattering amplitude is observed below threshold, no pole in the  bound region is found. The study of  the $dd \to \eta\, ^4\!He \to \pi^0 n\, ^3\!He$, $\pi^- p\, ^3\!He$ reactions in Refs.~\cite{WASA-at-COSY:2013sya,Adlarson:2016dme} also did not find any evidence of $\eta\, ^4\!He$ bound states.
 
 With this background concerning the $\eta\,^3\!He$ or $\eta\,^4\!He$ states, the possibility of an $\eta d$ bound state, with the deuteron formed by only two nucleons, and quite far away from each other, seems quite gloom. With this perspective, we proceed to do a thorough study of the $\gamma d \to \pi^0 \eta d$ reaction, using the dynamics that has proved realistic in the study of $\gamma p \to \pi^0 \eta p$ cross sections and polarization observables and draw our conclusions.
 
\section{Formalism}\label{formalism}
\subsection{Tree-level amplitude}
\begin{figure}[h!]
\includegraphics[width=0.6\textwidth]{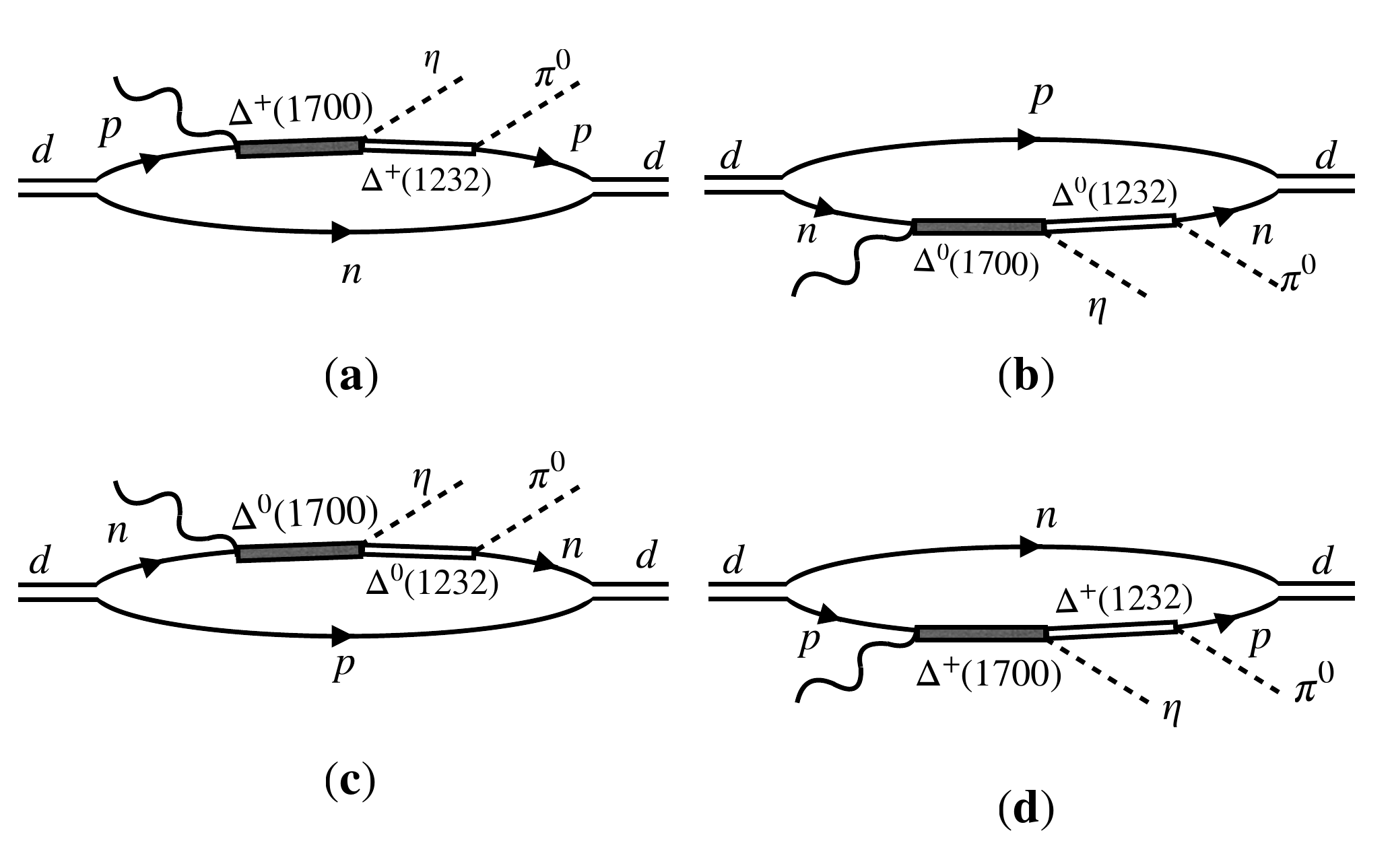} 
\caption{Different diagrams contributing to the $\gamma d \to \pi^0 \eta d$ process in the impulse approximation.}\label{Fig4}
\end{figure}

Inspired by the success of the formalism for the $\gamma p \to \pi^0 \eta p$ process in describing the relevant experimental data, as shown in Ref.~\cite{Doring:2005bx}, which proceeds through the excitation of $\Delta(1700)$ in the intermediate step, we can infer that the same mechanism gives the dominant contribution to $\gamma d \to \pi^0 \eta d$. Considering that the deuteron with $I=0$ has the wave function
\begin{align}
|d\rangle=\frac{1}{\sqrt{2}}[|pn\rangle -|np\rangle], 
\end{align}
 we have the diagrams shown in Fig.~\ref{Fig4}. The latter correspond to the tree-level contributions, also known as the impulse approximation. We  also consider the rescattering of the mesons off the spectator nucleon, as we will discuss in the next subsection. 
To write the amplitudes for the tree-level diagrams, we use the following Lagrangian for $\gamma N \to \Delta(1700)$~\cite{Debastiani:2017dlz}
\begin{align}
-it_{\gamma p\Delta^*}=g_{\gamma p\Delta^*} \vec S^\dagger\cdot\vec \epsilon,\label{photo}
\end{align}
where $\Delta^*$ stands for $\Delta(1700)$, $\vec{\epsilon}$ is the polarization vector for the photon and $\vec{S}$ represents the spin transition operator connecting states with spin 3/2 to 1/2.  The  $s$-wave coupling $g_{\gamma p\Delta^*}=0.188$, in Eq.~(\ref{photo}), as determined in Ref.~\cite{Debastiani:2017dlz}, reproduces the experimental data on the radiative decay width of $\Delta(1700)$~\cite{ParticleDataGroup:2020ssz}. The amplitude in Eq.~(\ref{photo}) is the same for the proton as well as the neutron since the photon must act as an isovector particle in the vertex in order to produce $\Delta(1700)$, an isospin 3/2  baryon. It must be added that it is not required to consider the isospin weight factor $\sqrt{2/3}$ for the $\gamma N\to \Delta^*$ transition since it is already embedded in the value of $g_{\gamma p\Delta^*}$.  

The vertex $\Delta(1700)\to \Delta(1232) \eta$ is described in terms of the coupling deduced in Ref.~\cite{Sarkar:2004jh}, where $\Delta(1700)$ was found to get generated from the pseudoscalar-decuplet baryon dynamics, as
\begin{align}
-it_{\eta \Delta\Delta^*} =-i g_{\eta \Delta\Delta^*}\label{deltastrcoup}
\end{align}
with $g_{\eta \Delta\Delta^*}=1.7-i 1.4$. Finally, for the $\Delta\to \pi N$ transition we write (as in Ref.~\cite{Ikeno:2021frl})
\begin{align}
-it_{\Delta \to \pi N}=-\frac{f^*}{m_\pi}\vec S\cdot\vec p_\pi ~T^\lambda,
\end{align}
where $\vec{p}_\pi \left(m_\pi\right)$ is the momentum (mass) of the pion, $f^*=2.13$ and $\vec S \left(T^\lambda\right)$ is the spin (isospin) transition operator acting on states with spin (isospin) $3/2$ and taking them to $1/2$. The action of the isospin operator leads to a factor $\sqrt{2/3}$ for  the two types of $\Delta\pi N$ vertices appearing in the diagrams in Fig.~\ref{Fig4}: $\Delta^+\pi^0 p$,  $\Delta^0\pi^0 n$. 
Thus, put together, for each of the diagrams we must consider isospin factors:  $1/2$ from the  $d \leftrightarrow pn$ vertices and $\sqrt{2/3}$ from the $\Delta\to \pi N$ vertex. In other words, the sum of the amplitudes of the diagrams in Fig.~\ref{Fig4} can be obtained by calculating the contribution of any one diagram and multiplying it by four times the isospin factor $1/\sqrt{6}$.

\begin{figure}[h!]
\includegraphics[width=0.55\textwidth]{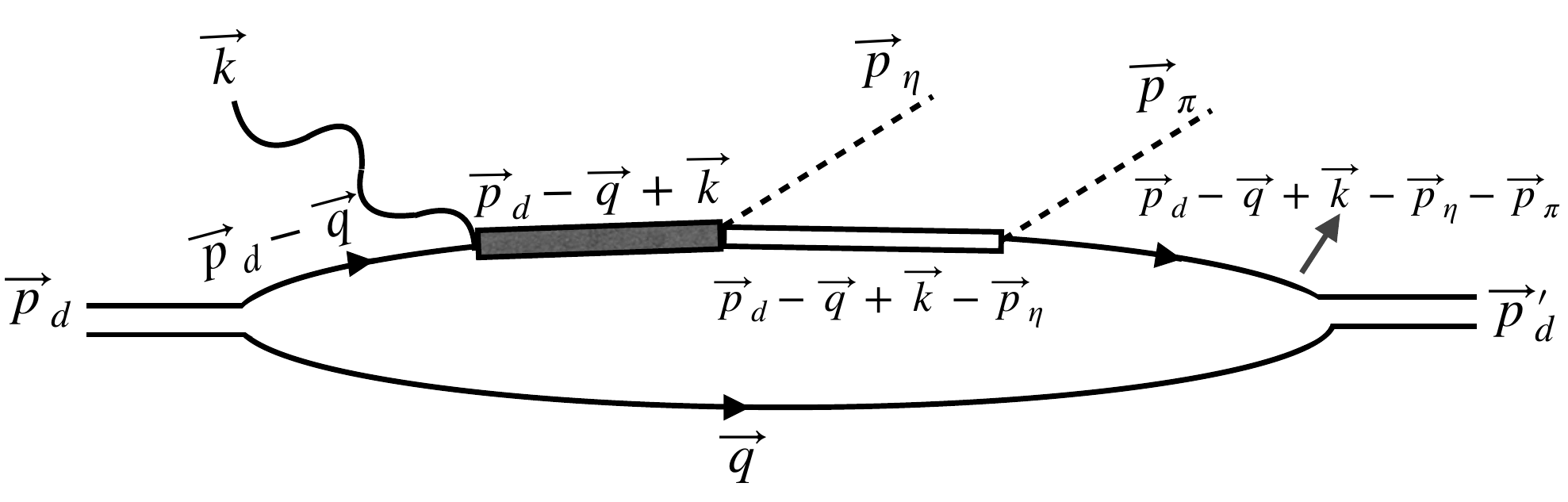} 
\caption{Momenta associated with the different particles appearing in the diagrams of Fig.~\ref{Fig4}. }\label{Fig5}
\end{figure}

To write the sum of the amplitudes in Fig.~\ref{Fig4} we show the momenta associated with each particle in Fig.~\ref{Fig5}. 
In this way, the sum of the amplitudes for the diagrams in Fig.~\ref{Fig4} can be written as
\begin{align}\nonumber
&-it_{4a+4b+4c+4d}=\frac{4}{\sqrt{6}}\int \frac{d^4 q}{\left(2\pi\right)^4} \left( -\frac{f^*}{m_\pi}\vec S\cdot\vec p_\pi \right) \left( g_{\gamma p\Delta^*} \vec S^\dagger\cdot\vec \epsilon \right)\left(-ig_{\eta\Delta\Delta^*}\right)\left[-i g_d\,\theta\!\left(q_{max}-|\vec p_{N}^{~d_i}|\right)\right]  \\ \nonumber
&\times\left[-i g_d\,\theta\!\left(q_{max}-|\vec p_{N}^{~d_f}|\right)\right] \frac{M_N}{E_N(\vec q)}\frac{i}{q^0-E_N(\vec q)+i\epsilon} \frac{M_N}{E_N(\vec p_d -\vec q)} ~\frac{i}{p_d^0-q^0-E_N(\vec p_d -\vec q)+i\epsilon}~ \\ \nonumber
&\times \frac{M_{\Delta^*}}{E_{\Delta^*}(\vec p_d -\vec q+\vec k)}~\frac{i}{p_d^0-q^0+k^0-E_{\Delta^*}(\vec p_d -\vec q+\vec k)+i\epsilon}~ \\ \nonumber &\times \frac{M_\Delta}{E_\Delta(\vec p_d -\vec q+\vec k-\vec p_\eta)}\frac{i}{p_d^0-q^0+k^0-p_\eta^0-E_\Delta(\vec p_d -\vec q+\vec k-\vec p_\eta)+i\epsilon} ~\\ 
&\times \frac{M_N}{E_N(\vec p_d -\vec q+\vec k-\vec p_\eta-\vec p_\pi)} \frac{i}{p_d^0-q^0+k^0-p_\eta^0-p_\pi^0-E_N(\vec p_d -\vec q+\vec k-\vec p_\eta -\vec p_\pi)+i\epsilon},\label{t1}
\end{align}
where $g_d$ is the $d\leftrightarrow pn$ coupling, with a value of $\left(2\pi\right)^{3/2} 2.68 \times 10^{-3}$ MeV$^{-1/2}$~\cite{Ikeno:2021frl} and  $\vec p_{N}^{~d_i} \left(\vec p_{N}^{~d_f}\right)$ is the momentum of the nucleon in the rest frame of the  deuteron in the initial (final) state. Within non-relativistic kinematics, which is suitable for the process, we can write
\begin{align}\nonumber
\vec p_{N}^{~d_i}&=\frac{\vec p_d}{2}-\vec{q},\\
\vec p_{N}^{~d_f}&=\frac{\vec p_d+\vec k-\vec p_\eta-\vec p_\pi}{2}-\vec{q}.
\end{align}

The integration on the $q^0$ variable, in Eq.~({\ref{t1}), can be done analytically, using Cauchy's theorem, to get
\begin{align}\nonumber
&-it_{4a+4b+4c+4d}=-2i\sqrt{\frac{2}{3}}\int \frac{d^3 q}{\left(2\pi\right)^3} \left( \frac{f^*}{m_\pi}\vec S\cdot\vec p_\pi \right) \left( g_{\gamma p\Delta^*} \vec S^\dagger\cdot\vec \epsilon \right)\left(g_{\eta\Delta\Delta^*}\right)\left[g_d\,\theta\Big(q_{max}-\Big|\frac{\vec p_d}{2}-\vec{q}\,\Big|\Big)\right]  \\ \nonumber
&\times\left[g_d\,\theta\Big(q_{max}-\Big|\frac{\vec p_d+\vec k-\vec p_\eta-\vec p_\pi}{2}-\vec{q}\,\Big|\Big)\right] \frac{M_N}{E_N(\vec q)} \frac{M_N}{E_N(\vec p_d -\vec q)} ~\frac{1}{p_d^0-E_N(\vec q)-E_N(\vec p_d -\vec q)+i\epsilon}~ \\ \nonumber
&\times \frac{M_{\Delta^*}}{E_{\Delta^*}(\vec p_d -\vec q+\vec k)}\frac{1}{p_d^0-E_N(\vec q)+k^0-E_{\Delta^*}(\vec p_d -\vec q+\vec k)+i\epsilon}\frac{M_\Delta}{E_\Delta(\vec p_d -\vec q+\vec k-\vec p_\eta)} \\ \nonumber &\times \frac{1}{p_d^0-E_N(\vec q)+k^0-p_\eta^0-E_\Delta(\vec p_d -\vec q+\vec k-\vec p_\eta)+i\epsilon} \frac{M_N}{E_N(\vec p_d -\vec q+\vec k-\vec p_\eta-\vec p_\pi)}\\ 
&\times \frac{1}{p_d^0-E_N(\vec q)+k^0-p_\eta^0-p_\pi^0-E_N(\vec p_d -\vec q+\vec k-\vec p_\eta -\vec p_\pi)+i\epsilon}.\label{t2}
\end{align}

One could proceed further by calculating Eq.~(\ref{t2}) numerically. However, a more realistic description can be accomplished with the following consideration: The result of Eq.~(\ref{t2}) with the $g_d\theta(\cdots)$ function for the deuteron is based implicitly on the solution of the Schr\"odinger equation obtained with a separable potential, $V(|\vec{p}\,|,|\vec{p}^{\,\prime}|)=f(|\vec{p}\,|)f(|\vec{p}^{\,\prime}|)v$. In such a case, one finds that the wave function in momentum space is
\begin{align}
\psi(|\vec{p}\,|)=g\frac{f(|\vec{p}\,|)}{E-E(\vec{p})}
\end{align}
with $g$ being the coupling of the state to the two particle component. This is general for works using separable potentials as those of Refs.~\cite{Alessandrini:1968zza,Kolybasov:1972bn,Fix:2001cz,Gamermann:2009uq,Egorov:2020xdt,Ikeno:2021frl} [see, for example, Eqs. (4.1) and (4.2) of Ref.~\cite{Alessandrini:1968zza}]. In our case the function $f(|\vec{p}\,|)$ is taken as a step function. Thus, we replace
\begin{align}
\left[g_d\,\theta\Big(q_{max}-\Big|\frac{\vec p_d}{2}-\vec{q}\,\Big|\Big)\right] \frac{M_N}{E_N(\vec q)}\frac{M_N}{E_N(\vec p_d -\vec q)}\frac{1}{p_d^0-E_N(\vec q)-E_N(\vec p_d -\vec q)+i\epsilon}\nonumber 
\end{align}
and
\begin{align}
&\left[g_d\,\theta\Big(q_{max}-\Big|\frac{\vec p_d+\vec k-\vec p_\eta-\vec p_\pi}{2}-\vec{q}\,\Big|\Big)\right] \frac{M_N}{E_N(\vec q)}\frac{M_N}{E_N(\vec p_d -\vec q+\vec k-\vec p_\eta-\vec p_\pi)}\nonumber\\
&\times\frac{1}{p_d^0-E_N(\vec q)+k^0-p_\eta^0-p_\pi^0-E_N(\vec p_d -\vec q+\vec k-\vec p_\eta -\vec p_\pi)+i\epsilon}\nonumber
\end{align}
by $-\left(2\pi\right)^{3/2}\psi\left(\frac{\vec p_d}{2}-\vec{q}\right)$ and $-\left(2\pi\right)^{3/2}\psi\left(\frac{\vec p_d+\vec k-\vec p_\eta-\vec p_\pi}{2}-\vec{q}\right)$, respectively, in the amplitude given by Eq.~(\ref{t2}), where $\psi$ represents the deuteron wave function normalized as $\int d^3q |\psi(q)|^2=1$. Note that for the aforementioned substitution we need four $M_N/E_N$ type terms, though there are only three such factors in Eq.~(\ref{t2}). However, consistently with the non-relativistic kinematics applicable in the present work, such $M_N/E_N$ ratios are expected to be close to unity and we can introduce one of these factors in Eq.~(\ref{t2}). It is, thus, reasonable to rewrite Eq.~(\ref{t2}) as
\begin{align}\nonumber
&t_{tree}=2\sqrt{\frac{2}{3}}g_{\gamma p\Delta^*}g_{\eta\Delta\Delta^*}\frac{f^*}{m_\pi}M_\Delta M_{\Delta^*} \int \frac{d^3 q}{\left(2\pi\right)^3}  \frac{(\vec S\cdot\vec p_\pi)(\vec S^\dagger\cdot\vec \epsilon)}{\left[E_{\Delta^*}(\vec p_d -\vec q+\vec k)\right]\left[E_\Delta(\vec p_d -\vec q+\vec k-\vec p_\eta)\right]}\\ \nonumber
&\times \frac{1}{p_d^0-E_N(\vec q\,)+k^0-E_{\Delta^*}(\vec p_d -\vec q+\vec k)+i\epsilon}~ \frac{1}{p_d^0-E_N(\vec q\,)+k^0-p_\eta^0-E_\Delta(\vec p_d -\vec q+\vec k-\vec p_\eta)+i\epsilon}\\
&\times \left(2\pi\right)^{3}\psi\Big(\frac{\vec p_d}{2}-\vec{q}\Big) \psi\Big(\frac{\vec p_d+\vec k-\vec p_\eta-\vec p_\pi}{2}-\vec{q}\Big),\label{t3}
\end{align}
where we will use different well known parametrizations for the deuteron wave function, such as those of Refs.~\cite{Machleidt:2000ge,Lacombe:1981eg,Reid:1968sq,Adler:1975ga}. 

\subsection{Rescattering amplitudes}
\subsubsection{Pion rescattering}\label{pires}
We now discuss the different possible rescattering diagrams which can contribute to the formalism by considering, based on the results obtained in Refs.~\cite{Doring:2005bx,Debastiani:2017dlz}, that the mechanism which plays the main role is the photoexcitation of one of the nucleons to $\Delta(1700)$, followed by $\Delta(1700)\to \eta\Delta(1232)$. We first discuss the rescattering of the pion produced at the $\Delta(1232)\to\pi N$ vertex, as shown in Fig.~\ref{Fig6}.
\begin{figure*}[h!]
\begin{tabular}{ll}
\includegraphics[width=0.4\textwidth]{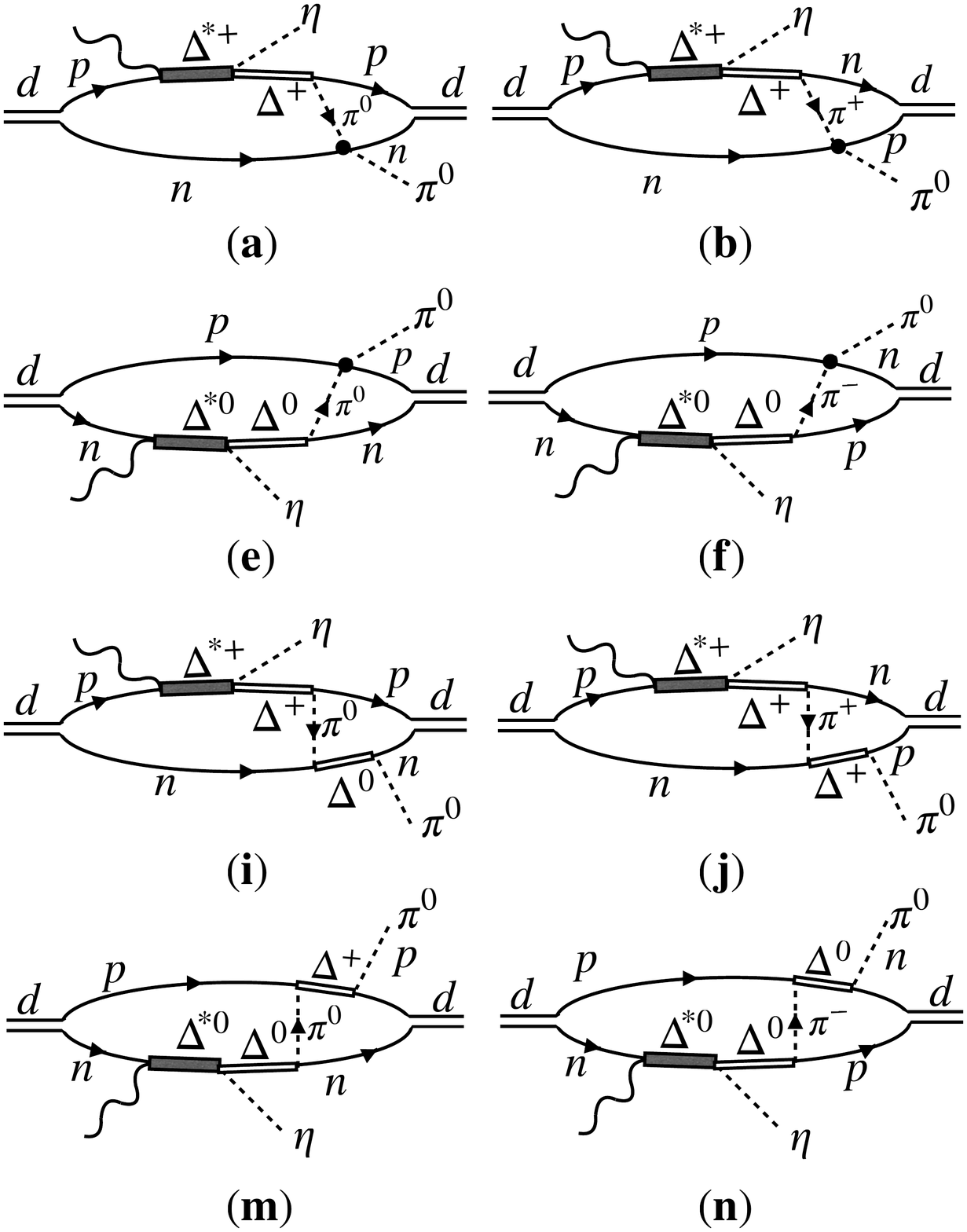}&\includegraphics[width=0.4\textwidth]{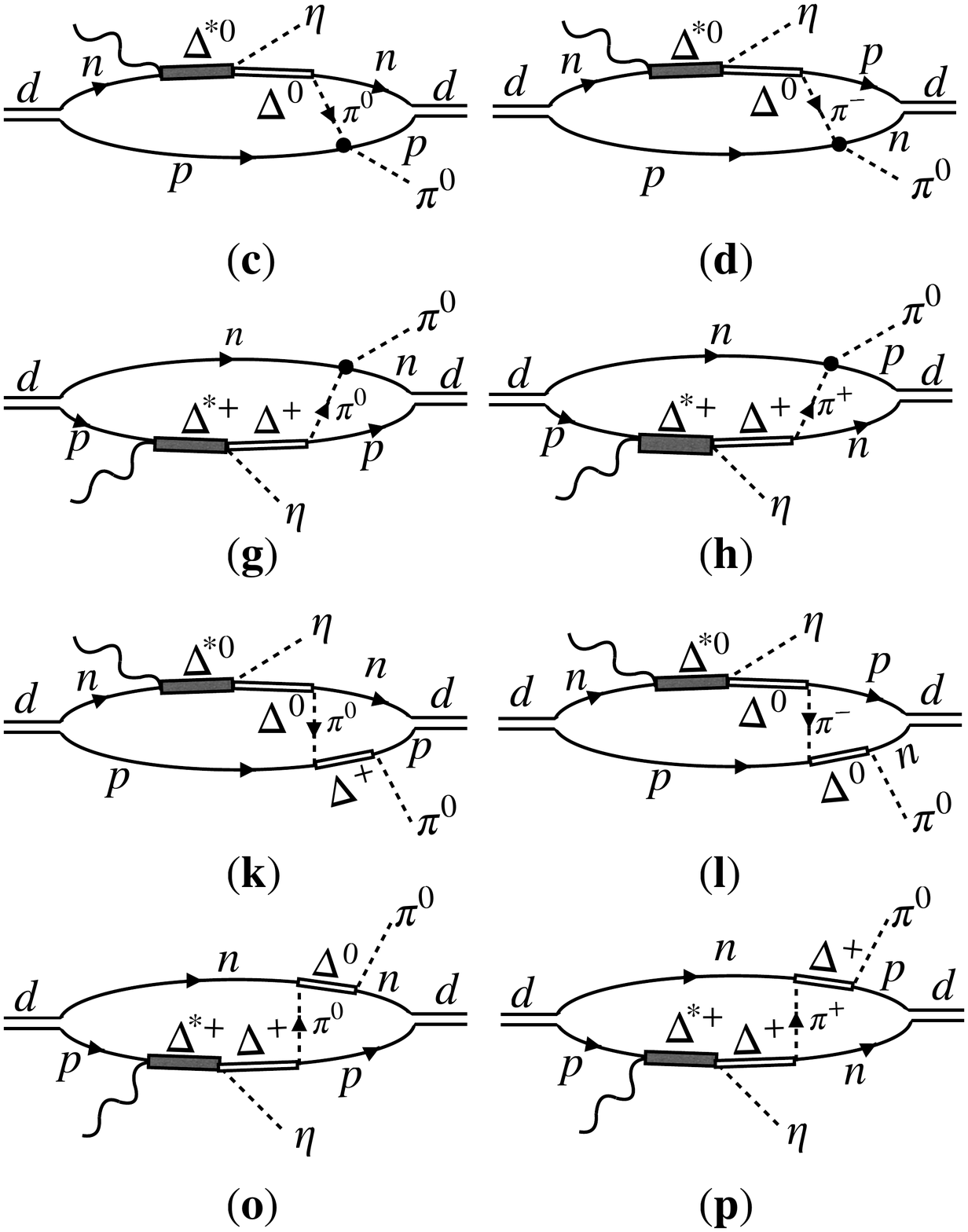}
\end{tabular}
\caption{Different diagrams  contributing to the rescattering of the pion in the intermediate state. The thick dot stands for the $s$-wave $\pi N \to \pi N$ interaction. }\label{Fig6}
\end{figure*}
The  pion produced in the process of  excitation, and the subsequent de-excitation, of one of the nucleons can interact with the spectator nucleon in $s$-wave or in $p$-wave. The diagrams corresponding to the $s$-wave interaction of the rescattered pion with the spectator nucleon are shown in the first and the second rows of Fig.~\ref{Fig6}. The diagrams in the third and  fourth rows depict the $p$-wave interaction of the rescattered pion with the spectator nucleon. We have already discussed the different interaction vertices which are required to write the amplitudes for the diagrams shown in Fig.~\ref{Fig6}, except for the $s$-wave $\pi N$ interaction vertex (shown as a filled circle in the diagrams in the first and second rows). To describe such a vertex, we follow Refs.~\cite{Koltun:1965yk,Oset:1985wt} and use the Lagrangian
\begin{align}
\mathcal{L}_{\pi N}=-4\pi\left[\frac{\lambda_1}{\mu}\bar \psi~\vec \phi\cdot\vec \phi~\psi+\frac{\lambda_2}{\mu^2}~\bar \psi ~\vec \tau\cdot\left(\vec \phi\times\partial_t\vec \phi\right)\psi\right],\label{LpiN_swave}
\end{align}
where $\lambda_1=0.0075$, $\lambda_2=0.053$, $\mu$ represents the pion mass, $\tau_i$'s denote the Pauli matrices, $\bar \psi =\left(\bar p, \bar n\right)$ and $\psi=\left( p, n\right)^T$ and $\vec{\phi}$ is related to the pion fields (in the  cartesian basis). The $\pi N$ $s$-wave amplitudes obtained from Eq.~(\ref{LpiN_swave}) have a common structure
\begin{align}
t^{l=0}_{\pi N}= 4\pi \left(\frac{\lambda_1}{m_\pi} \mathcal{A}_{ij}+\frac{\lambda_2}{m_\pi^2} \mathcal{B}_{ij}\right),\label{tpiN_swave}
\end{align}
with the values for $\mathcal{A}_{ij}$ and $\mathcal{B}_{ij}$ summarized in Table~\ref{Tab:1} for the different  $\pi N$ processes appearing in Fig.~\ref{Fig6}. We shall also consider the model of Ref.~\cite{Inoue:2001ip} to describe the $\pi N$ interaction in the $s$-wave, where the $t$-matrices are obtained by solving the Bethe-Salpeter equation within coupled channels and relevant data are reproduced. As we shall show, the results obtained within the two types of inputs are almost identical.
\begin{table}[h!]
\caption{Values of the $\mathcal{A}_{ij}$ and $\mathcal{B}_{ij}$ parts of the $\pi N$ amplitudes given by Eq.~(\ref{tpiN_swave}). The variables $p_{1\pi}^0$ and $p_{2\pi}^0$, in this table, refer to the energy of the pion in the initial and final state, respectively. We will follow the momenta label shown in Fig.~\ref{Fig7} to write the amplitudes for the different rescattering diagrams of Fig.~\ref{Fig6}. In this case, we will have $p_{1\pi}^0=q^{\prime 0}$ and $p_{2\pi}^0=p_{\pi}^0$.}\label{Tab:1}\vspace{0.5cm}
\begin{tabular}{ccc}
\hline\hline
Processes& $\mathcal{A}_{ij}$ & $\mathcal{B}_{ij}$  \\\hline\hline
$\pi^0 n \to \pi^0 n$, $\pi^0 p \to \pi^0 p$& 2&0\\
$\pi^- p \to \pi^0 n$&0 &$~\sqrt{2}\left(p_{1\pi}^0+p_{2\pi}^0\right)$\\
$\pi^+ n \to \pi^0 p$&0 &$-\sqrt{2}\left(p_{1\pi}^0+p_{2\pi}^0\right)$
\\\hline\hline
\end{tabular}
\end{table}
 
To proceed with writing the amplitudes for the different diagrams shown in Fig.~\ref{Fig6} we assign momenta to different particles as shown in Fig.~\ref{Fig7}.
\begin{figure}[h!]
 \begin{tabular}{cc}
\includegraphics[width=0.48\textwidth]{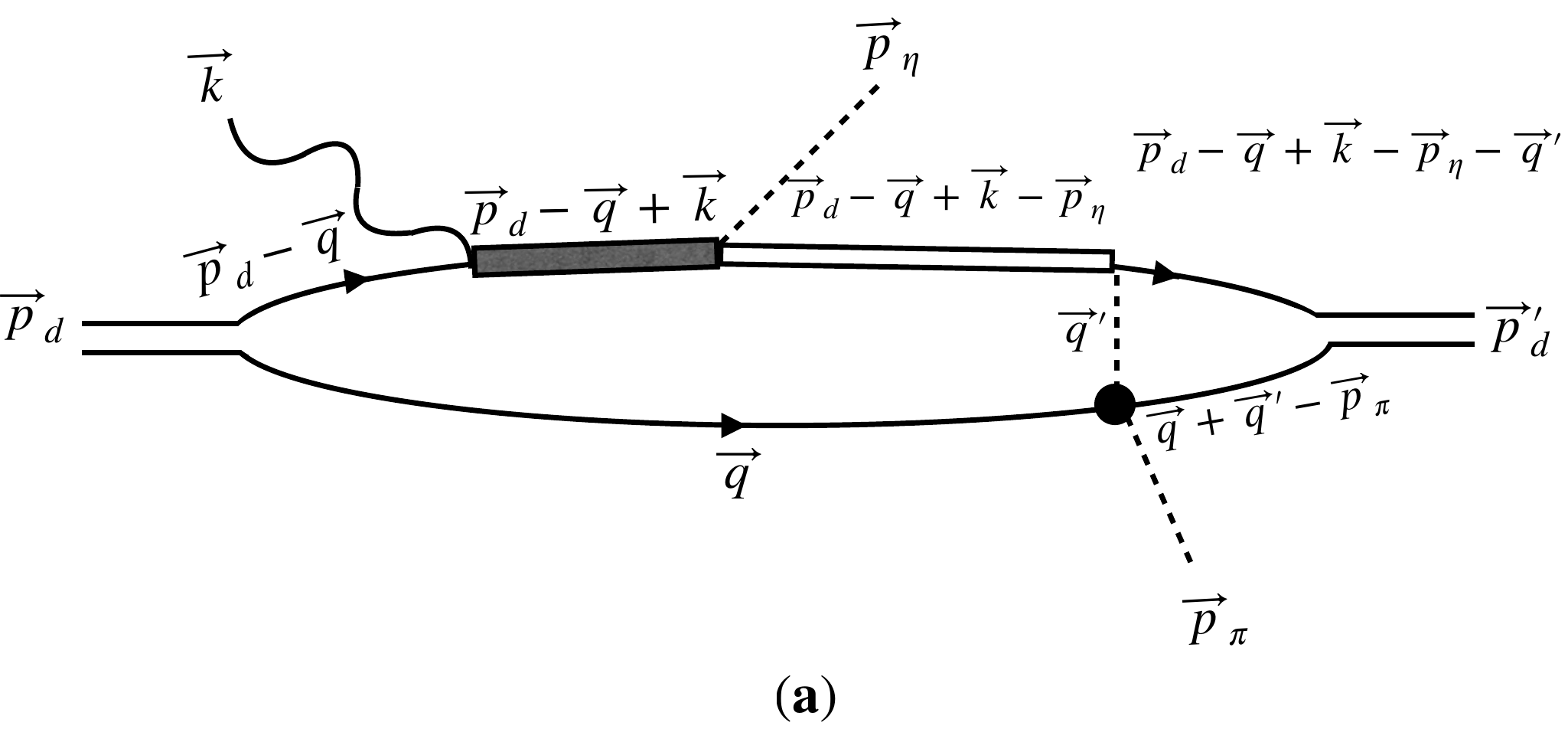}\\\includegraphics[width=0.48\textwidth]{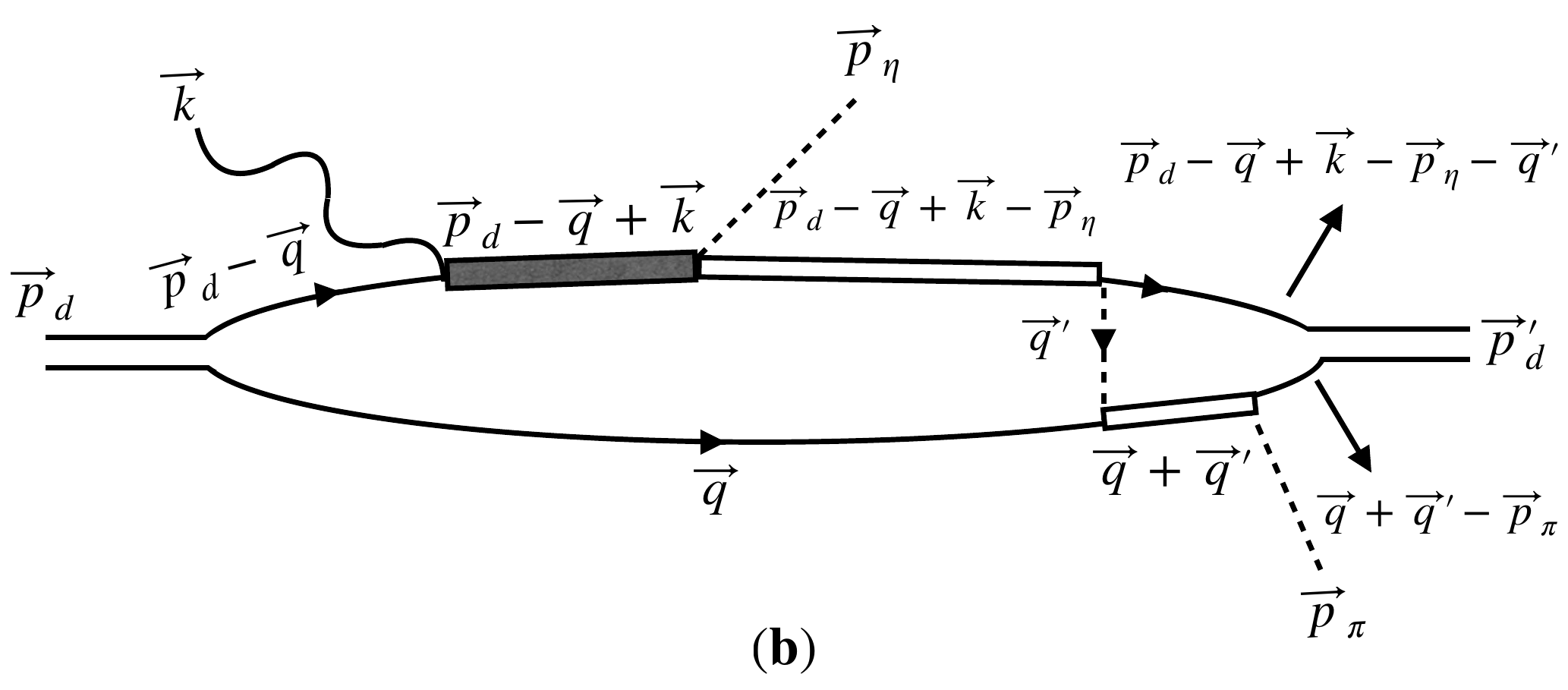} 
\end{tabular}
\caption{Momenta associated with the different particles appearing in the two types of diagrams involving rescattering of the pion (as shown in Fig.~\ref{Fig6}). The filled (empty) rectangle represents a virtual $\Delta(1700)$ [$\Delta(1232)$]. The diagram (a) corresponds to an  $s$-wave interaction at the pion rescattering vertex while the diagram (b) depicts the $p$-wave interaction of the rescattered pion [leading to formation of $\Delta(1232)$].}\label{Fig7}
\end{figure}
The diagram on the left panel  in Fig.~\ref{Fig7} corresponds to the $s$-wave interaction of the rescattered pion with the nucleon, while the one on the right panel shows the possibility that the rescattered  pion leads to the excitation  of the spectator nucleon to $\Delta(1232)$. 
\begin{table}
\caption{Values of the isospin coefficients appearing in Eq.~(\ref{tpi_s_res}), $\mathcal{I}_{dpn}$, $\mathcal{I}^\prime_{dpn}$ and $\mathcal{I}_{\Delta\pi N}$, for the diagrams in Fig.~\ref{Fig6}. We use the phase convention $|\pi^+\rangle=-|I=1,I_3=1\rangle$. }\label{Tab:2}
\begin{tabular}{cccc}
\hline\hline
 vertex & Coefficient & vertex & Coefficient \\\hline\hline
$d\leftrightarrow pn$&$\frac{1}{\sqrt{2}}$&$\Delta^+\leftrightarrow \pi^0 p$&$\sqrt{\frac{2}{3}}$\\
$d\leftrightarrow np$&$-\frac{1}{\sqrt{2}}$&$\Delta^0\leftrightarrow \pi^0 n$&$\sqrt{\frac{2}{3}}$\\
&&$\Delta^+\leftrightarrow \pi^+ n$&$-\sqrt{\frac{1}{3}}$\\
&&$\Delta^0\leftrightarrow \pi^- p$&$\sqrt{\frac{1}{3}}$
\\\hline\hline
\end{tabular}
\end{table}

Let us start by writing the amplitude for the diagrams corresponding to the $s$-wave interaction of the rescattered pion [shown as Fig.~\ref{Fig6}(a)-(h)]. All such diagrams have a common structure 
\begin{align}\nonumber
&-it_{\pi \text{res},l=0}=\int\!\!\frac{d^4 q}{\left(2\pi\right)^4}\int\!\!\frac{d^4 q^\prime}{\left(2\pi\right)^4} \left[-i ~ \mathcal{I}_{dpn} ~g_d\,\theta\!\left(q_{max}-\biggr|\frac{\vec p_{d}}{2}-\vec q~\biggr|\right)\right] \left(\!-\mathcal{I}_{\Delta\pi N}\frac{f^*}{m_\pi}\vec S\cdot\vec q^{\,\prime} \right) \left(-ig_{\eta\Delta\Delta^*}\right) \\ \nonumber
&\times \left( g_{\gamma p\Delta^*} \vec S^\dagger \cdot\vec \epsilon \right)\left[-i~\mathcal{I}^\prime_{dpn}~g_d\,\theta\!\left(q_{max}-\biggr|\frac{-\vec p_\eta+\vec p_\pi +\vec p_d+\vec k}{2}-\vec q-\vec q^{\,\prime} \biggr|\right)\right] \\\nonumber 
&\times\!\left[-i4\pi \left(\frac{\lambda_1}{m_\pi} \mathcal{A}_{ij}+\frac{\lambda_2}{m_\pi^2} \mathcal{B}_{ij}\right)\right]\frac{M_N}{E_N(\vec q\,)}\frac{i}{q^0-E_N(\vec q\,)+i\epsilon} \frac{M_N}{E_N(\vec p_d -\vec q\,)} ~\frac{i}{p_d^0-q^0-E_N(\vec p_d -\vec q\,)+i\epsilon}\\ \nonumber
&\times \frac{M_{\Delta^*}}{E_{\Delta^*}(\vec p_d -\vec q+\vec k)}~\frac{i}{p_d^0-q^0+k^0-E_{\Delta^*}(\vec p_d -\vec q+\vec k)+i\epsilon}~ \frac{M_\Delta}{E_\Delta(\vec p_d -\vec q+\vec k-\vec p_\eta)}\\ \nonumber 
&\times \frac{i}{p_d^0-q^0+k^0-p_\eta^0-E_\Delta(\vec p_d -\vec q+\vec k-\vec p_\eta)+i\epsilon} ~\frac{1}{2\omega_\pi(\vec q^{\,\prime})}~\frac{i}{q^{\prime 0}-\omega_\pi(\vec q^{\,\prime})+i\epsilon}\\ \nonumber
&\times \frac{M_N}{E_N(\vec p_d -\vec q+\vec k-\vec p_\eta-\vec q^{\,\prime})} \frac{i}{p_d^0-q^0+k^0-p_\eta^0-q^{\prime 0}-E_N(\vec p_d -\vec q+\vec k-\vec p_\eta - \vec q^{\,\prime})+i\epsilon}\\
&\times
\frac{M_N}{E_N( \vec q + \vec q^{\,\prime}-\vec p_\pi)}\frac{i}{q^0+q^{\prime 0}-p^0_\pi -E_N( \vec q + \vec q^{\,\prime}-\vec p_\pi)+i\epsilon},\label{tpi_s_res}
\end{align}
where $\mathcal{I}_{dpn}$, $\mathcal{I}^\prime_{dpn}$ and $\mathcal{I}_{\Delta\pi N}$ are the isospin coefficients for the $d \to pn$ (initial and final states) and $\Delta(1232) \to \pi N$ vertices, respectively. The values of these $\mathcal{I}$ coefficients are listed in Table~\ref{Tab:2} for different diagrams [Fig.~\ref{Fig6}(a)-(h)]. As already mentioned, $g_{\gamma p\Delta^*}$ is same for proton as well as neutron. Considering the values  in Table~\ref{Tab:2}, it can be seen that the sum of the amplitudes for the diagrams in Fig.~\ref{Fig6}(a)-(h) leads to a global factor $2\sqrt{2/3}$. Further,  integration on the variables $q^0$ and $q^{\prime 0}$ can be done analytically, to get the  total contribution as
 \begin{align}\nonumber
&t^\text{total}_{\pi \text{res},l=0}=8\pi\sqrt{\frac{2}{3}}\int \frac{d^3 q}{\left(2\pi\right)^3}\int \frac{d^3 q^\prime}{\left(2\pi\right)^3} ~g_d^2 ~ \theta\!\left(q_{max}-\biggr|\frac{-\vec p_\eta+\vec p_\pi+\vec p_d+\vec k}{2}-\vec q-\vec q^{\,\prime} \biggr|\right)  \\ \nonumber
&\times~\theta\!\left(q_{max}-\biggr|\frac{\vec p_{d}}{2}-\vec q~\biggr|\right) g_{\gamma p\Delta^*} ~g_{\eta\Delta\Delta^*} ~\frac{f^*}{m_\pi}~(\vec S\cdot\vec q^{\,\prime})(\vec S^\dagger\cdot\vec \epsilon)\nonumber\\
&\times~\left(2\frac{\lambda_1}{m_\pi}-\frac{\lambda_2}{m_\pi^2} [p^0+k^0-p^0_\eta+p^0_\pi-E_N(\vec{q}\,)-E_N(\vec p_d -\vec q+\vec k-\vec p_\eta-\vec q^{\,\prime})]\right) \frac{M_N}{E_N(\vec q\,)} \frac{M_N}{E_N(\vec p_d -\vec q)}\nonumber \\\nonumber 
&\times\frac{1}{2\omega_\pi(\vec q^{\,\prime})}\frac{M_{\Delta^*}}{E_{\Delta^*}(\vec p_d -\vec q+\vec k)}\frac{M_\Delta}{E_\Delta(\vec p_d -\vec q+\vec k-\vec p_\eta)}\frac{M_N}{E_N(\vec p_d -\vec q+\vec k-\vec p_\eta-\vec q^{\,\prime})}\\ \nonumber
&\times \frac{M_N}{E_N(\vec q + \vec q^{\,\prime}-\vec p_\pi)}~ \frac{1}{p_d^0-E_N(\vec q\,)-E_N(\vec p_d -\vec q)+i\epsilon}\frac{1}{p_d^0- E_N(\vec q\,)+k^0-E_{\Delta^*}(\vec p_d -\vec q+\vec k)+i\epsilon}\\ \nonumber 
&\times \frac{1}{p_d^0-E_N(\vec q\,)+k^0-p_\eta^0-E_\Delta(\vec p_d -\vec q+\vec k-\vec p_\eta\,)+i\epsilon}  \\ \nonumber
&\times \frac{1}{p_d^0-E_N(\vec q\,)+k^0-p_\eta^0-E_N(\vec p_d -\vec q+\vec k-\vec p_\eta - \vec q^{\,\prime})-\omega_\pi(\vec q^{\,\prime})+i\epsilon}\\
&\times
\frac{1}{p_d^0+k^0-p^0_\pi - p_\eta^0-E_N(\vec q + \vec q^{\,\prime}-\vec p_\pi)-E_N(\vec p_d -\vec q+\vec k-\vec p_\eta - \vec q^{\,\prime})+i\epsilon}.\label{tpi_s_res2}
\end{align}
Note that all possible cuts related to the diagrams of Fig.~\ref{Fig6} are accounted for in Eq.~(\ref{tpi_s_res2}). In particular, a $\pi NN$ cut can be noticed in the second last term of the formula. The $d^3 q$ and $d^3 q^\prime$ integrations are done numerically keeping explicitly the $i\epsilon$ small and testing the convergence when $\epsilon\to 0$. If unstable particles are involved in the propagators, the $i\epsilon$ is substituted by $i\Gamma/2$, with $\Gamma$ being the width of the particle.

Consistently with the calculations at the tree level, we substitute in Eq.~(\ref{tpi_s_res2}),
\begin{align}\nonumber
& \frac{g_d^2~\theta(...)\theta(...)}{p_d^0+k^0-p^0_\pi - p_\eta^0-E_N(\vec q + \vec q^{\,\prime}-\vec p_\pi)-E_N(\vec p_d -\vec q+\vec k-\vec p_\eta - \vec q^{\,\prime})+i\epsilon}\\
&\times \frac{M_N}{E_N(\vec q\,)} \frac{M_N}{E_N(\vec p_d -\vec q)}\frac{M_N}{E_N(\vec p_d -\vec q+\vec k-\vec p_\eta-\vec q^{\,\prime}\,)}\frac{M_N}{E_N(\vec q + \vec q^{\,\prime}-\vec p_\pi)}\nonumber\\
&\times\frac{1}{p_d^0-E_N(\vec q\,)-E_N(\vec p_d -\vec q)+i\epsilon}
\to \left(2\pi\right)^{3}\psi\Big(\frac{\vec p_d}{2}-\vec{q}\Big) \psi\Big(\frac{\vec p_d+\vec k-\vec p_\eta+\vec p_\pi}{2}-\vec q-\vec q^{\,\prime}\Big),\label{wavesub}
\end{align}
 to consider more realistic and well-known deuteron wave functions.

Next, we can write the amplitude for the remaining diagrams for pion rescattering,  which involve the excitation of the spectator nucleon to $\Delta(1232)$ due to the interaction with  the rescattered pion [shown as Fig.~\ref{Fig6}(i)-(p)]. Following the momenta labels depicted in Fig.~\ref{Fig7}b, we can write
\begin{align}\nonumber
-it_{\pi \text{res},l=1}&=\int \frac{d^4 q}{\left(2\pi\right)^4}\int \frac{d^4 q^\prime}{\left(2\pi\right)^4} \left[-i ~ \mathcal{I}_{dpn} ~g_d\,\theta\Big(q_{max}-\Big|\frac{\vec p_{d}}{2}-\vec q\,\Big|\Big)\right]\left(-\mathcal{I}_{\Delta\pi N}\frac{f^*}{m_\pi}\vec S_1\cdot\vec q^{\,\prime} \right) \left(-ig_{\eta\Delta\Delta^*}\right) \\ \nonumber
&\times\left( g_{\gamma p\Delta^*} \vec S^\dagger_1\cdot\vec \epsilon \right)\,\left(\mathcal{I}^\prime_{\Delta\pi N}\mathcal{I}^{\prime\prime}_{\Delta\pi N}\left[\frac{f^*}{m_\pi}\right]^2\vec S_2\cdot\vec p_\pi\vec S_2^{\,\dagger}\cdot\vec q^{\,\prime} \right) \Biggr\{-i\,\mathcal{I}^\prime_{dpn}\,g_d \Biggl.\\\nonumber
&\Biggl.\times\theta\Big(q_{max}-\Big|\frac{-\vec p_\eta+\vec p_\pi+\vec p_d+\vec k}{2}-\vec q-\vec q^{\,\prime} \Big|\Big)\Biggr\} \frac{M_N}{E_N(\vec q\,)}\frac{M_N}{E_N(\vec p_d -\vec q)} ~\frac{M_{\Delta^*}}{E_{\Delta^*}(\vec p_d -\vec q+\vec k)} \\\nonumber 
&\times \frac{M_\Delta}{E_\Delta(\vec p_d -\vec q+\vec k-\vec p_\eta)} \frac{1}{2\omega_\pi(\vec q^{\,\prime})}\frac{M_\Delta}{E_\Delta( \vec q + \vec q^{\,\prime})} \frac{M_N}{E_N( \vec q + \vec q^{\,\prime}-\vec p_\pi)} \\ \nonumber
&\times \frac{M_N}{E_N(\vec p_d -\vec q+\vec k-\vec p_\eta-\vec q^{\,\prime})}\frac{i}{q^0-E_N(\vec q\,)+i\epsilon}~\frac{i}{p_d^0-q^0-E_N(\vec p_d -\vec q)+i\epsilon} \\\nonumber
&\times \frac{i}{p_d^0-q^0+k^0-E_{\Delta^*}(\vec p_d -\vec q+\vec k)+i\epsilon}~\frac{i}{q^{\prime 0}-\omega_\pi(\vec q^{\,\prime})+i\epsilon}\\ \nonumber 
&\times  \frac{i}{p_d^0-q^0+k^0-p_\eta^0-E_\Delta(\vec p_d -\vec q+\vec k-\vec p_\eta)+i\epsilon}\\ \nonumber
&\times  \frac{i}{p_d^0-q^0+k^0-p_\eta^0-q^{\prime 0}-E_N(\vec p_d -\vec q+\vec k-\vec p_\eta - \vec q^{\,\prime})+i\epsilon}\\
&\times
\frac{i}{q^0+q^{\prime 0}-E_\Delta(\vec q + \vec q^{\,\prime})+i\epsilon}
\frac{i}{q^0+q^{\prime 0}-p^0_\pi -E_N(\vec q + \vec q^{\,\prime}-\vec p_\pi)+i\epsilon},\label{tpi_p_res}
\end{align}
where $\mathcal{I}_{dpn}$ $\left(\mathcal{I}^\prime_{dpn}\right)$ denote the isospin coefficient for the initial (final) $d p n$ vertex, while $\mathcal{I}_{\Delta\pi N}$, $\mathcal{I}^\prime_{\Delta\pi N}$ and $\mathcal{I}^{\prime\prime}_{\Delta\pi N}$ represent the same for the first $\Delta\to \pi N$ vertex, the $\pi N \to \Delta$ transition vertex involving the spectator nucleon and for the vertex of de-excitation of $\Delta$ leading to the on-shell $\pi^0$ emission, i.e., $\Delta\to \pi^0 N$, respectively.  The subscript ``1" on the spin operator $\vec S$ signifies its action on the 
upper (lower) nucleon, while the operator with subscript ``2" acts on the lower (upper) nucleon in Figs.~\ref{Fig6}i-\ref{Fig6}l (in Figs.~\ref{Fig6}m-\ref{Fig6}p).

Considering the values of the isospin coefficients given in Table~\ref{Tab:2}, integrating over $q^0$, $q^{\prime\,0}$, and following the procedure given by Eq.~(\ref{wavesub}), we obtain
{\small \begin{align}\nonumber
t^\text{total}_{\pi \text{res},l=1}&=\frac{2}{3}\sqrt{\frac{2}{3}} g_{\eta\Delta\Delta^*}g_{\gamma p\Delta^*}\left(\frac{f^*}{m_\pi}\right)^{3}\int d^3 q\int \frac{d^3 q^\prime}{\left(2\pi\right)^3} \psi\Big(\frac{-\vec p_\eta+\vec p_\pi+\vec p_d+\vec k}{2}-\vec q-\vec q^{\,\prime}\Big)   \\ \nonumber
&\times\psi\Big(\frac{\vec p_{d}}{2}-\vec q\Big) ~\vec S_1\cdot\vec q^{\,\prime}\vec S^\dagger_1\cdot\vec \epsilon~\vec S_2\cdot\vec p_\pi \vec S_2^\dagger\cdot\vec q^{\,\prime} \frac{M_{\Delta^*}}{E_{\Delta^*}(\vec p_d -\vec q+\vec k)}  \frac{M_\Delta}{E_\Delta(\vec p_d -\vec q+\vec k-\vec p_\eta)}  \\ \nonumber
&\times\frac{1}{2\omega_\pi(\vec q^{\,\prime})} \frac{M_\Delta}{E_\Delta( \vec q + \vec q^{\,\prime})}\frac{1}{p_d^0+k^0-E_N(\vec q\,)- p_\eta^0- E_\Delta(\vec p_d -\vec q+\vec k-\vec p_\eta)+i\epsilon} \\\nonumber
&\times \frac{1}{p_d^0+k^0-p_\eta^0-E_N(\vec q+\vec q^{\,\prime}+\vec p_\eta)-E_\Delta( \vec q + \vec q^{\,\prime})+i\epsilon}\\\nonumber
&\times\frac{1}{p_d^0+k^0-p_\eta^0-E_N(\vec q\,)-E_N(\vec p_d -\vec q+\vec k-\vec p_\eta - \vec q^{\,\prime})-\omega_\pi(\vec q^{\,\prime})+i\epsilon}\\
&\times \frac{1}{p_d^0+k^0-E_N(\vec q\,)-E_{\Delta^*}(\vec p_d -\vec q+\vec k)+i\epsilon}.\label{tpi_p_res2}
\end{align}}
\subsubsection{Additional pion rescattering mechanism}
So far we have considered $\gamma N\to\Delta^*(1700)\to\eta\Delta(1232)\to \pi\eta N$ as the primary mechanism to describe the $\gamma d\to\pi^0\eta d$ reaction. There could be, however, other intermediate processes which can produce a pion first (instead of an $\eta$), that could further rescatter with the spectator nucleon of the deuteron and lead to the final $\pi^0\eta d$ state. Indeed, in Refs.~\cite{GomezTejedor:1993bq,GomezTejedor:1995pe}, the $\gamma N\to \pi\pi N$ reaction was studied considering, among others, contributions from the production of the $\Delta(1232)$, $N^*(1440)$ and $N^*(1520)$ resonances, and the data on the cross sections were described fairly well. Following these former works, we could also consider, for example, a process like $\gamma N\to N^*(1440)\to \Delta(1232)\pi\to\pi\pi N$ and one of the pions in the final state could rescatter with one of the nucleons of the deuteron. However, it was shown in Refs.~\cite{GomezTejedor:1993bq,GomezTejedor:1995pe}, that for energies of the photon $E_\gamma\gtrsim800$ MeV, contributions to the cross section of $\gamma N\to\pi\pi N$ from the production of $N^*(1440)$ or $N^*(1520)$ are small and the dominant contribution comes from the processes $\gamma p\to\pi^+\Delta^0\to \pi^+\pi^-p,\pi^+\pi^0 n$, $\gamma n\to \pi^-\Delta ^+\to\pi^-\pi^0p,\pi^-\pi^+n$, in which $\Delta(1232)$ is excited. The latter reactions, as shown in Refs.~\cite{GomezTejedor:1993bq,GomezTejedor:1995pe}, involve a kind of Kroll-Ruderman vertex for the $\gamma N\to \pi \Delta(1232)$ reaction. Such Kroll-Ruderman vertices are obtained by demanding gauge invariance of the amplitudes. At the end, this Kroll-Ruderman term is found largely dominant and we rely upon this term.

In this way, we consider the additional diagrams for pion rescattering shown in Fig.~\ref{Fig8}.
\begin{figure}
\centering
\includegraphics[width=0.55\textwidth]{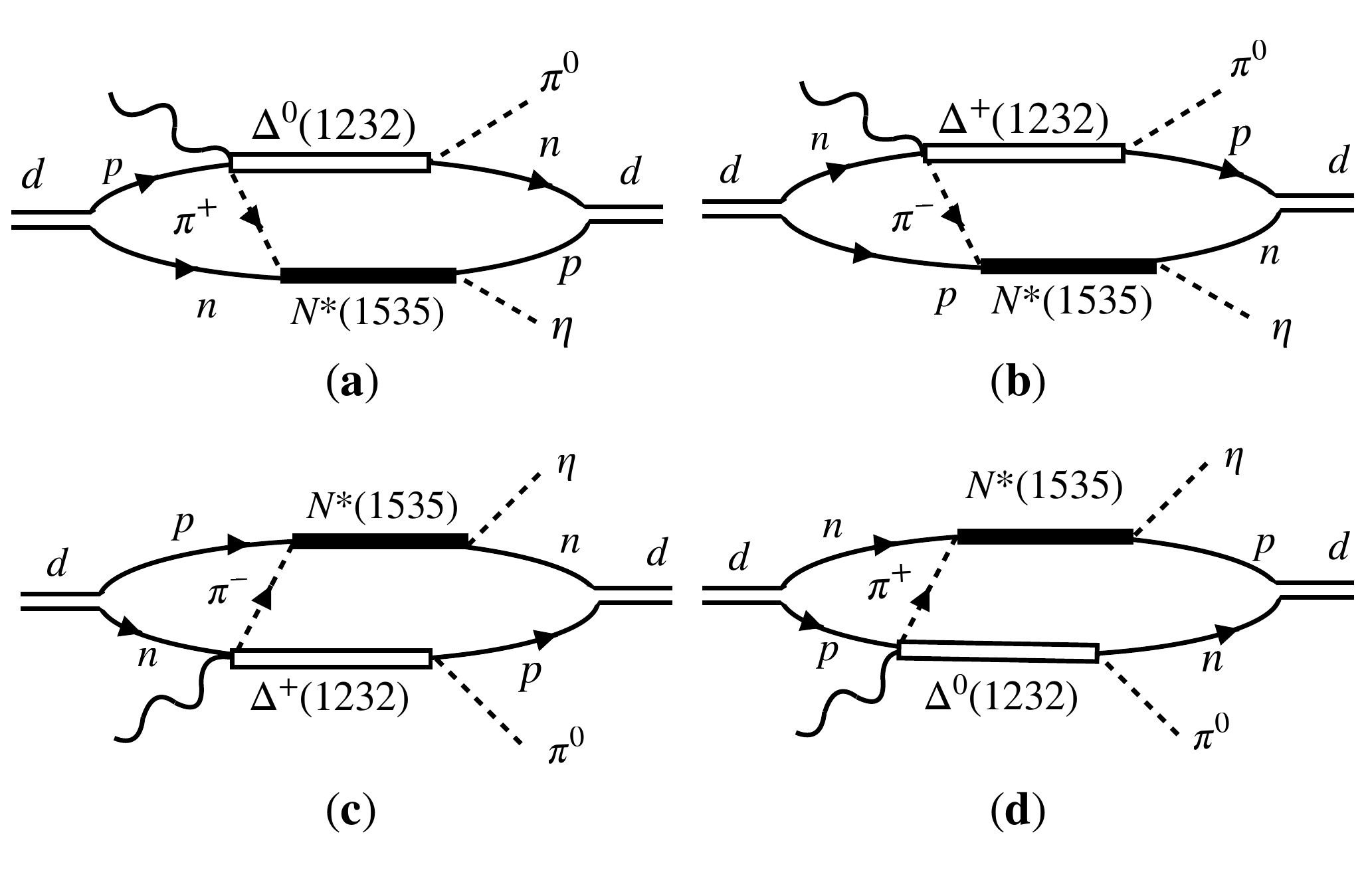}
\caption{Additional diagrams for the $\gamma d\to \pi^0\eta d$ process involving the rescattering of a pion.}\label{Fig8}
\end{figure}
Following Refs.~\cite{GomezTejedor:1993bq,GomezTejedor:1995pe}, the $\gamma N\to\pi N$ Kroll-Ruderman type of vertex is described by the amplitude
\begin{align}
-it^\text{KR}_{\gamma N\to\pi\Delta}=\alpha \vec{S}^\dagger\cdot\vec{\epsilon},
\end{align}
where $\alpha$ is a constant determined from the aforementioned gauge invariance condition. In this way, the amplitudes for the $\gamma N\to \pi \pi N$ processes shown in Fig.~\ref{Fig8} are
\begin{align}
-it_{\gamma p\to\pi^+\pi^0 n}&=\mathcal{I}_{\Delta^0\pi^- p}\frac{f^*}{m_\pi}e\vec{S}\cdot\vec{\epsilon},\nonumber\\
-it_{\gamma n\to\pi^-\pi^0 p}&=-\mathcal{I}_{\Delta^+\pi^+ n}\frac{f^*}{m_\pi}e\vec{S}\cdot\vec{\epsilon},
\end{align}
where $\mathcal{I}_{\Delta^{0(+)}\pi^{-(+)} N^{+(0)}}$ represents the corresponding isospin coefficients for  $\Delta^{0(+)}\to\pi^{-(+)} N^{+(0)}$ (with $N^+\equiv p$, $N^0\equiv n$) given in Table~\ref{Tab:2}, and $e=\sqrt{4\pi\alpha}$, with $\alpha\simeq 1/137$. Using the preceding expressions, we can now evaluate the contributions from the diagrams shown in Fig.~\ref{Fig8}. For example, in case of the diagram in Fig.~\ref{Fig8}(a), using the momenta assignment shown in Fig.~\ref{Fig9},
\begin{figure}
\centering
\includegraphics[width=0.55\textwidth]{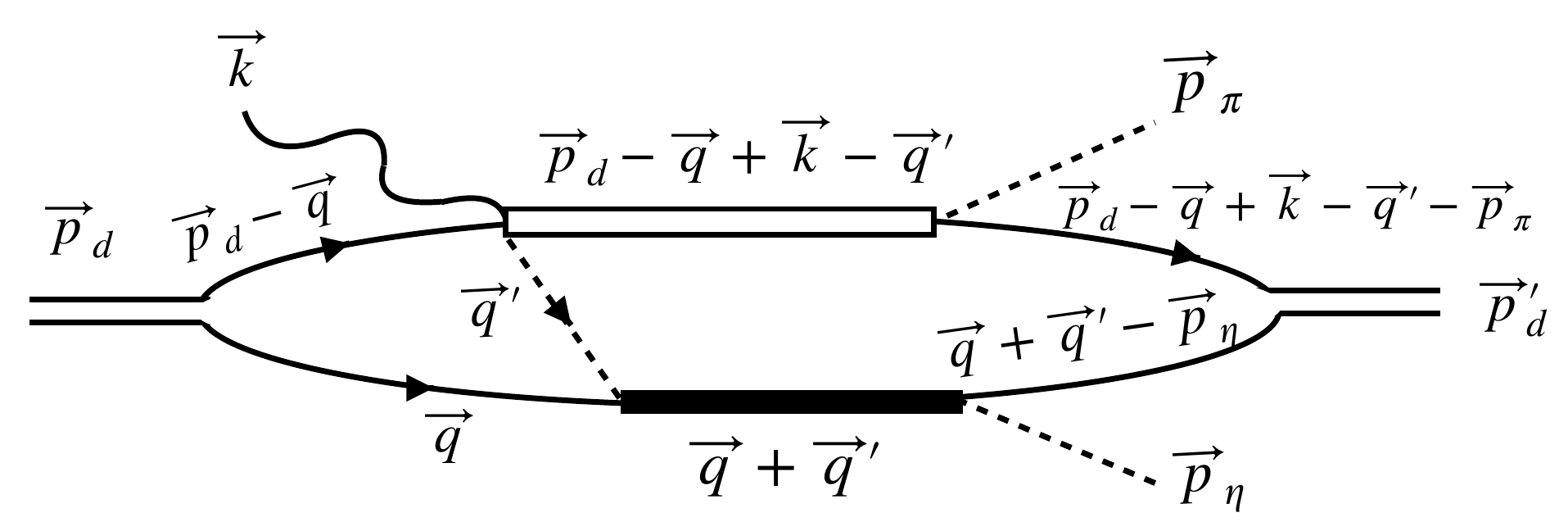}
\caption{Momenta labels associated with the particles of the diagram in Fig.~\ref{Fig8}(a).}\label{Fig9}
\end{figure}
we have 
\begin{align}
t^\text{KR}_{\pi\text{res},l=0}&=\mathcal{I}_{dpn}\mathcal{I}^\prime_{dpn}\mathcal{I}_{\Delta^0\pi^0 n}\mathcal{I}_{\Delta^0\pi^- p}\left(\frac{f^*}{m_\pi}\right)^2e\,g^2_dm^4_N m_\Delta g_{\eta N N^*}g_{\pi^+n N^{*+}}\nonumber\\
&\quad\times\int\frac{d^4q}{(2\pi)^4}\int\frac{d^4 q^{\,\prime}}{(2\pi)^4}\frac{1}{E_N(\vec{p}_d-\vec{q}+\vec{k}-\vec{q}^{\,\prime}-\vec{p}_\pi)}\frac{1}{E_N(\vec{q}+\vec{q}^{\,\prime}-\vec{p}_\eta)}\nonumber\\
&\quad\times\frac{1}{p^0_d-q^0+k^0-q^{\,\prime\, 0}-p^0_\pi-E_N(\vec{p}_d-\vec{q}+\vec{k}-\vec{q}^{\,\prime}-\vec{p}_\pi)+i\epsilon}\nonumber\\
&\quad\times\frac{1}{q^0+q^{\,\prime\,0}-p^0_\eta-E_N(\vec{q}+\vec{q}^{\,\prime}-\vec{p}_\eta)+i\epsilon}(\vec{S}\cdot\vec{p}_\pi)(\vec{S}^\dagger\cdot\vec{\epsilon})\nonumber\\
&\quad\times\frac{1}{q^0+q^{\prime0}-E_{N^*}(\vec{q}+\vec{q}^{\,\prime})+i\epsilon}\frac{1}{E_\Delta(\vec{p}_d-\vec{q}+\vec{k}-\vec{q}^{\,\prime})}\nonumber\\
&\quad\times\frac{1}{p^0_d-q^0+k^0-q^{\,\prime\,0}-E_\Delta(\vec{p}_d-\vec{q}+\vec{k}-\vec{q}^{\,\prime})+i\epsilon}\frac{1}{E_\pi(\vec{q}^{\,\prime})}\frac{1}{q^{\,\prime\,0}-E_\pi(\vec{q}^{\,\prime})+i\epsilon}\frac{1}{E_N(\vec{p}_d-\vec{q})}\nonumber\\
&\quad\times\frac{1}{p^0_d-q^0-E_N(\vec{p}_d-\vec{q})+i\epsilon}\frac{1}{E_N(\vec{q}\,)}\frac{1}{q^0-E_N(\vec{q}\,)+i\epsilon}\nonumber\\
&\quad\times\theta\Big(q_\text{max}-\Big|\frac{\vec{p}_d}{2}-\vec{q}\,\Big|\Big)\theta\Big(q_\text{max}-\Big|\frac{\vec{p}_d+\vec{k}-\vec{p}_\pi+\vec{p}_\eta}{2}-\vec{q}-\vec{q}^{\,\prime}\Big|\Big).\label{tKR}
\end{align}
In Eq.~(\ref{tKR}), $g_{\eta N N^*}=g_{\eta p N^{*+}}=g_{\eta n N^{*0}}$ represents the coupling of $N^{*}(1535)$ to $\eta N$ and $g_{N^{*+}\pi^+n}$ is the coupling of $N^{*+}(1535)$ to $\pi^+n$. The corresponding values are obtained with the model of Ref.~\cite{Inoue:2001ip}, in which $N^{*}(1535)$ is generated from the pseudoscalar meson-baryon interaction, and are:
\begin{align}
g_{\eta N N^*}=1.46-i0.43,\quad g_{\pi^+nN^{*+}}=-0.47-i0.27.\label{NstaretaN}
\end{align}
It should be mentioned that the $M_{N^*}/E_{N^*}$ factor related to the $N^*$ propagator is included in the couplings, which is consistent with the  Breit-Wigner parametrization of the $\eta N$ and $\pi N$ amplitudes  in Ref.~\cite{Inoue:2001ip}. For instance,
\begin{align}
t_{\eta N}=\frac{g^2_{\eta N N^*(1535)}}{E_{\eta N}-M_{N^*}+i\Gamma_{N^*}/2},
\end{align}
with $E_{\eta N}$ representing the total energy of the $\eta N$ system.

Next, we can now integrate on the $q^0$ and $q^{\,\prime\, 0}$ variables appearing in Eq.~(\ref{tKR}) using Cauchy's theorem and consider Eq.~(\ref{wavesub}) to introduce the deuteron wave functions. We can repeat the same procedure for all the diagrams shown in Fig.~(\ref{Fig8}) and sum the contributions to get the following expression in the $\gamma d$ rest frame,
\begin{align}
t^{\text{KR},\text{total}}_{\pi\text{res},l=0}&=-e\frac{\sqrt{2}}{3}\left(\frac{f^*}{m_\pi}\right)^2g_{N^{*+}\to\pi^+\eta}g_{N^{*+}\to\eta p}M_\Delta\int\frac{d^3q}{(2\pi)^3}\int d^3 q^\prime\frac{1}{E_\pi(\vec{q}^{\,\prime})}\frac{1}{E_\Delta(\vec{q}+\vec{q}^{\,\prime}}\frac{1}{E_{N^*}(\vec{q}+\vec{q}^{\,\prime})}\nonumber\\
&\quad\times\Bigg[\frac{1}{\sqrt{s}-E_\Delta(\vec{q}+\vec{q}^{\,\prime})-E_N(\vec{q}\,)-E_\pi(\vec{q}^{\,\prime})+i\epsilon}\frac{1}{\sqrt{s}-E_\Delta(\vec{q}+\vec{q}^{\,\prime})-E_{N^*}(\vec{q}+\vec{q}^{\,\prime})+i\frac{\Gamma_\Delta}{2}+i\epsilon}\nonumber\\
&\quad\times\frac{1}{\sqrt{s}-E_\Delta(\vec{q}+\vec{q}^{\,\prime})-p^0_\eta-E_N(\vec{q}+\vec{q}^{\,\prime}-\vec{p}_\eta)+i\epsilon}\nonumber\\
&\quad+\frac{1}{\sqrt{s}-p^0_\pi-E_N(\vec{q}+\vec{q}^{\,\prime}+\vec{p}_\pi)-E_{N^*}(\vec{q}+\vec{q}^{\,\prime})+i\epsilon}\nonumber\\
&\quad\times\Bigg(\frac{1}{\sqrt{s}-p^0_\pi-E_N(\vec{q}+\vec{q}^{\,\prime}+\vec{p}_\pi)-E_N(\vec{q}\,)-E_\pi(\vec{q}^{\,\prime})+i\epsilon}\nonumber\\
&\quad\times\frac{1}{\sqrt{s}-E_N(\vec{q}\,)-E_\pi(\vec{q}^{\,\prime})-E_\Delta(\vec{q}+\vec{q}^{\,\prime})+i\epsilon}+\frac{1}{\sqrt{s}-E_\Delta(\vec{q}+\vec{q}^{\,\prime})-E_N(\vec{q}\,)-E_\pi(\vec{q}^{\,\prime})+i\epsilon}\nonumber\\
&\quad\times\frac{1}{\sqrt{s}-E_\Delta(\vec{q}+\vec{q}^{\,\prime})-E_{N^*}(\vec{q}+\vec{q}^{\,\prime})+i\epsilon}\Bigg)\Bigg](\vec{S}\cdot\vec{p}_\pi)(\vec{S}^\dagger\cdot\vec{\epsilon}).
\end{align}
\subsubsection{Eta rescattering}
Having considered pion rescattering through s- and $p$-wave $\pi N$ interactions, 
\begin{figure}[h!]
\includegraphics[width=0.55\textwidth]{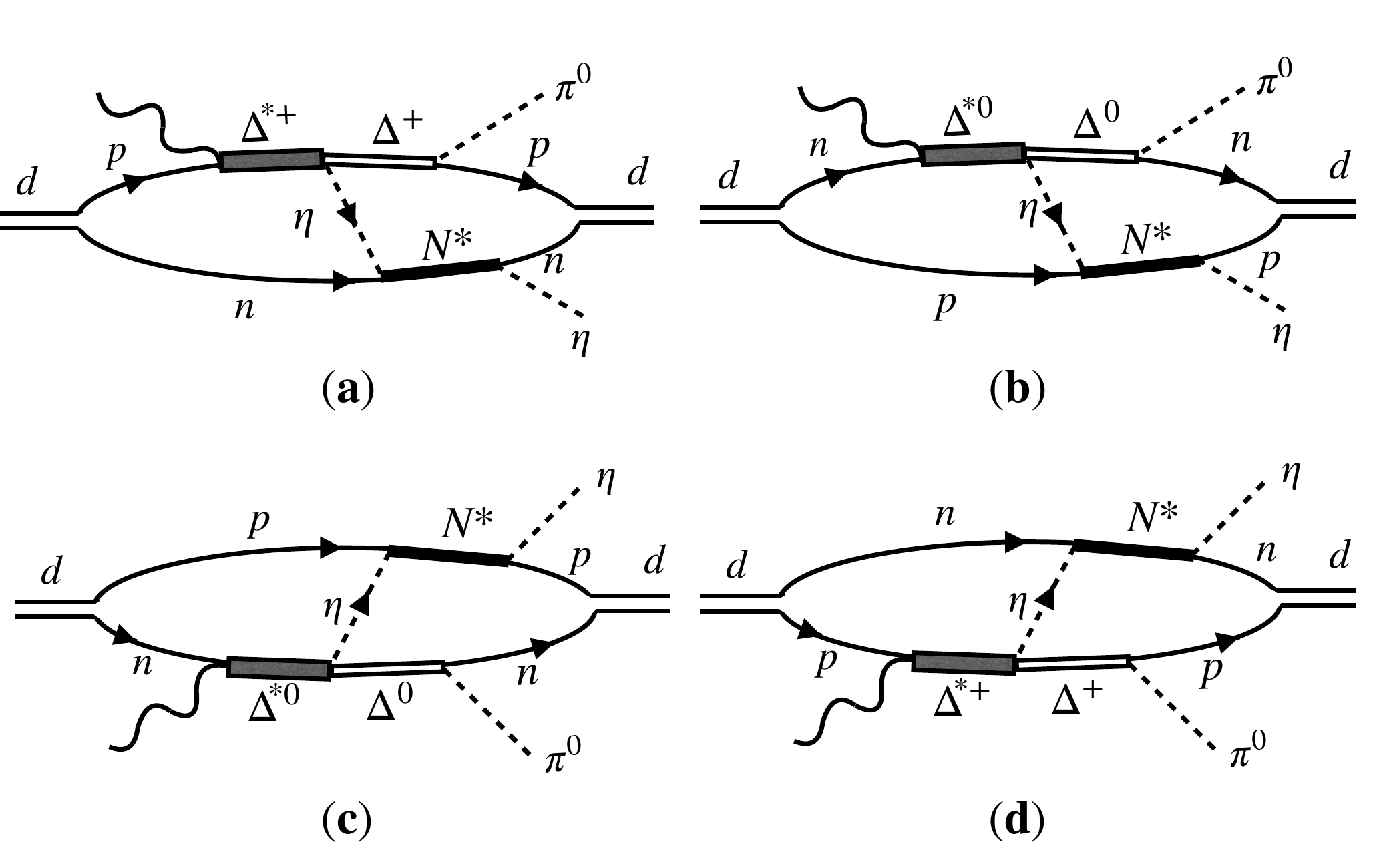}
\caption{Different diagrams contributing to the rescattering of $\eta$ in the intermediate state.}\label{Fig10}
\end{figure}
it is important to assess the possible contributions from the rescattering of $\eta$ too. We consider the rescattering of $\eta$ through the mechanisms shown in Fig.~\ref{Fig10}. It is well known that  the $\eta N$ interaction is attractive in the $s$-wave and is related to the formation of $N^*(1535)$. Indeed, as shown in Ref.~\cite{Inoue:2001ip} the $\eta N$ channel has a large coupling to this $S_{11}$ resonance [see Eq.~(\ref{NstaretaN})].

We  follow Ref.~\cite{Inoue:2001ip} to account for the $\eta N\leftrightarrow N^*(1535)$ vertices while writing the amplitudes for the diagrams shown in Fig.~\ref{Fig10}.
To do this, we label different lines with  momenta as shown in Fig.~\ref{Fig11}.
\begin{figure}
\includegraphics[width=0.55\textwidth]{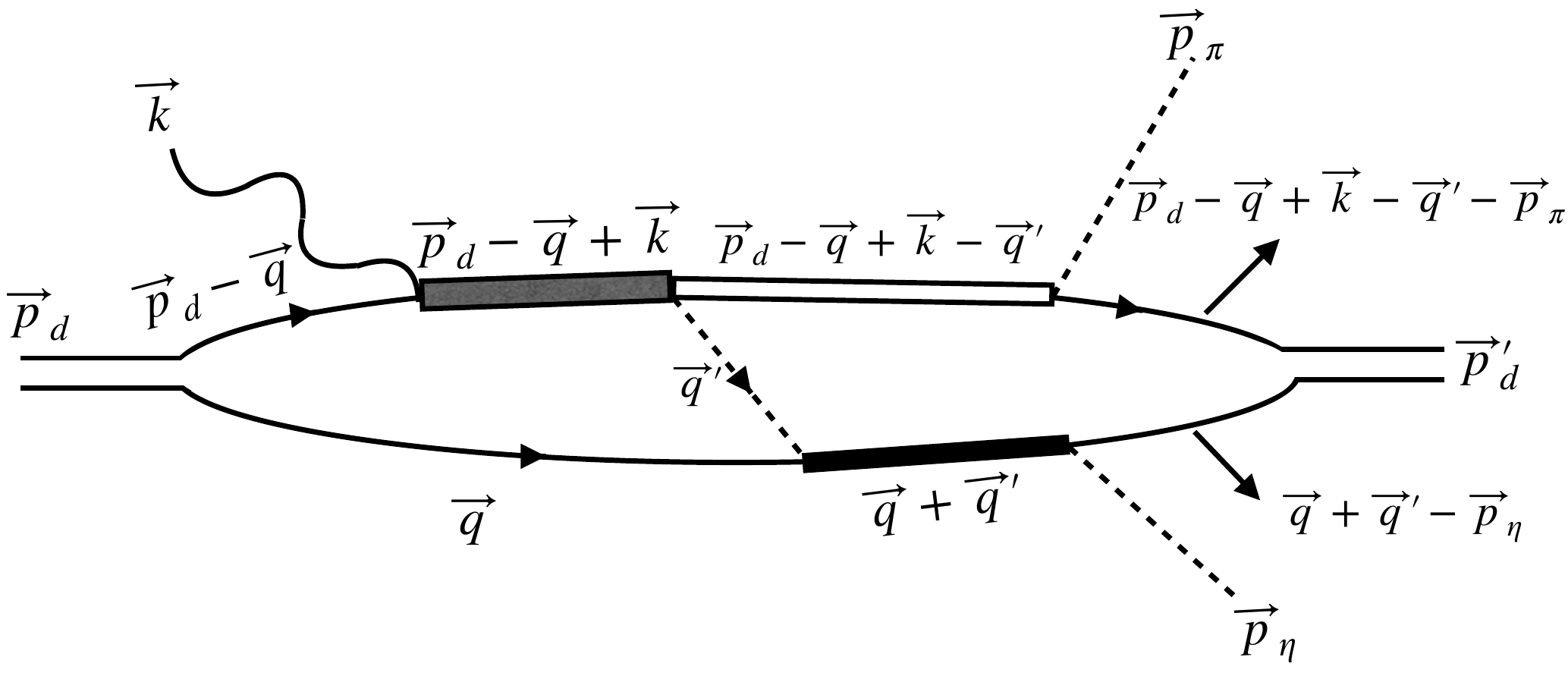}
\caption{Momenta associated with the different particles shown in Fig.~\ref{Fig10}a.}\label{Fig11}
\end{figure}
The amplitudes for all the diagrams shown in Fig.~\ref{Fig10} have a common structure and can be written as
\begin{align}\nonumber
-it_{\eta res,l=0}&=\int\!\!\frac{d^4 q}{\left(2\pi\right)^4}\int\,\frac{d^4 q^\prime}{\left(2\pi\right)^4} \left[-i\mathcal{I}_{dpn}g_d\,\theta\Big(q_{max}-\Big|\frac{\vec p_{d}}{2}-\vec q\,\Big|\Big)\right]\left(-\mathcal{I}_{\Delta\pi N}\frac{f^*}{m_\pi}\vec S\cdot\vec p_\pi\right) \left(-ig_{\eta\Delta\Delta^*}\right) \\ \nonumber
&\times\left( g_{\gamma p\Delta^*} \vec S^\dagger\cdot\vec \epsilon \right) (-ig_{\eta NN^*(1535)})^2 \left[-i\mathcal{I}^\prime_{dpn}g_d\,\theta\Big(q_{max}-\Big|\frac{-\vec p_\eta+\vec p_\pi+\vec p_d+\vec k}{2}-\vec q-\vec q^{\,\prime} \Big|\Big)\right] \\\nonumber 
&\times\frac{M_N}{E_N(\vec q\,)}\frac{i}{q^0-E_N(\vec q\,)+i\epsilon} \frac{M_N}{E_N(\vec p_d -\vec q)}\frac{i}{p_d^0-q^0-E_N(\vec p_d -\vec q)+i\epsilon}\frac{M_{\Delta^*}}{E_{\Delta^*}(\vec p_d -\vec q+\vec k)}\\ \nonumber
&\times \frac{i}{p_d^0-q^0+k^0-E_{\Delta^*}(\vec p_d -\vec q+\vec k)+i\epsilon}\frac{M_\Delta}{E_\Delta(\vec p_d -\vec q+\vec k-\vec q^{\,\prime})}\\ \nonumber 
&\times \frac{i}{p_d^0-q^0+k^0-q^{\prime 0}-E_\Delta(\vec p_d -\vec q+\vec k-\vec q^{\,\prime})+i\epsilon} \frac{1}{2\omega_\eta(\vec q^{\,\prime})}\frac{i}{q^{\prime 0}-\omega_\eta(\vec q^{\,\prime})+i\epsilon}\\ \nonumber
&\times \frac{M_N}{E_N(\vec p_d -\vec q+\vec k-\vec p_\pi-\vec q^{\,\prime})} \frac{i}{p_d^0-q^0+k^0-p_\pi^0-q^{\prime 0}-E_N(\vec p_d -\vec q+\vec k-\vec p_\pi - \vec q^{\,\prime})+i\epsilon},\\
&\times
\frac{i}{q^0+q^{\prime 0}-E_{N^*}( \vec q + \vec q^{\,\prime})+i\epsilon}
\frac{M_N}{E_N( \vec q + \vec q^{\,\prime}-\vec p_\eta)} \frac{i}{q^0+q^{\prime 0}-p^0_\eta -E_N( \vec q + \vec q^{\,\prime}-\vec p_\eta)+i\epsilon}.\label{teta_s_res}
\end{align}

Using the values of the isospin coefficients given in Table~\ref{Tab:2}, integrating over $q^0$ and $q^{\prime 0}$, and following the procedure consistent with Eq~(\ref{wavesub}), we obtain
\begin{align}\nonumber
t^\text{total}_{\eta \text{res},l=0}&=2\sqrt{\frac{2}{3}}\frac{f^*}{m_\pi}g_{\eta\Delta\Delta^*}g_{\gamma p\Delta^*}g^2_{\eta N N^*}\int d^3 q\int \frac{d^3q^\prime}{\left(2\pi\right)^3}
\psi\Big(\frac{\vec p_{d}}{2}-\vec q\Big)\psi\Big(\frac{\vec p_\eta-\vec p_\pi+\vec p_d+\vec k}{2}-\vec q-\vec q^{\,\prime}\Big) \\\nonumber
&\times\vec S\cdot\vec p_\pi \vec S^\dagger\cdot\vec \epsilon  \frac{M_\Delta}{E_\Delta( \vec p_d +\vec k-\vec q - \vec q^{\,\prime})}\frac{M_{\Delta^*}}{E_{\Delta^*}(\vec p_d -\vec q+\vec k)} \frac{1}{2\omega_\eta(\vec q^{\,\prime})}\\\nonumber
&\times\frac{\mathcal{N}}{p_d^0+k^0-E_N(\vec q\,)- E_{\Delta^*}(\vec p_d -\vec q+\vec k)+i\epsilon}\\\nonumber
&\times  \frac{1}{p_d^0+k^0-E_{N^*}(\vec q+\vec q^{\,\prime})- E_\Delta(\vec p_d +\vec k-\vec q - \vec q^{\,\prime})+i\epsilon}\\\nonumber
&\times \frac{1}{p_d^0+k^0- E_\Delta(\vec p_d +\vec k-\vec q - \vec q^{\,\prime})-E_N(\vec q\,)-E_\eta(\vec q^{\,\prime})+i\epsilon}\\\nonumber
&\times  \frac{1}{p_d^0+k^0- p_\eta^0-E_\Delta(\vec p_d +\vec k-\vec q - \vec q^{\,\prime})-E_N(\vec q+\vec q^{\,\prime}-\vec p_\eta)+i\epsilon}\\\nonumber
&\times\frac{1}{p_d^0+k^0- p_\pi^0-E_{N^*}(\vec q+\vec q^{\,\prime})-E_N(\vec p_d+\vec k-\vec q-\vec q^{\,\prime}-\vec p_\pi)+i\epsilon}\\
&\times\frac{1}{p_d^0+k^0- p_\pi^0-E_\eta(\vec q^{\,\prime})-E_N(\vec p_d+\vec k-\vec q-\vec q^{\,\prime}-\vec p_\pi)-E_N(\vec q\,)+i\epsilon},\label{teta_s_res2}
\end{align}
where the expression for $\mathcal{N}$ is
{\small\begin{align}\nonumber
&\mathcal{N}=\left[E_{N^*}(\vec q+\vec q^{\,\prime})+E_N(\vec p_d+\vec k-\vec q-\vec q^{\,\prime}-\vec p_\pi)\right]E_N(\vec p_d+\vec k-\vec q-\vec q^{\,\prime}-\vec p_\pi)\\\nonumber
&+\left[E_\Delta(\vec p_d+\vec k-\vec q-\vec q^{\,\prime})+E_\eta(\vec q^{\,\prime})\right]\left[E_{N^*}(\vec q+\vec q^{\,\prime})+E_N(\vec p_d+\vec k-\vec q-\vec q^{\,\prime}-\vec p_\pi)+E_\Delta(\vec p_d+\vec k-\vec q-\vec q^{\,\prime})\right]\\\nonumber
&+p_\eta^0\left[E_{N^*}(\vec q+\vec q^{\,\prime})+E_N(\vec p_d+\vec k-\vec q-\vec q^{\,\prime}-\vec p_\pi)+E_\Delta(\vec p_d+\vec k-\vec q-\vec q^{\,\prime})+E_\eta(\vec q^{\,\prime})\right]\\\nonumber
&+p_\pi^0\left[E_{N^*}(\vec q+\vec q^{\,\prime})+2E_N(\vec p_d+\vec k-\vec q-\vec q^{\,\prime}-\vec p_\pi)\!+\!E_\Delta(\vec p_d+\vec k-\vec q-\vec q^{\,\prime})+E_\eta(\vec q^{\,\prime})+p_\eta^0+p_\pi^0\right]\\\nonumber
&+E_N(\vec q+\vec q^{\,\prime}-\vec p_\eta)\left[E_N(\vec q\,)+E_{N^*}(\vec q+\vec q^{\,\prime})+E_N(\vec p_d+\vec k-\vec q-\vec q^{\,\prime}-\vec p_\pi)+E_\Delta(\vec p_d+\vec k-\vec q-\vec q^{\,\prime})\right.\\\nonumber
&\left.+E_\eta(\vec q^{\,\prime})+p_\pi^0-2p_d^0-2k^0\right]+3\left(p_d^0+k^0\right)^2+E_N(\vec q\,)\left[E_{N^*}(\vec q+\vec q^{\,\prime})+E_N(\vec p_d+\vec k-\vec q-\vec q^{\,\prime}-\vec p_\pi)\right.\\\nonumber
&\left.+E_\Delta(\vec p_d+\vec k-\vec q-\vec q^{\,\prime})+p_\eta^0+p_\pi^0-2p_d^0-2k^0\right] -\left[2E_{N^*}(\vec q+\vec q^{\,\prime})+3E_N(\vec p_d+\vec k-\vec q-\vec q^{\,\prime}-\vec p_\pi)\right.\\
&\left.+3E_\Delta(\vec p_d+\vec k-\vec q-\vec q^{\,\prime})+2\left(E_\eta(q^\prime)+p_\eta^0\right)+3p_\pi^0\right]\left(p_d^0+k^0\right).
\end{align}}

Finally, as mentioned before, the unstable nature of states like $\Delta^*(1700)$, $\Delta(1232)$ and $N^*(1535)$ is taken into account by replacing $E_R-i\epsilon$ by $E_R-i\Gamma_R/2$ in the different amplitudes, where $R$ stands for a resonance. In the case of $\Delta(1232)$, we consider an energy dependent width
\begin{align}
\Gamma_\Delta\left(M_{\Delta inv}\right)=\Gamma_\Delta \frac{M_\Delta}{M_{\Delta inv}}\left(\frac{q_\pi}{q_{\pi on}}\right)^3,\label{deltawidth}
\end{align}
where,
\begin{align}
M_{\Delta inv}^2= E_\Delta^2-|\vec p_\Delta |^2.
\end{align}
For example, for the impulse approximation,  we can determine $M_{\Delta inv}$, using the kinematic labels shown in Fig.~\ref{Fig5}, as
\begin{align}
M_{\Delta inv}^2= \left(p_d^0+k^0-E_N(\vec q)-p^0_\eta\right)^2-\left(\vec p_d+\vec k-\vec q- \vec p_\eta\,\right)^2.
\end{align}
Further, $q_\pi$ and $q_{\pi on}$, in Eq.~(\ref{deltawidth}), are defined as
\begin{align}\nonumber
q_\pi&=\frac{\lambda^{1/2}\left(M_{\Delta inv}^2, M_N^2, m_\pi^2\right)}{2 M_{\Delta inv}},\\\nonumber
q_{\pi on}&=\frac{\lambda^{1/2}\left(M_{\Delta}^2, M_N^2, m_\pi^2\right)}{2 M_{\Delta}}.
\end{align}

\section{Results and discussions}
With the amplitudes discussed in the previous section we calculate the invariant mass distributions for $\eta d$ and $\pi^0 d$ in the final state as
\begin{align}
\frac{d\sigma}{dM_{\eta d}}&=\frac{M_d^2}{8 \big|\vec k\big| s}\frac{1}{\left(2\pi\right)^4} \big|\vec p_\pi\big|\big|\vec p^{\,R\eta d}_\eta\big|\int d \text{cos}\theta_\pi \int d\Omega^{R\eta d}_\eta \overline {\sum\limits_{\mu,\lambda}} \sum_{\mu^\prime} \big| t^\lambda_{\mu,\mu^\prime}\big|^2,\label{invmass1}\\
\frac{d\sigma}{dM_{\pi^0 d}}&=\frac{M_d^2}{8 \big|\vec k\big| s}\frac{1}{\left(2\pi\right)^4} \big|\vec p_\eta\big|\big|\vec p^{\,R\pi d}_\pi\big|\int d \text{cos}\theta_\eta \int d\Omega^{R\pi d}_\pi \overline {\sum\limits_{\mu,\lambda}} \sum_{\mu^\prime} \big| t^\lambda_{\mu,\mu^\prime}\big|^2,\label{invmass2}
\end{align}
where $\vec k$ is the momentum of the photon, $s$ is the standard Mandelstam variable, $\vec p_\pi \left(\vec p_\eta\right)$ is the pion (eta) momentum in the global center of mass frame, and $\vec p^{\,R\eta d}_\eta \left(\vec p^{\,R\pi d}_\pi\right)$ denotes the eta (pion) momentum in the rest frame of $\eta d \left(\pi d\right)$.
\begin{align}
&\big| \vec p_\pi\big|=\frac{\lambda^{1/2}\left(s, m_\pi^2, M_{\eta d}^2\right)}{2\sqrt{s}},\hspace{1.3cm}\big| \vec p_\eta\big|=\frac{\lambda^{1/2}\left(s, m_\eta^2, M_{\pi^0 d}^2\right)}{2\sqrt{s}},\label{phsp1}\\
&\big| p^{\,R\pi d}_\pi\big|=\frac{\lambda^{1/2}\left(M_{\pi^0 d}^2, m_\pi^2, M_d^2\right)}{2M_{\pi^0 d} },~~\big| p^{\,R\eta d}_\eta\big|=\frac{\lambda^{1/2}\left(M_{\eta d}^2, m_\eta^2, M_d^2\right)}{2M_{\eta d}}.\label{phsp2}
\end{align}
The variable $\Omega^{R\eta d}_\eta~\left(\Omega^{R\pi d}_\pi\right)$ in Eq.~(\ref{invmass1}) [Eq.~(\ref{invmass2})] denotes the solid angle of $\eta~\left(\pi\right)$ in the $\eta d \left(\pi d\right)$ rest frame.

The summation signs in Eqs.~(\ref{invmass1}) and (\ref{invmass2}) indicate the sum over the polarizations of the particles in the initial and final states, with the bar over the  sign representing averaging over the initial state polarizations. The subscript in $t^\lambda_{\mu,\mu^\prime}$ indicates the dependence of  the amplitudes on the spin projections of the deuteron in the initial ($\mu$) and final ($\mu^\prime$) states, while the superscript denotes the dependence on the transverse polarization of the photon. The contributions from the different spin transitions for the different amplitudes are summarized in Appendix~\ref{appA}. 

Further, we calculate the amplitudes in the global center of mass frame. Thus, we must boost $\vec p^{\,R\pi d}_\pi$ and $\vec p^{\,R\eta d}_\eta$ to the global center of mass frame. The boosted $\eta$ momentum is 
\begin{align}
\vec p_\eta=\vec p^{\,R\eta d}_\eta+\vec p_\pi^{\,R\eta d}\left[\frac{ \vec p_\pi^{\,R\eta d}\cdot\vec p_\eta^{\,R\eta d}}{|\vec p_\pi^{\,R\eta d}|^2}\left(\frac{E_{\gamma d}^{\,R\eta d}}{\sqrt{s}}-1\right)-\frac{ E_\eta^{\,R\eta d}}{M_{\eta d}}\right],\label{etaboost}
\end{align}
where $E_{\gamma d}^{\,R\eta d}=\sqrt{s+|\vec p_\pi^{\,R\eta d}|^2}$ is the total energy of $\gamma d$ in the $\eta d$ rest frame, $\vec p_\pi^{\,R\eta d}$ is the pion momentum in the $\eta d$ rest frame, which is related to the pion momentum in the global center of mass frame as 
\begin{align}
\vec p_\pi^{\,R\eta d}=\frac{\sqrt{s}}{M_{\eta d}}\vec p_\pi.
\end{align}
The expression for $\vec{p}_\pi$  is analogous to Eq.~(\ref{etaboost}), and can be obtained by interchanging the $\pi$, $\eta$ subscripts  in Eq.~(\ref{etaboost}).

Since we carry out the calculation of the amplitudes in the global center of mass frame, $\vec p_d +\vec k=0$ and $p_d^0+k^0$ is taken as $\sqrt{s}$.

 Further, it can be useful to specify the directions chosen for the different momenta in our formalism. We choose the photon momentum to be parallel to the $z$-axis, such that $\vec k =\left(0,~0,~|\vec k|\right)$. When calculating the $\eta d$ invariant mass distribution, we write
\begin{align}
\vec p_\pi=&|\vec p_\pi|\bigl(\text{ sin}~\theta_\pi,~0,~\text{cos}~\theta_\pi\bigr),\\
\vec p_\eta^{\, R\eta d}=&|\vec p_\eta^{\, R\eta d}|\bigl(\text{ sin}~\theta_\eta^{R\eta d} \text{cos}~\phi_\eta^{R\eta d},~\text{ sin}~\theta_\eta^{R\eta d} \text{sin}~\phi_\eta^{R\eta d},~\text{cos}~\theta_\eta^{R\eta d}\bigr).
\end{align}
For the calculations of  the $\pi d$ invariant mass distribution, we choose
\begin{align}
\vec p_\eta=&|\vec p_\eta|\bigl(\text{ sin}~\theta_\eta,~0,~\text{cos}~\theta_\eta\bigr),\\
\vec p_\pi^{\, R\pi d}=&|\vec p_\eta^{\, R\pi d}|\bigl(\text{ sin}~\theta_\pi^{R\pi d} \text{cos}~\phi_\pi^{R\pi d},~\text{ sin}~\theta_\pi^{R\pi d} \text{sin}~\phi_\pi^{R\pi d},~\text{cos}~\theta_\pi^{R\pi d}\bigr).
\end{align}

We are now in a position to start discussing the results. Before beginning, though, we must remind the reader that the experimental data on $\eta d$ and $\pi^0 d$ invariant mass spectra are presented for two different set of beam energies in Ref.~\cite{Ishikawa:2021yyz}: (1) 950-1010~MeV (2) 1010-1150~MeV. To compare our results with the experimental data, we calculate the $\eta d$ and $\pi^0 d$ mass distributions for  different beam energies in each range and calculate the average of the results obtained. In particular, we consider the energies $E_{\gamma_i}=950$, $980$ and $1010$ MeV for the first energy range and $E_{\gamma_i}=1010$, $1050$, $1100$, and $1150$ MeV for the second energy range. Then, the differential cross sections are determined as
\begin{align}
\frac{d\sigma}{dM_\text{inv}}=\frac{1}{n}\sum\limits_{i=1}^n\frac{d\sigma(E_{\gamma_i})}{dM_\text{inv}},\label{dsigEi}
\end{align}
where $n$ is the number of photon energies considered in the specified intervals and $d\sigma(E_{\gamma_i})/dM_\text{inv}$ corresponds to the differential cross sections calculated for a certain value of $E_{\gamma_i}$. Note that the physical region associated with $M_\text{inv}$ changes with $E_{\gamma_i}$, thus, when calculating Eq.~(\ref{dsigEi}), a value of zero is attributed to the differential cross section whenever we are outside of the corresponding $M_\text{inv}$ physical region for the given value of $E_{\gamma_i}$. This procedure reproduces the phase space distributions obtained in Ref.~\cite{Ishikawa:2021yyz}.

Besides, we must also keep in mind that an input required for the calculations is the deuteron wave function.  As mentioned earlier, there are several parametrizations available in the literature~\cite{Machleidt:2000ge,Lacombe:1981eg,Reid:1968sq,Adler:1975ga}, which have all been determined by fitting the data on $NN$ scattering, $e^-~d$ scattering. Modern calculations of the deuteron wave function have been done by using effective field theories describing the $NN$ interaction at next-to-next-to-next-to-leading order~\cite{Epelbaum:2014efa}. In view of such findings, we consider the different descriptions of Refs.~\cite{Machleidt:2000ge,Lacombe:1981eg,Reid:1968sq,Adler:1975ga,Epelbaum:2014efa} and study the consequently arising uncertainties in the model. With this motivation, we show the $\eta d$ and $\pi^0d$ mass distributions obtained within the impulse approximation, in Fig~\ref{Fig12}, when considering the deuteron wave functions from Refs.~\cite{Machleidt:2000ge,Lacombe:1981eg,Reid:1968sq,Adler:1975ga,Epelbaum:2014efa}. We focus first on evaluating the contributions to the cross sections from the $s$-wave part of the deuteron wave function.
\begin{figure}[h!]
\begin{tabular}{cc}
\includegraphics[width=0.42\textwidth]{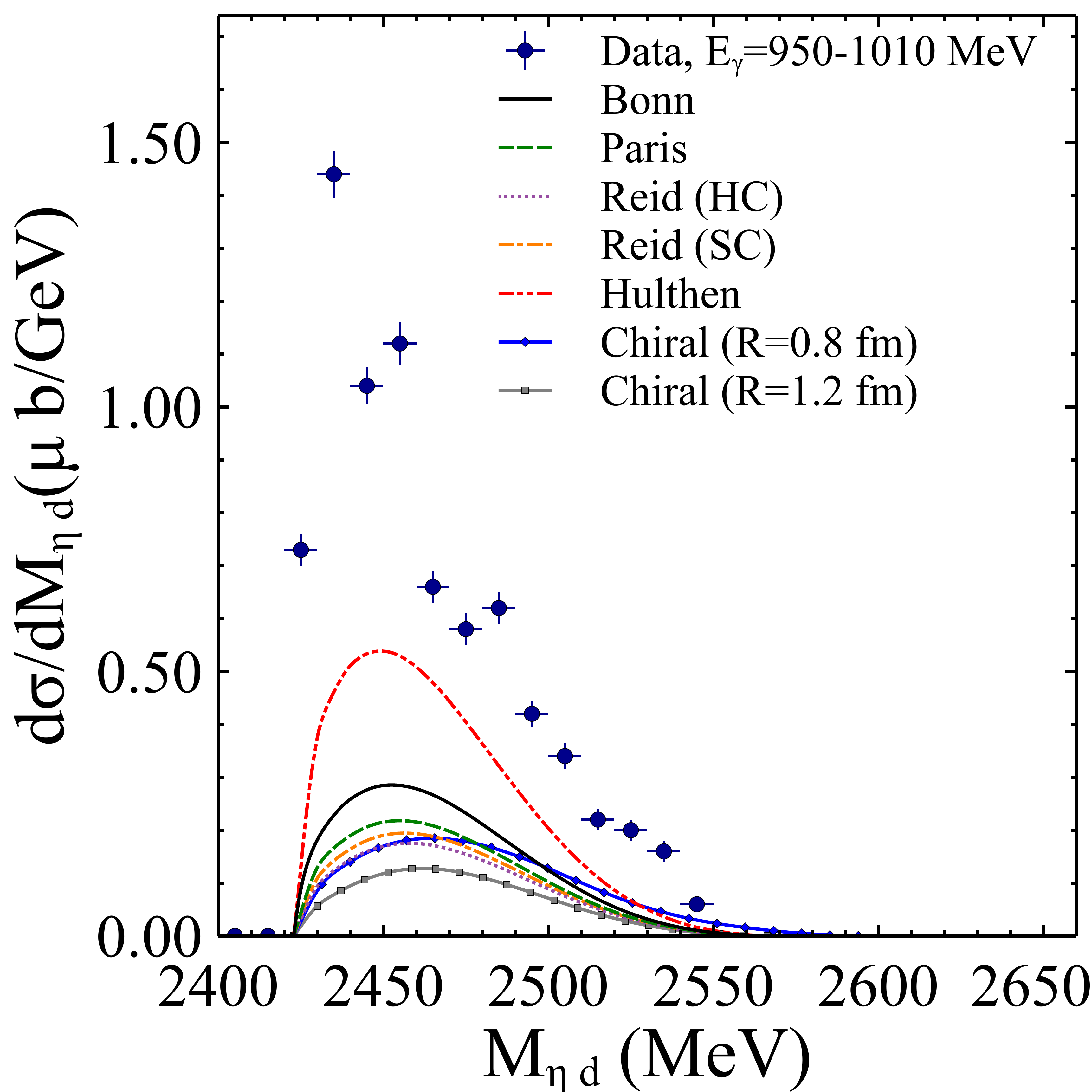}&\includegraphics[width=0.42\textwidth]{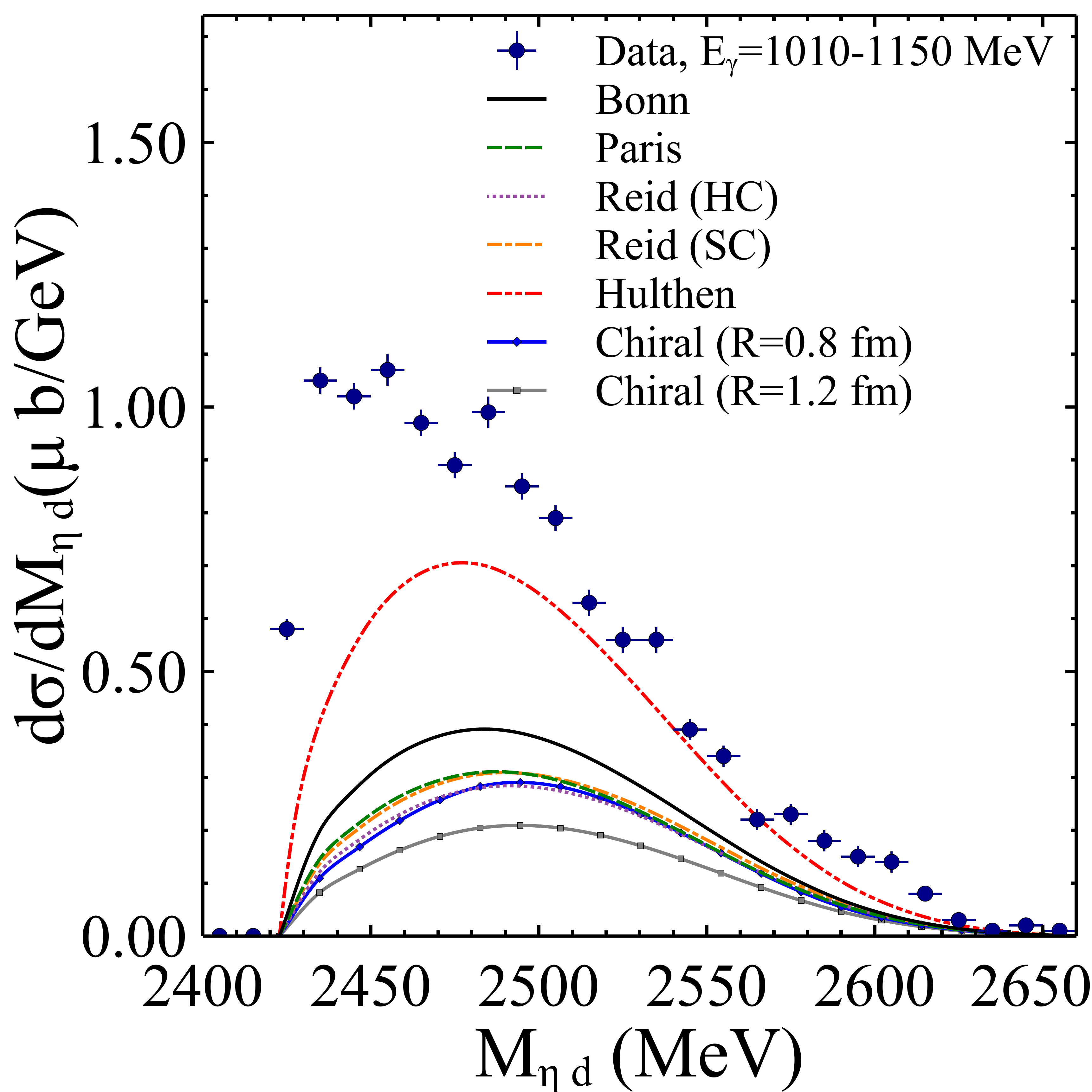}\\
\includegraphics[width=0.42\textwidth]{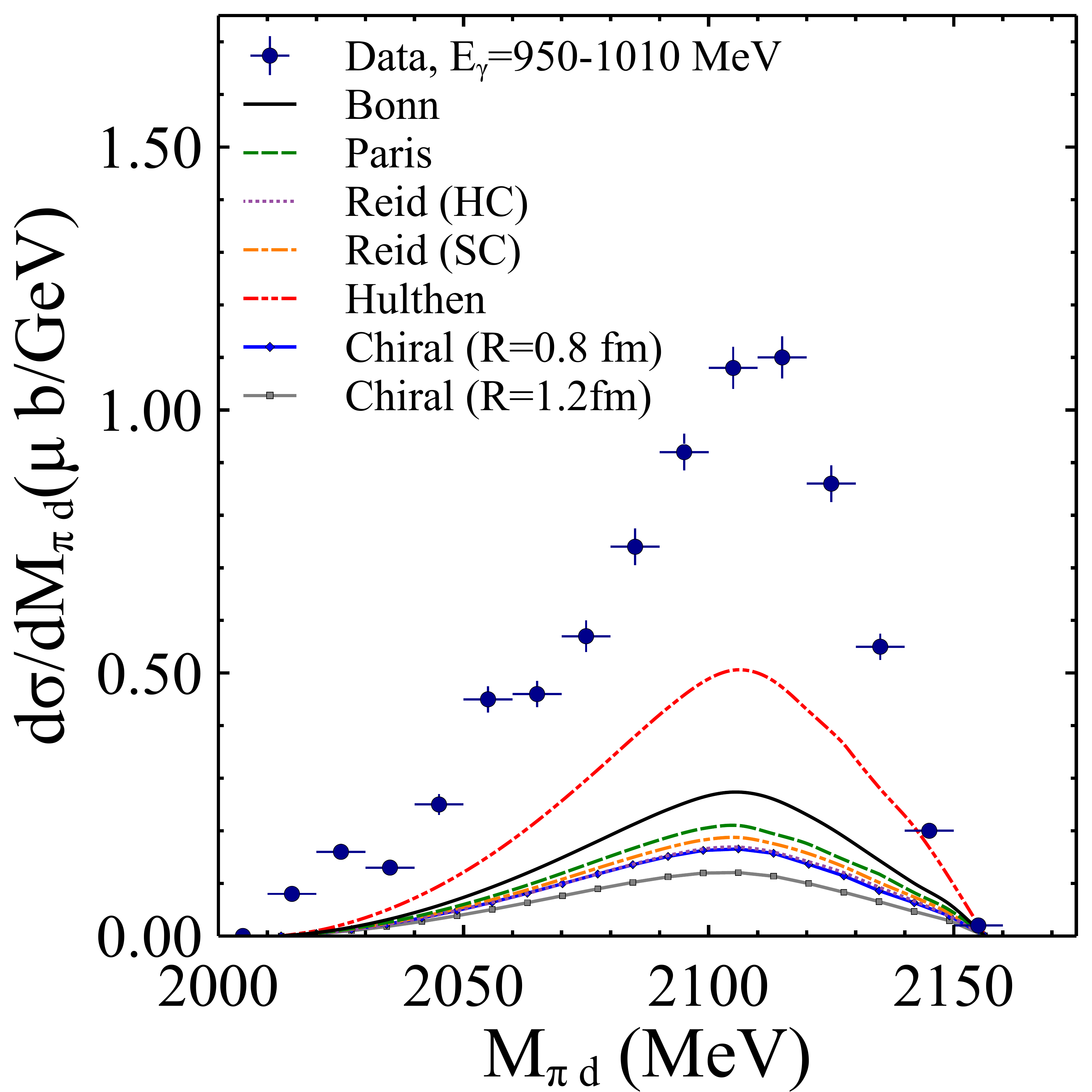}&\includegraphics[width=0.42\textwidth]{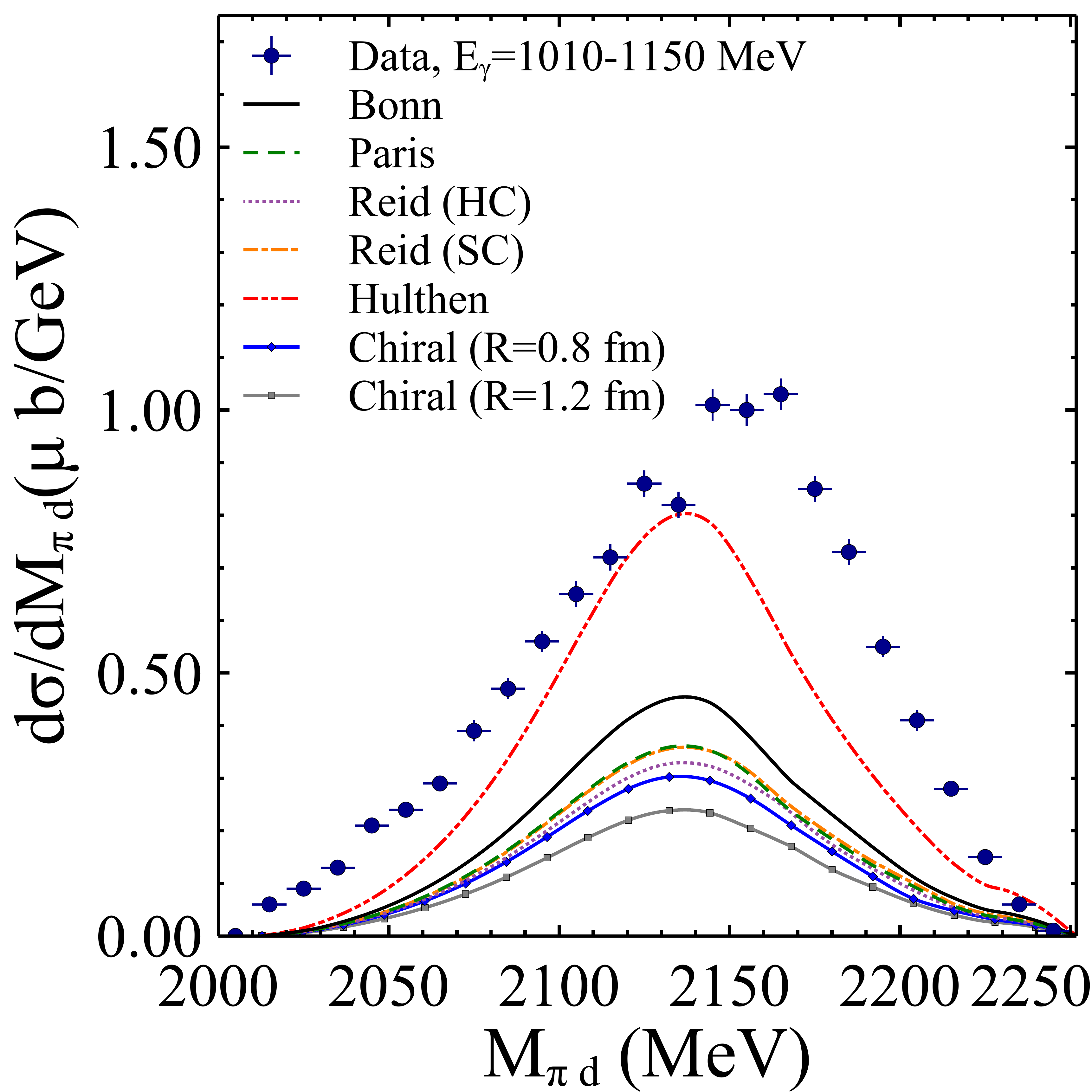}
\end{tabular}
\caption{Differential cross sections obtained in the impulse approximation as a function of the $\eta d$ (upper panels) and $\pi^0 d$ (lower panels) invariant masses. The left (right) side figures show average cross sections for the beam energy range $E_\gamma=950-1010$ MeV ($E_\gamma=1010-1150$ MeV). Experimental data, shown as filled circles, are taken from Ref.~\cite{Ishikawa:2021yyz}. The deuteron wave functions considered in the calculations are based on the following parametrizations for the $NN$ potentials: Bonn~\cite{Machleidt:2000ge}, Paris~\cite{Lacombe:1981eg}, Reidt Hard-Core (HC) and Soft-Core (SC)~\cite{Reid:1968sq}, Hulth\'en~\cite{Adler:1975ga}, and chiral effective field theories~\cite{Epelbaum:2014efa}. In the latter case, we show the results obtained with the wave function determined by using the hardest (softest) cutoff $R=0.8$ fm ($R=1.2$ fm) considered in Ref.~\cite{Epelbaum:2014efa} at which the low energy constants appearing in the Lagrangian at next-to-next-to-next-to-leading order are determined.}\label{Fig12}
\end{figure}

It can be seen in Fig.~\ref{Fig12} that the shape of the data~\cite{Ishikawa:2021yyz} on the differential cross section can already be reproduced with the impulse approximation, and that the magnitude is substantially sensitive to the choice of the wave function considered in the calculations. 
The sensitivity of the results to the different parametrizations of the deuteron wave function implies that they must differ in the momentum range relevant to the process.

To understand such differences, it can be useful to investigate how the momentum gets distributed among the nucleons in the deuteron (in the initial and final states). For this purpose, we generate random numbers when calculating the phase-space integration for the differential cross sections, and collect the events which satisfy the condition 
\begin{align}
\theta\Big(q_{max}-\Big|\frac{\vec p_d}{2}-\vec{q}\,\Big|\Big)\times\theta\Big(q_{max}-\Big|\frac{\vec p_d+\vec k-\vec p_\eta-\vec p_\pi}{2}-\vec{q}\,\Big|\Big)
=1,\label{condition}
\end{align}
while changing $q_\text{max}$ from $10$ to $1000$ MeV, in steps of 10 MeV, with $q_\text{max}$ being a cut-off for the loop variable $|\vec{q}|$ in Eq.~(\ref{t3}). In this way, if we call $R_i$ the number found for the $i$th value of $q_\text{max}$, the difference $R_{i+1}-R_i$ provides the fraction of events where either $|\vec p_d/2-\vec{q}|$ or $|(\vec p_d+\vec k-\vec p_\eta-\vec p_\pi)/2-\vec{q}|$ are between $q_\text{max}$ and $q_\text{max}+10$ MeV. Such an analysis gives us the information on the typical momentum value picked by the deuteron wave function. The result is depicted in Fig.~\ref{Fig13} for three different beam energies, chosen as an example.  It can be deduced from Fig.~\ref{Fig13} that the deuteron wave function gets determined, most frequently, in the momentum range 300-400 MeV.
\begin{figure}[h!]
\includegraphics[width=0.45\textwidth]{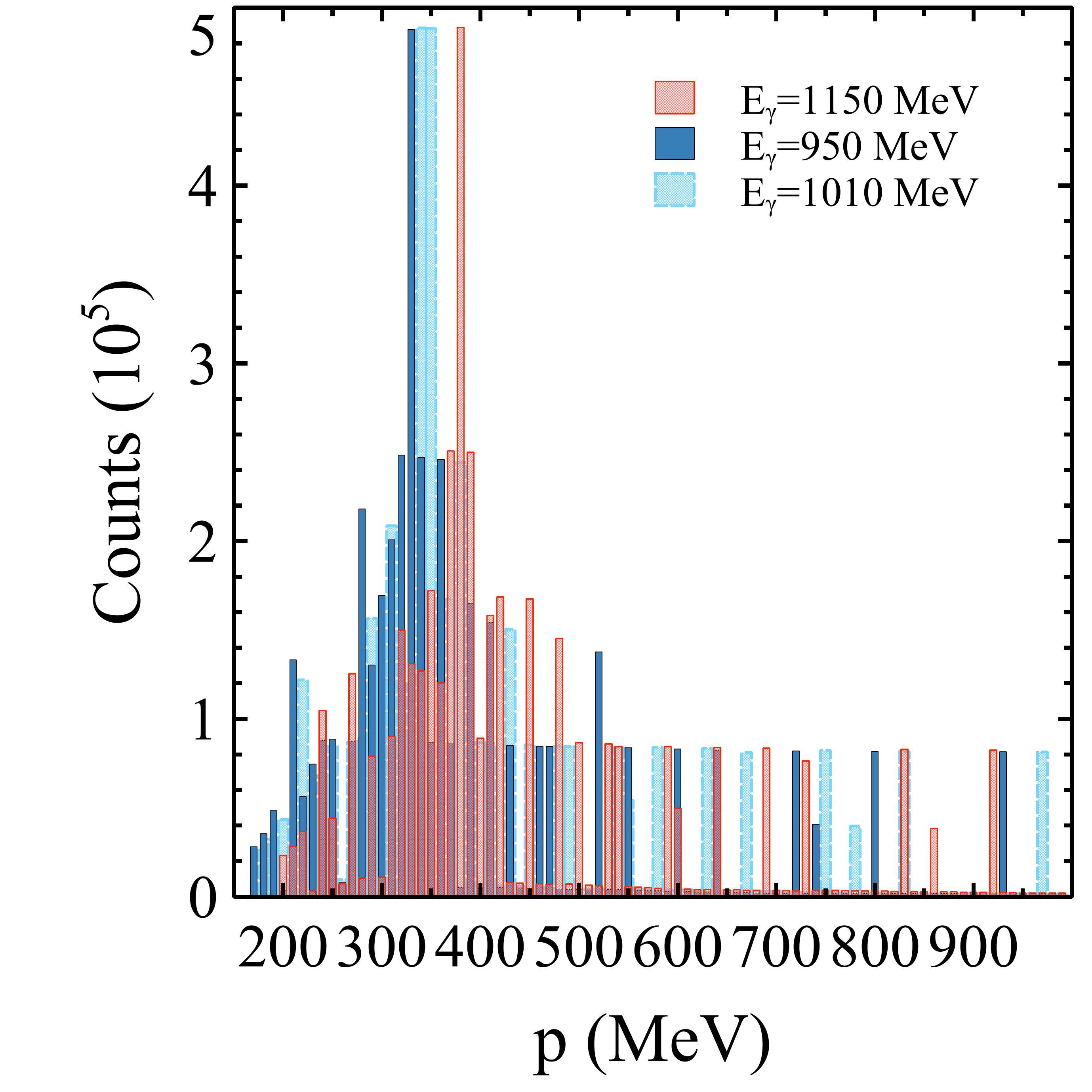}
\caption{Accumulation of events satisfying the condition in Eq.~(\ref{condition}) for values of $q_\text{max}$ in the range $0-1000$ MeV.}\label{Fig13}
\end{figure}

Let us now look at the different wave functions, with the focus on the momentum region 300-400 MeV (shown as an inset in Fig.~\ref{Fig14}). 
\begin{figure}[h!]
\includegraphics[width=0.5\textwidth]{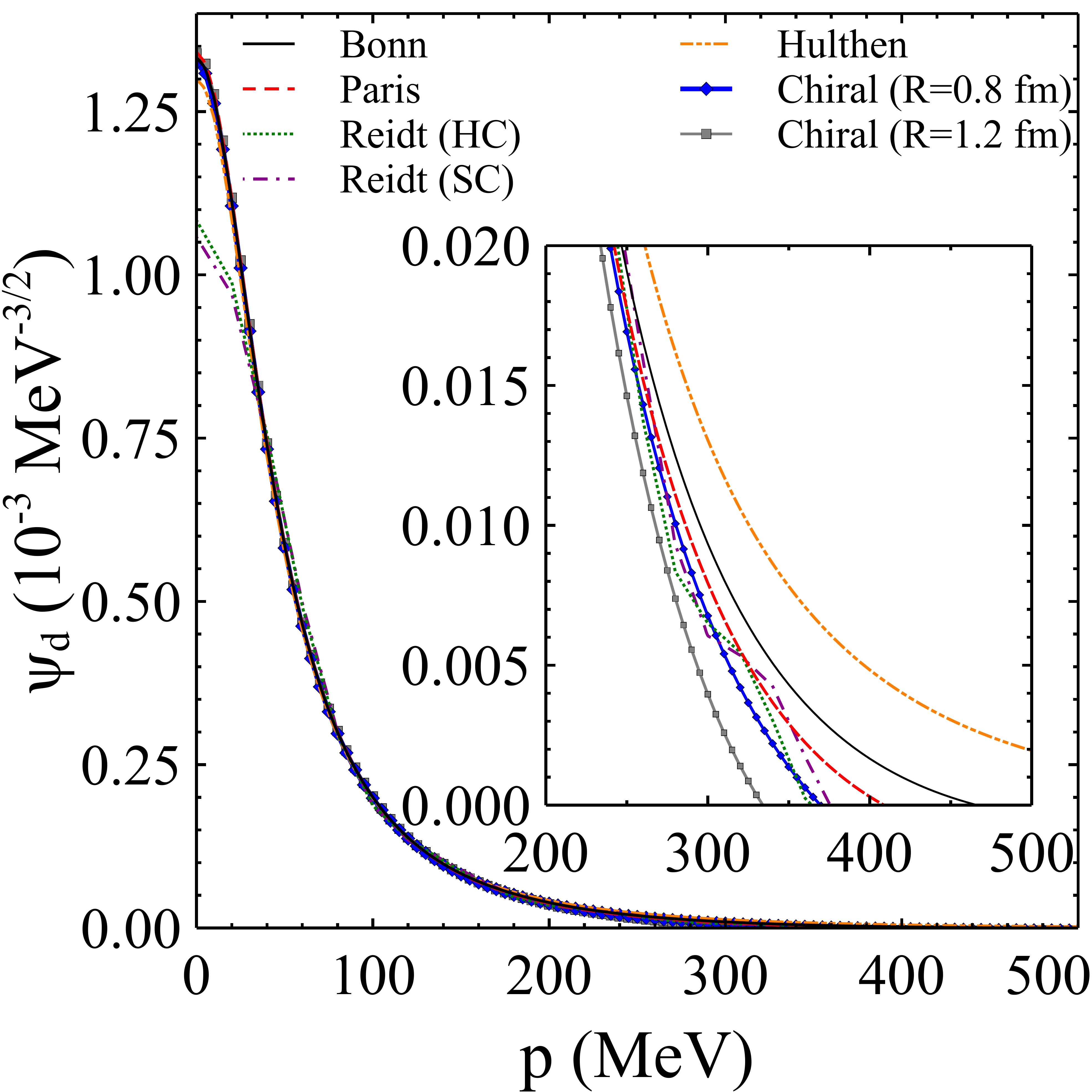}
\caption{Deuteron wave functions ($s$-wave part) based on the following parametrizations for the $NN$ potentials: Bonn~\cite{Machleidt:2000ge}, Paris~\cite{Lacombe:1981eg}, Reidt Hard-Core (HC) and Soft-Core (SC)~\cite{Reid:1968sq} and Hulth\'en~\cite{Adler:1975ga}.}\label{Fig14}
\end{figure}
Before further discussions, we should recall that in our approach the wave function of the deuteron has been normalized as 
\begin{align}
\int d^3p |\langle \vec p\,|\psi\rangle|^2=1, 
\end{align}
which is consistent with the value of the $g_d$ coupling appearing in the expressions which have been identified as the deuteron wave function. As can be seen in Fig.~\ref{Fig14}, the different parametrizations of the deuteron wave function agree well in the 50-250 MeV region. However, there are significant differences in the momentum region relevant for the calculations, which should not come as a surprise.
The different parametrizations of Refs.~\cite{Machleidt:2000ge,Lacombe:1981eg,Reid:1968sq,Adler:1975ga} for the $NN$ potential are based on meson exchange potentials and, thus, should be expected to work at distances where the nucleons do not overlap. The same can be said for the model of Ref.~\cite{Epelbaum:2014efa}. However, at the momentum values falling in the range 300-400 MeV, a significant overlap between the nucleons is expected and the $NN$ scattering models of Refs.~\cite{Machleidt:2000ge,Lacombe:1981eg,Reid:1968sq,Adler:1975ga,Epelbaum:2014efa} cannot provide precise descriptions for the deuteron wave function.

We must now proceed and show the contributions from the rescattering diagrams (shown in Figs.~\ref{Fig6} and \ref{Fig10}). The results on the differential cross sections are shown in Fig.~\ref{Fig15}, as a function of the $\eta d$ and $\pi^0 d$ invariant masses. Experimental data are taken from Ref.~\cite{Ishikawa:2021yyz}. Since we have already discussed the uncertainties with different deuteron wave function parametrizations, we find it sufficient to show the results obtained with the Bonn~\cite{Machleidt:2000ge} and Hulth\'en models~\cite{Adler:1975ga}, which differ appreciably in the momentum region of interest.
\begin{figure}[h!]
\includegraphics[width=0.8\textwidth]{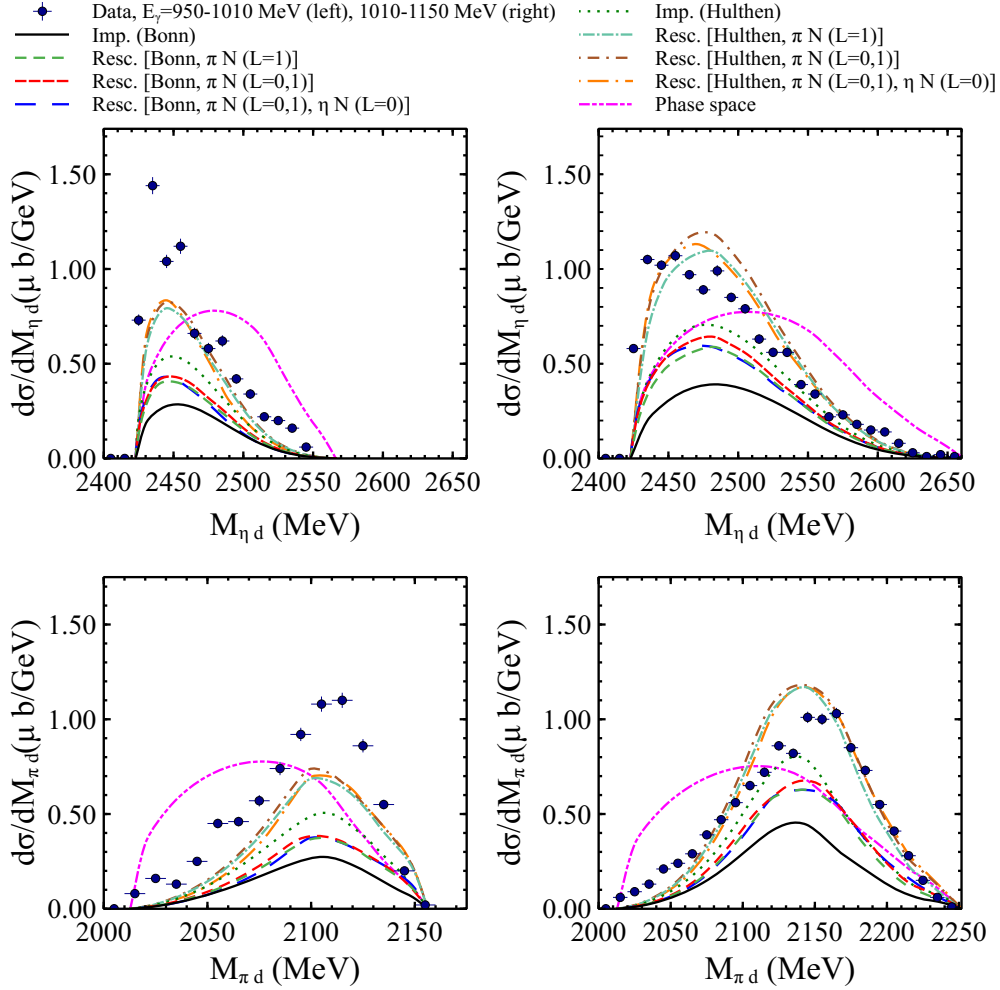}
\caption{Differential cross sections as a function of the $\eta d$ (upper panels) and $\pi^0 d$ (lower panels) invariant masses, as obtained in the impulse approximation and by considering the rescattering of $\pi^0$ in the $p$-wave (orbital angular momentum $L=1$), as well as in the $s$-wave ($L=0$), and the rescattering of $\eta$ in the $s$-wave ($L=0$). The left (right) side figures show average cross sections for the beam energy range $E_\gamma=950-1010$ MeV ($E_\gamma=1010-1150$ MeV). Experimental data, shown as filled circles, are taken from Ref.~\cite{Ishikawa:2021yyz}.}\label{Fig15}
\end{figure}
The results in Fig.~\ref{Fig15} show that the contribution from the rescattering processes depends on the deuteron wave function and can describe most characteristics of the data, especially for the beam energy range 1010-1150 MeV.  The uncertainties arising from the parametrizations of $NN$ potentials are unavoidable and inherent to the process. More precise calculations are not feasible since the reaction mechanism attributes momenta values at which the deuteron wave function can not be determined in terms of meson exchange potentials.

We can see that the effect of rescattering is relevant and leads to an increase of the strength of the mass distribution of about 50$\%$, with the rescattering of a pion in the $p$-wave, through the mechanism $\pi N\to \Delta(1232)\to \pi N$, producing the dominant contribution. Dynamically, an extra scattering weakens the contribution to the amplitude in general, but in this case the rescattering mechanism helps sharing the momentum transfer between the two nucleons of the deuteron and involves the deuteron wave function at smaller momenta, where it is bigger. 

It is interesting to see that our calculations differ appreciably from phase space. It is easy to trace that back to our dynamical model. If we look at Fig.~\ref{Fig4}, the mechanism favors the $\pi^0$ to go with as high energy as possible to place the $\Delta(1232)$ on-shell. This leaves less energy for the $\eta$ and the $\eta d$ invariant mass becomes smaller, something clearly seen in the experimental data. Conversely, the $\pi^0$ goes out with larger energy than  expected from phase space leading to a $\pi d$ invariant mass bigger than for phase space.

Next, we determine the contribution to the differential cross sections of the rescattering mechanisms illustrated in Fig.~\ref{Fig8}. We show in Fig.~\ref{Fig16} the results obtained when including such rescattering contributions.
\begin{figure}
\begin{tabular}{cc}
\includegraphics[width=0.45\textwidth]{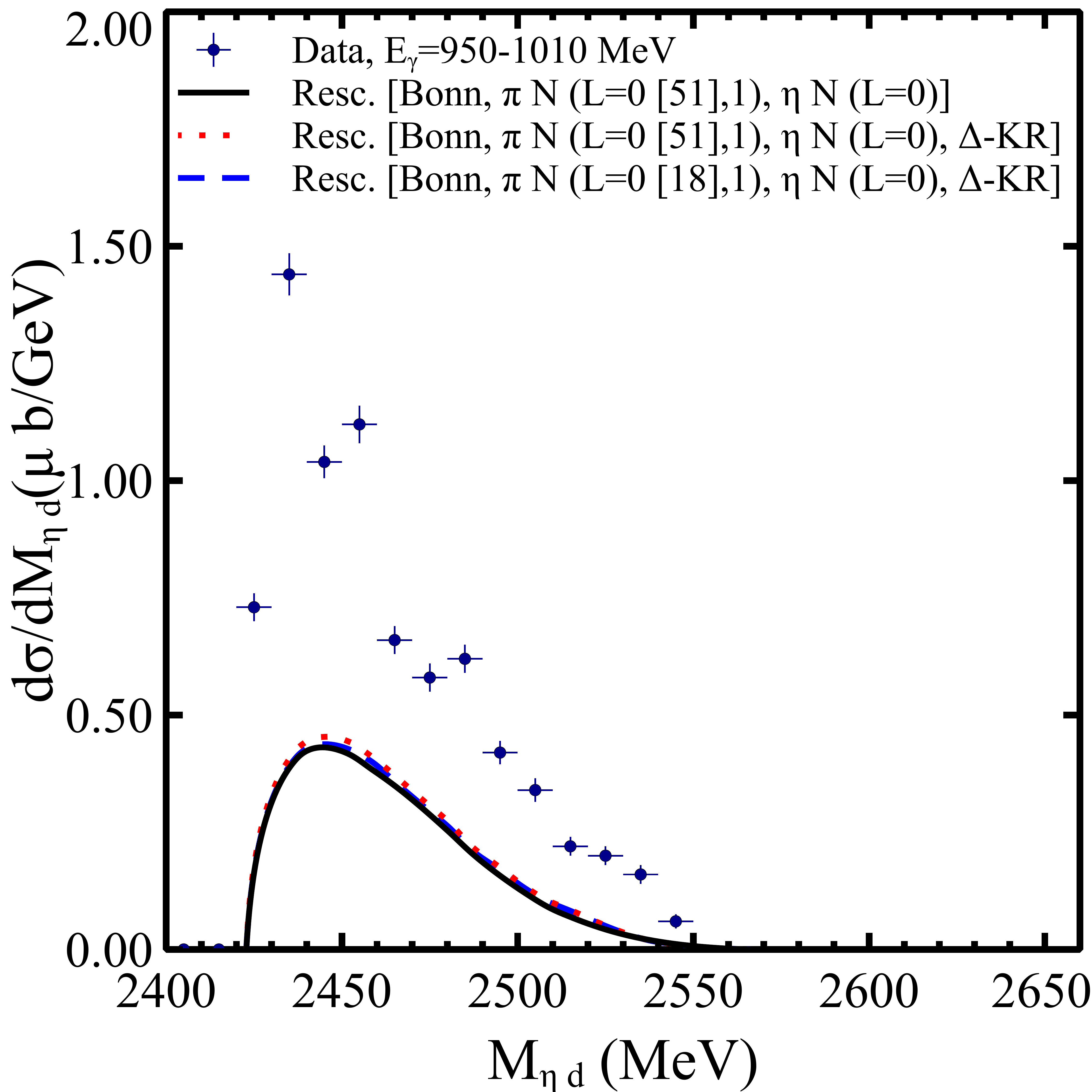}&\includegraphics[width=0.45\textwidth]{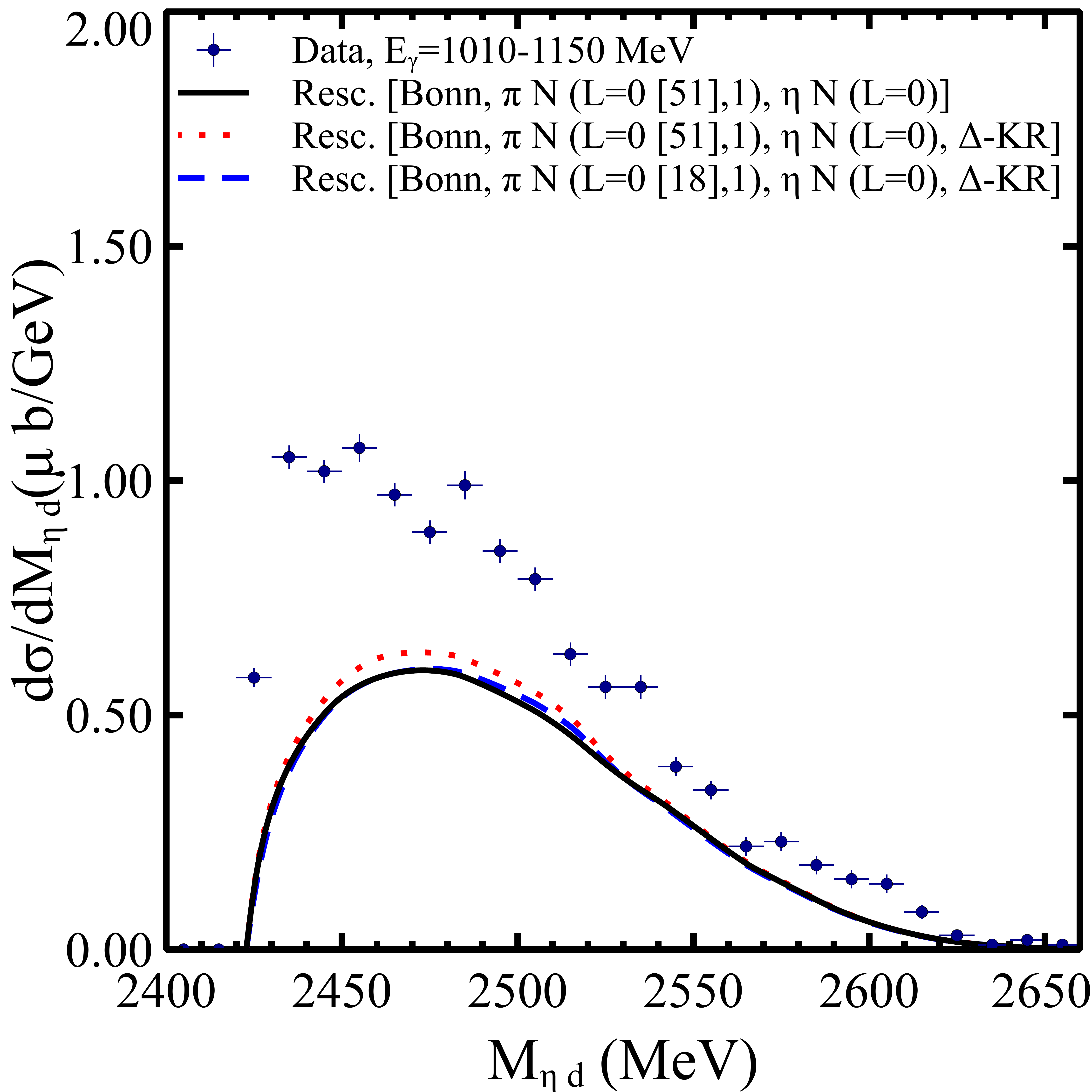}\\
\includegraphics[width=0.45\textwidth]{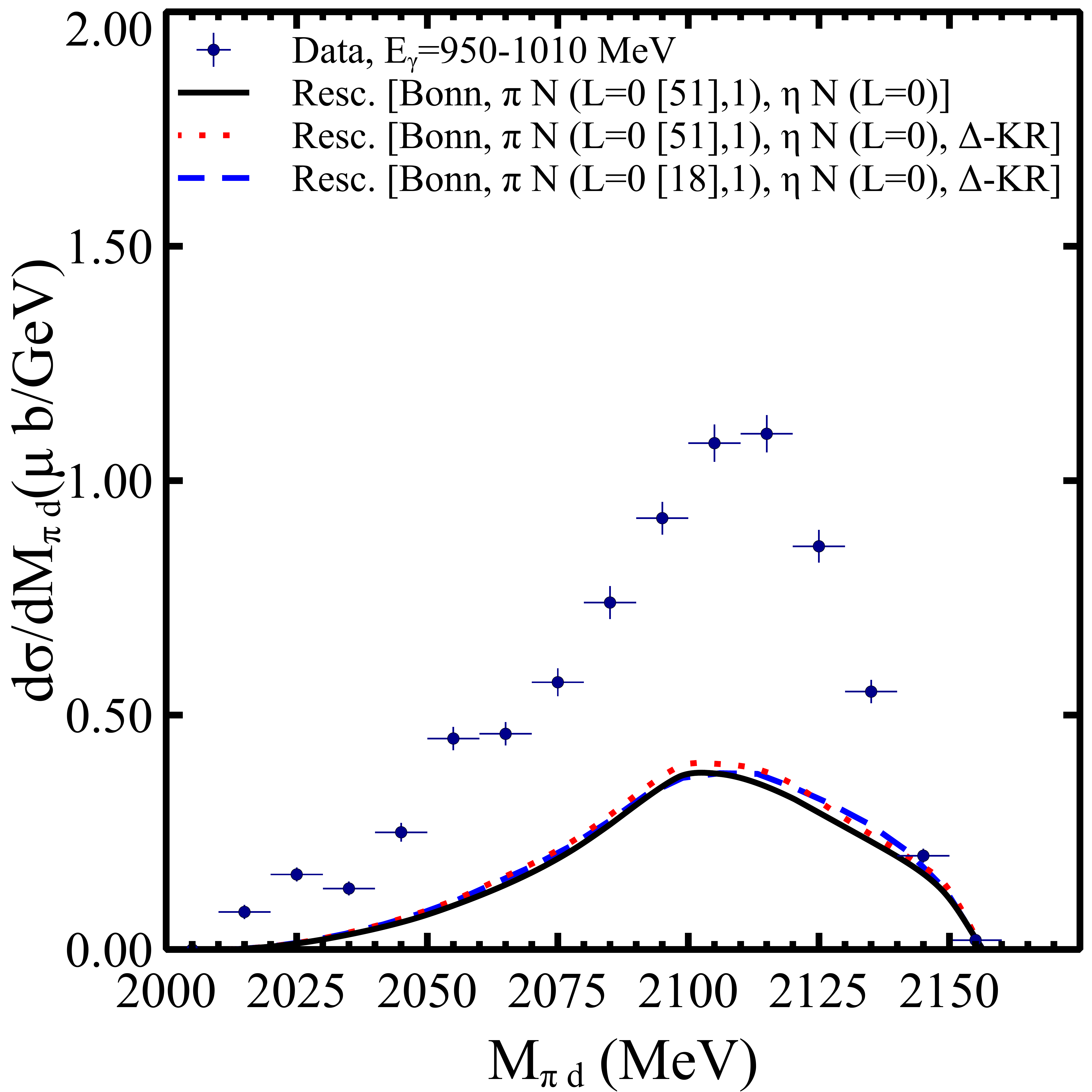}&\includegraphics[width=0.45\textwidth]{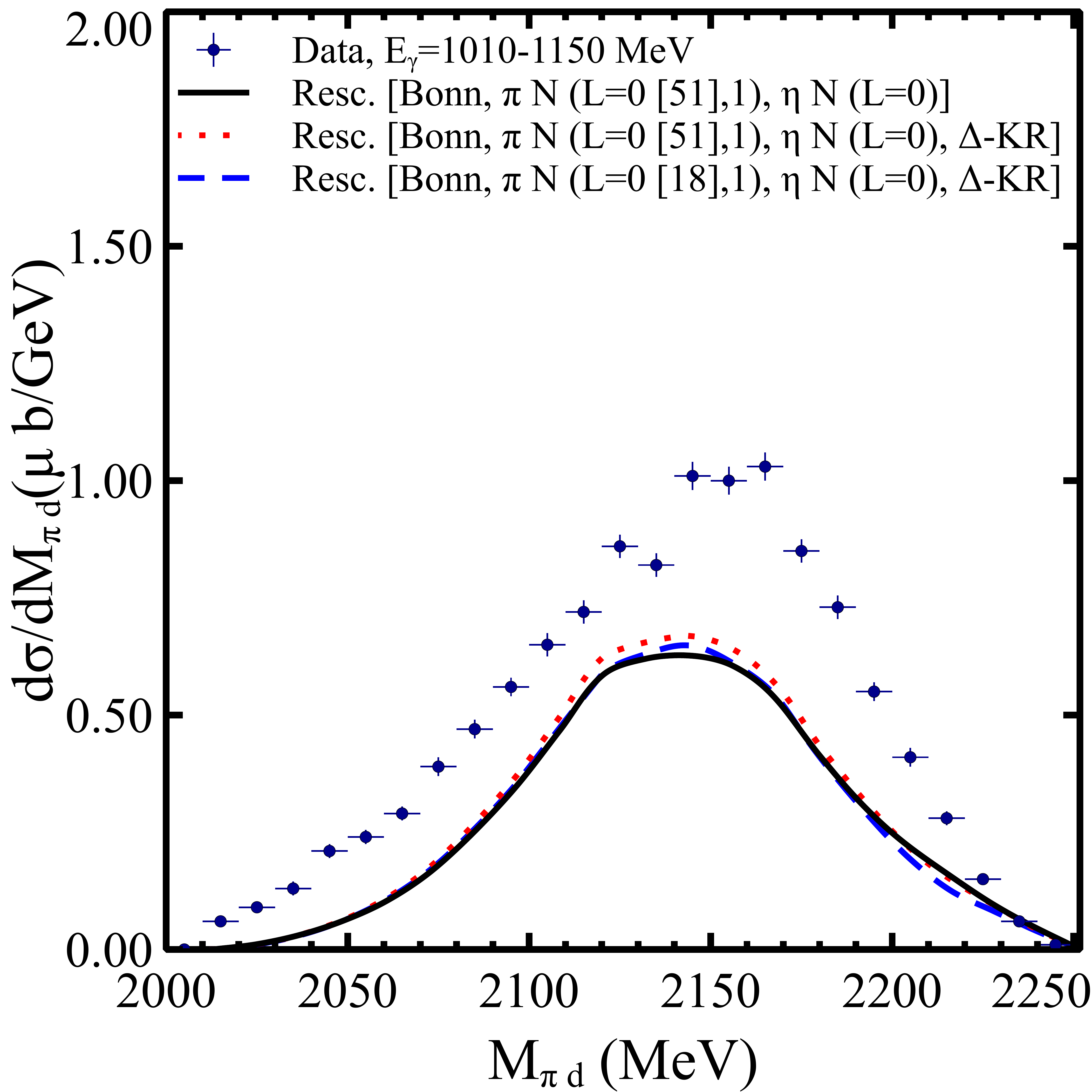}
\end{tabular}
\caption{Differential cross sections determined by considering the rescattering of a pion, including contributions from the processes shown in Fig.~\ref{Fig8}, and $\eta$ rescattering. We also show the differences obtained within two models for the description of the $\pi N$ interaction in the $s$-wave: the one of Ref.~\cite{Oset:1985wt} (dotted line) and the one of Ref.~\cite{Inoue:2001ip} (dashed line), where $N^*(1535)$ is generated from the pseudoscalar-baryon interaction.}\label{Fig16}
\end{figure}
As can be seen from the figure, the mechanisms shown in Fig.~\ref{Fig8} give a small contribution to the differential cross sections and can be neglected. This is in line with our finding that the contributions related to the rescattering of the $\eta$ are quite small (see Fig.~\ref{Fig15}). There we had the $\eta N\to N^*(1535)\to\eta N$ amplitudes in the rescattering, while now we have the $\pi N\to N^*(1535)\to\eta N$ amplitude, and from Ref.~\cite{Inoue:2001ip} the coupling of $N^*(1535)$ to $\eta N$ is, in modulus, $\approx3.1$ (4.5) times the coupling of $N^*(1535)$ to $\pi^- p$ ($\pi^0 n$).

It is also relevant to show the changes produced in the differential cross sections when a more detailed model for describing the $\pi N$ interaction is considered. The model explained in Sec.~\ref{pires} for describing the $\pi N$ interaction in the $s$-wave does not involve coupled channels and resolution of the Bethe-Salpeter equation to determine the scattering matrix. Thus, resonance contributions, which will change the energy dependence considered for the $\pi N\to \pi N$ amplitude,  are not implemented. In Ref.~\cite{Inoue:2001ip}, the interaction between pseudoscalar mesons and baryons from the octet were studied in the zero strangeness sector within a coupled channel approach. The Bethe-Salpeter equation was solved and the $\pi N$ scattering matrices obtained, together with the corresponding phase shifts and inelasticities, were compared with those extracted from partial wave analysis. Compatible results were found for energies of the $\pi N$ system $\approx1100-1600$ MeV. In Fig.~\ref{Fig16} we also show (as a dashed line) the results obtained considering the model of Ref.~\cite{Inoue:2001ip}. The agreement with the results obtained by using the model of Ref.~\cite{Oset:1985wt} is remarkable.

Continuing with the estimation of uncertainties, in Ref.~\cite{Ishikawa:2022mgt}, the Hulth\'en wave function with different parameters to those considered in Ref.~\cite{Adler:1975ga}, and which reproduces the momentum distribution of nucleons in a deuteron derived from the $d(e,e^\prime p)n$ reaction, was used to determine the Fermi momentum of the initial bound proton. It is then interesting to quantify the differential cross section obtained with such a wave function. We show the results in Fig.~\ref{Fig17}. As can be seen, the magnitude obtained with the wave function of Ref.~\cite{Ishikawa:2022mgt} is between the one found with the $s$-wave component of the deuteron wave function of Ref.~\cite{Machleidt:2000ge} and that determined with the one of Ref.~\cite{Adler:1975ga}.
\begin{figure}
\centering
\begin{tabular}{cc}
\includegraphics[width=0.4\textwidth]{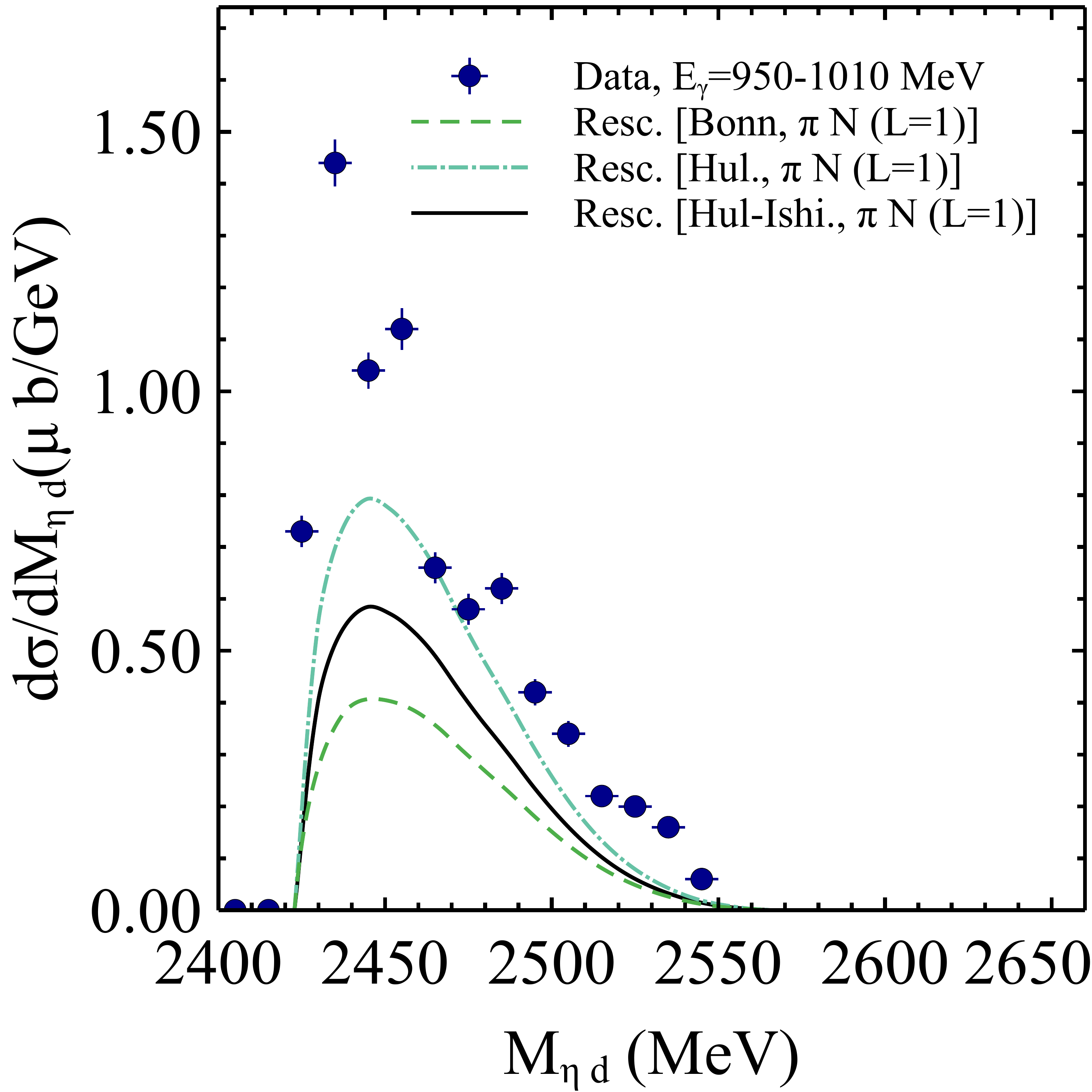}&\includegraphics[width=0.4\textwidth]{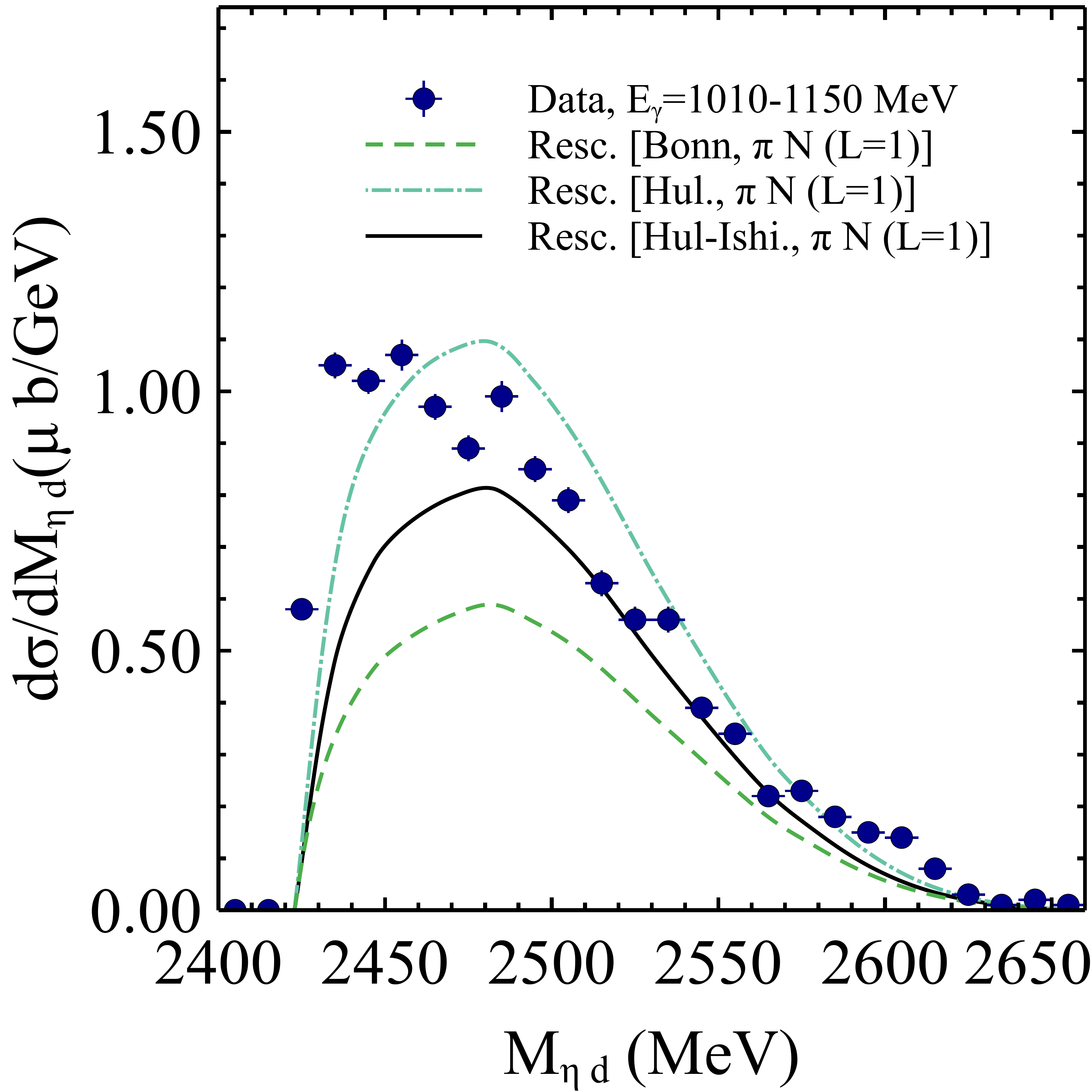}\\
\includegraphics[width=0.4\textwidth]{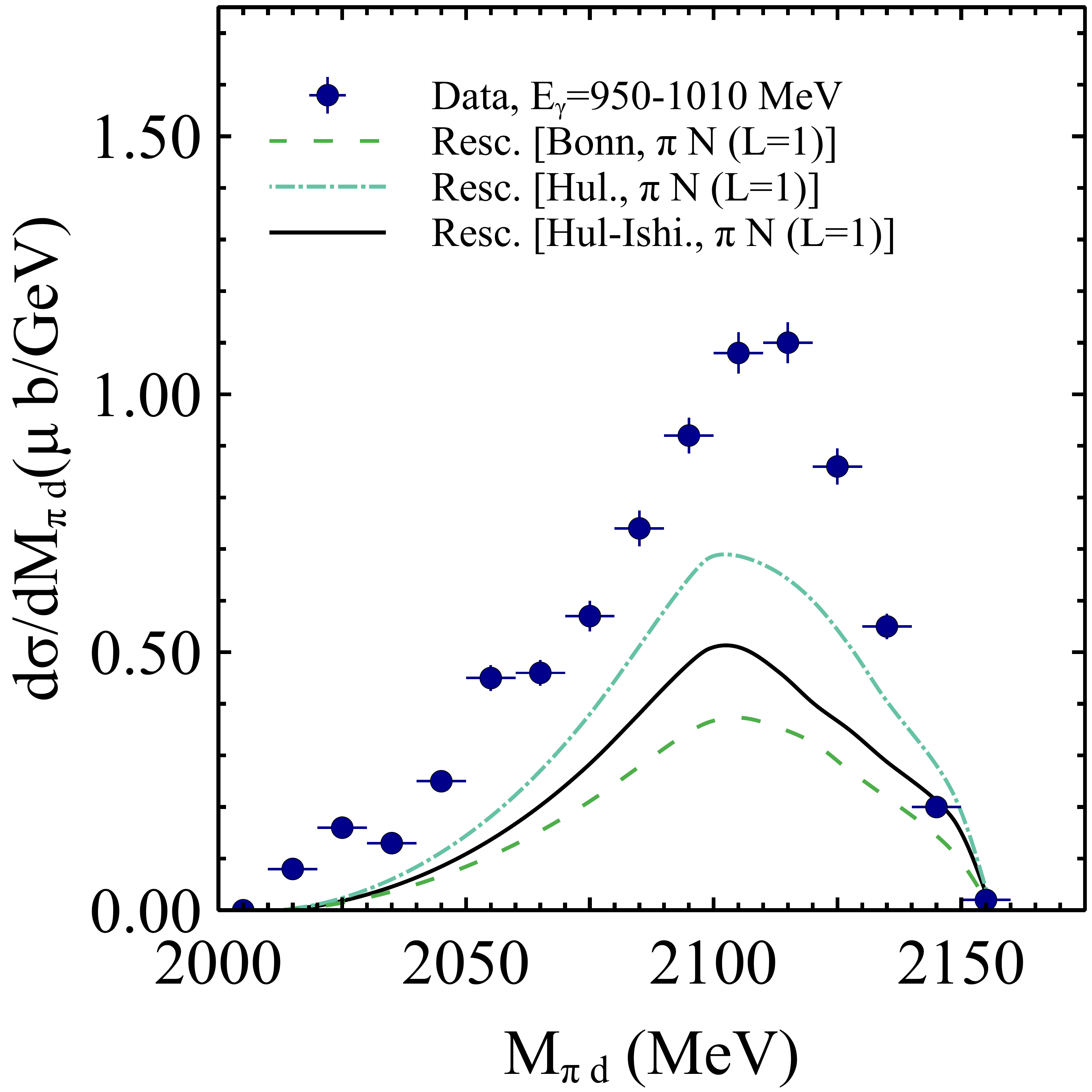}&\includegraphics[width=0.4\textwidth]{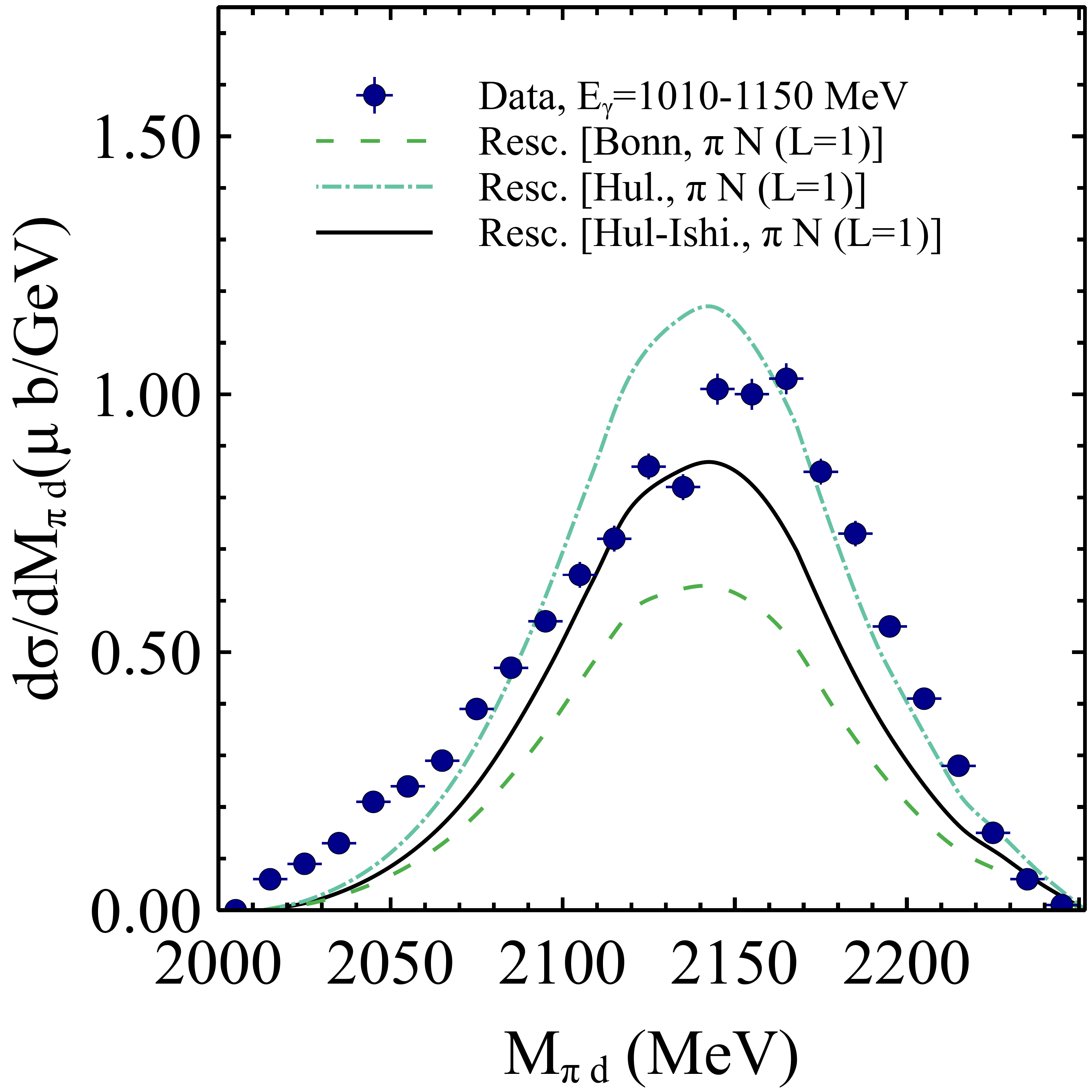}\\
\end{tabular}
\caption{Differential cross sections obtained considering the pion rescattering in the $p$-wave and the wave function of Ref.~\cite{Ishikawa:2022mgt}.}\label{Fig17}
\end{figure}

Note, however, that so far, all the results found have been obtained by considering only the $s$-wave component of the deuteron wave function. In view of the result found in Fig.~\ref{Fig13}, the $d$-wave component of the deuteron wave function can be important. To estimate the relevance of including the $d$-wave component of the deuteron wave function in the results for the differential cross section, we calculate the tree level amplitude by considering the s- and $d$-wave contributions of the deuteron wave function with the parametrization of Ref.~\cite{Machleidt:2000ge} and the wave function of Ref.~\cite{Epelbaum:2014efa}, which is determined from chiral effective field theories. Details of this calculation are provided in Appendix~\ref{appB}. In Fig.~\ref{Fig18}, we show the results obtained for the differential cross sections considering the new tree level amplitudes.
\begin{figure}
\begin{tabular}{cc}
\includegraphics[width=0.45\textwidth]{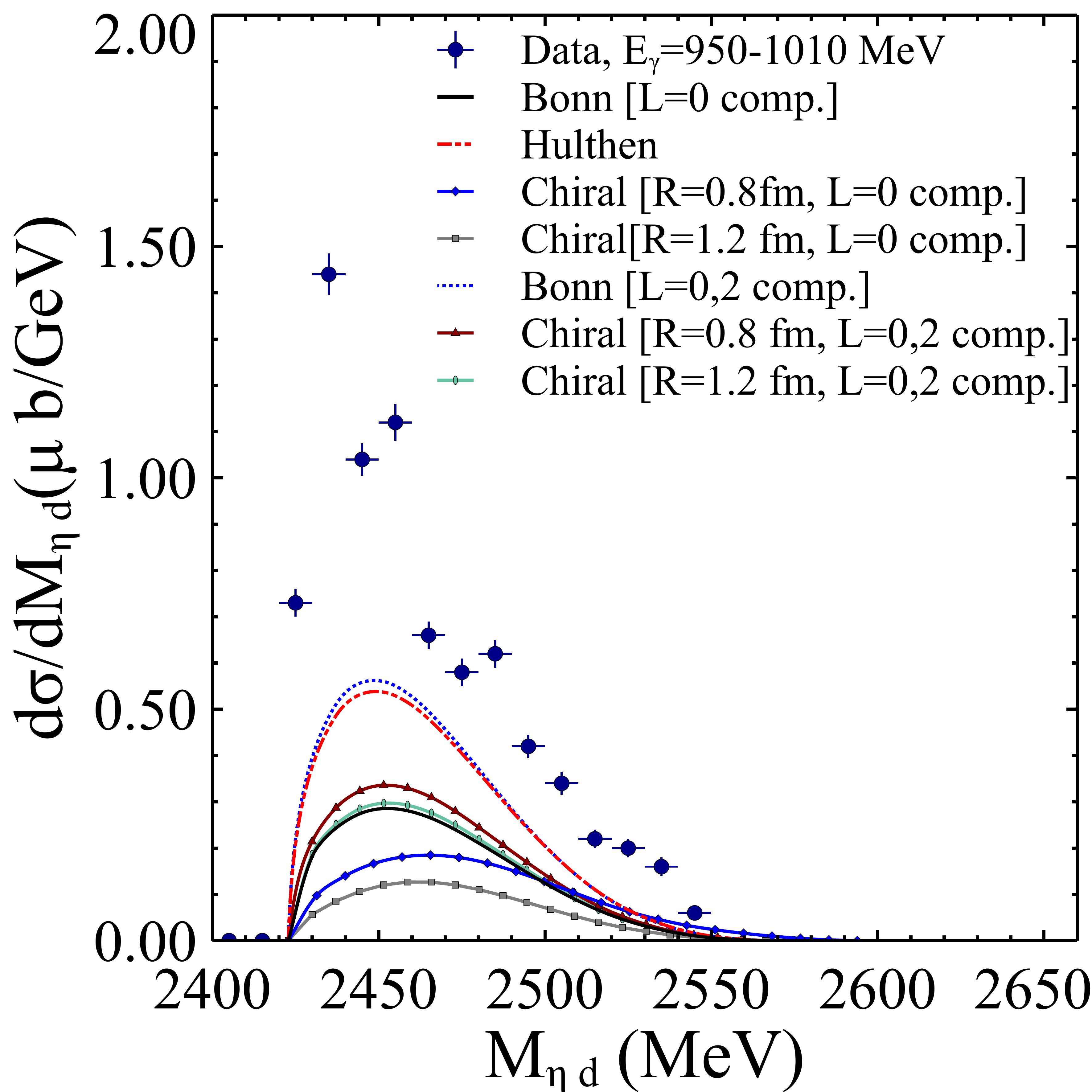}&\includegraphics[width=0.45\textwidth]{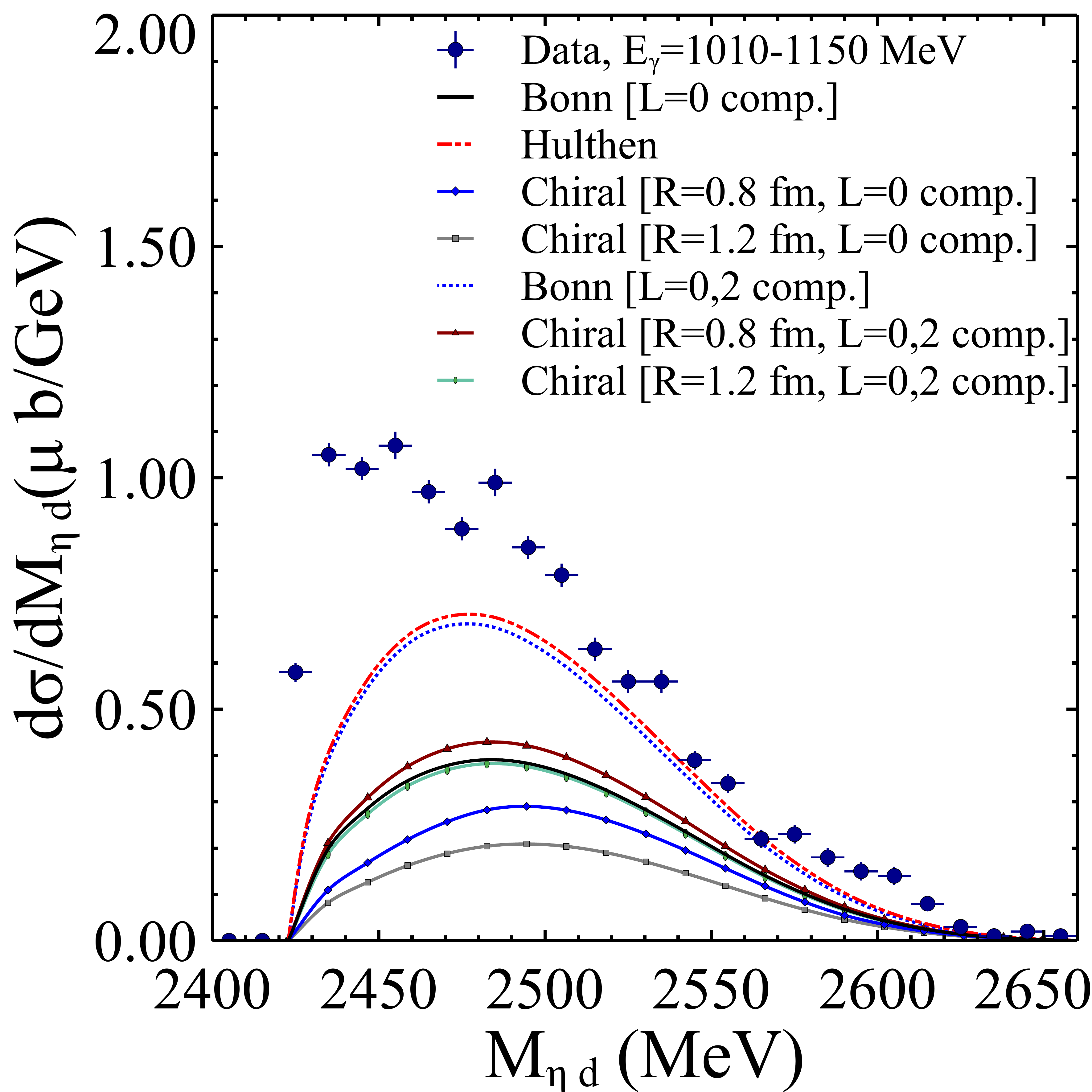}\\
\includegraphics[width=0.45\textwidth]{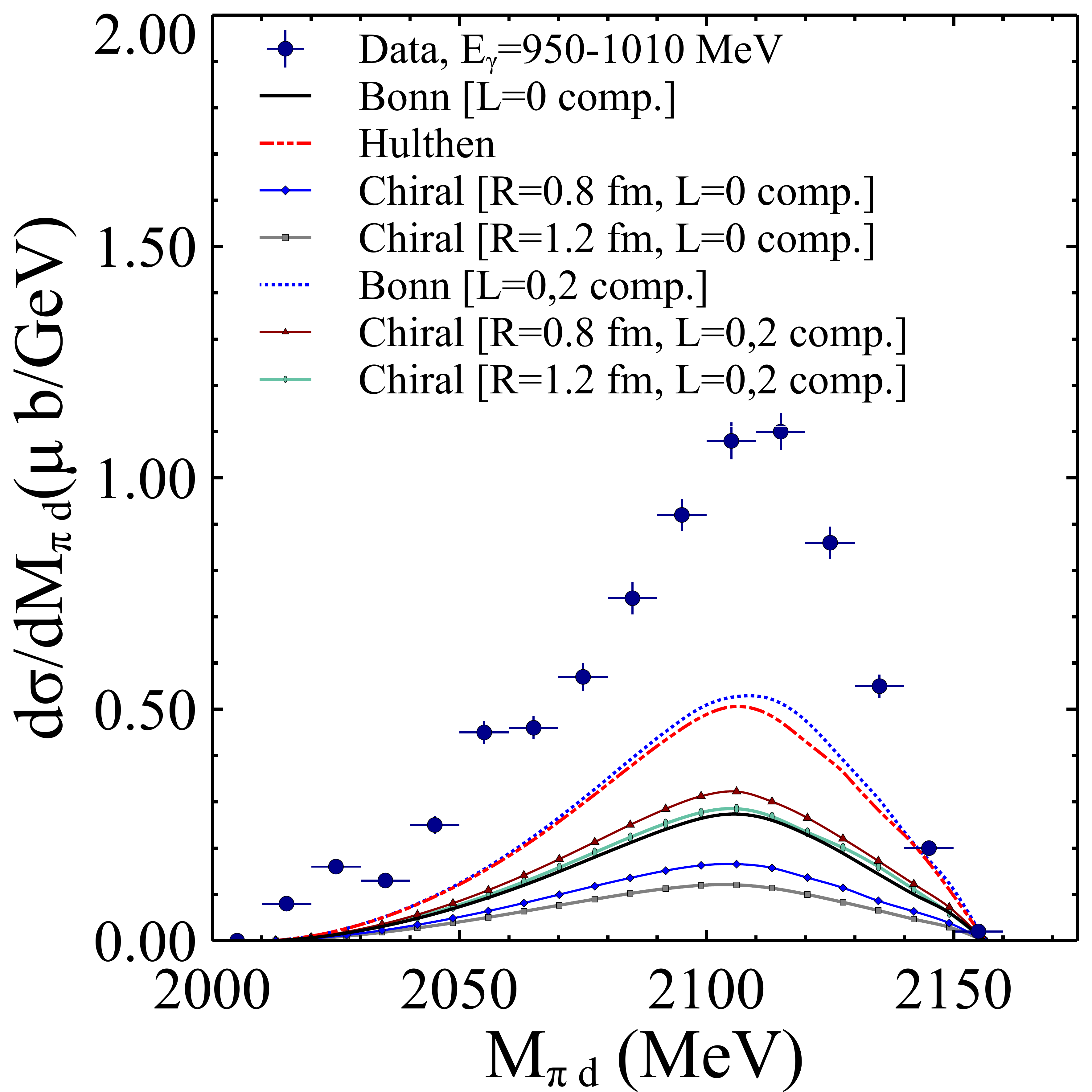}&\includegraphics[width=0.45\textwidth]{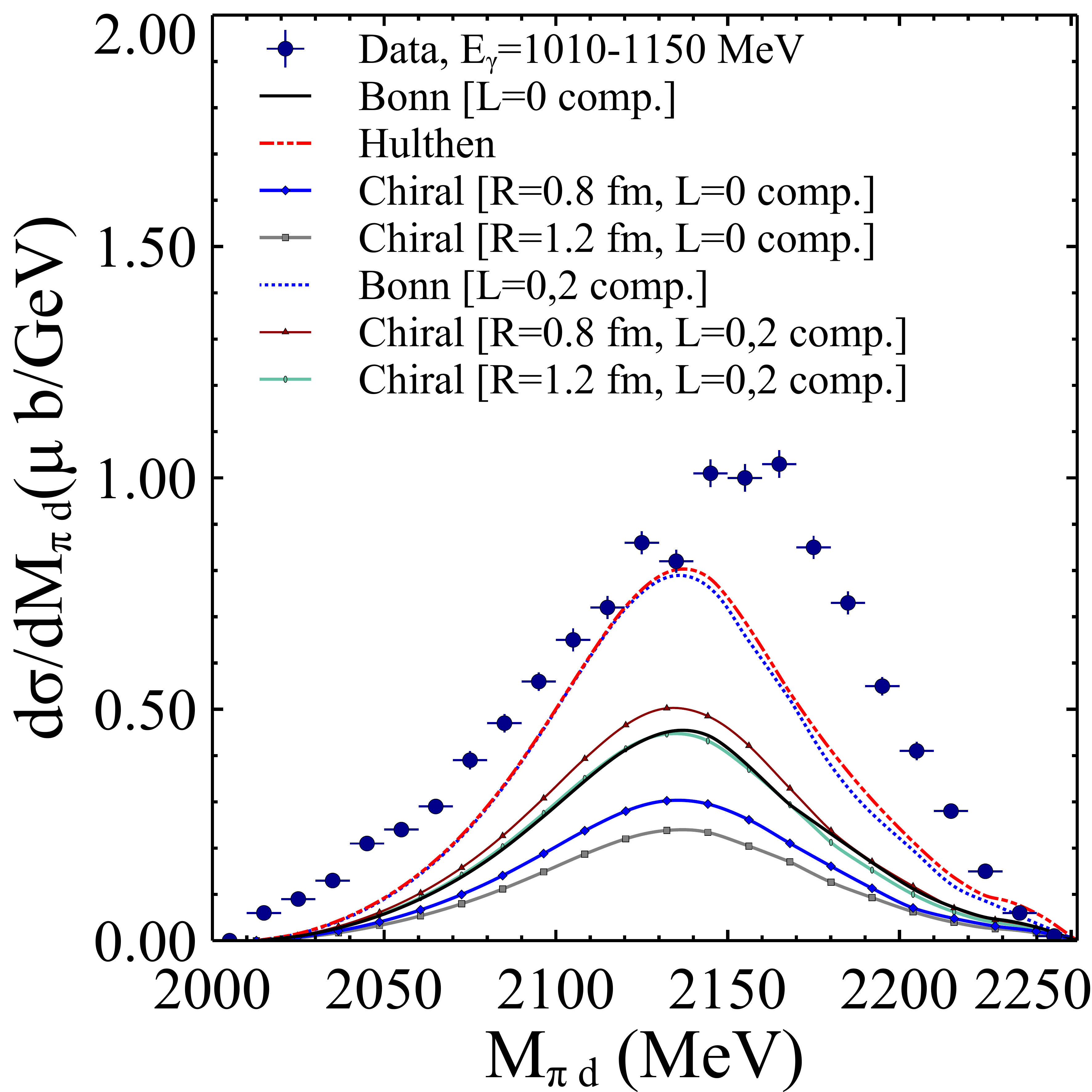}
\end{tabular}
\caption{Differential cross sections obtained in the impulse approximation by considering the s- ($L=0$) and $d$-wave ($L=2$) components of the deuteron wave function within the parametrization of Ref.~\cite{Machleidt:2000ge} and the wave function obtained from chiral effective field theories~\cite{Epelbaum:2014efa}. To facilitate the estimation of the effect of including the $d$-wave component of the deuteron wave function, we also show the results obtained with the Hulth\'en wave function~\cite{Adler:1975ga}, which only takes into account a $L=0$ component.}\label{Fig18}
\end{figure}
As can be seen, including the $d$-wave component of the deuteron wave function produces an important enhancement of the magnitude of the differential cross section to an extent that (1) the results found with the model of Ref.~\cite{Machleidt:2000ge} almost coincide with those obtained with the Hulth\'en parametrization of the deuteron wave function~\cite{Adler:1975ga}, which only considers the $s$-wave component. (2) In the case of the model of Ref.~\cite{Epelbaum:2014efa}, including the $d$-wave component produces a result which is close to the one obtained with the $s$-wave component of the wave function of Ref.~\cite{Machleidt:2000ge}. 

As to the contribution of this $d$-wave component in the rescattering mechanisms, we do not calculate it, but argue here that it should be small. This is because the rescattering mechanisms redistributes the momenta transfer and the momenta involved in the deuteron wave function in this case are substantially smaller than those in the impulse approximation. In view of the results obtained, and taking as reference the wave function of Ref.~\cite{Machleidt:2000ge} (which is the one commonly used in a large number of works involving the deuteron), our results with the $d$-wave component of the deuteron wave function and including the rescattering mechanisms should be very close to those obtained with the Hulth\'en wave function of Ref.~\cite{Adler:1975ga} and rescattering (long-dash-dotted line in Fig.~\ref{Fig15}). All together we see a fair, though not perfect, reproduction of the invariant mass distributions, underestimating the data at lower photon beam energies.

\begin{figure}[h!]
\includegraphics[width=0.4\textwidth]{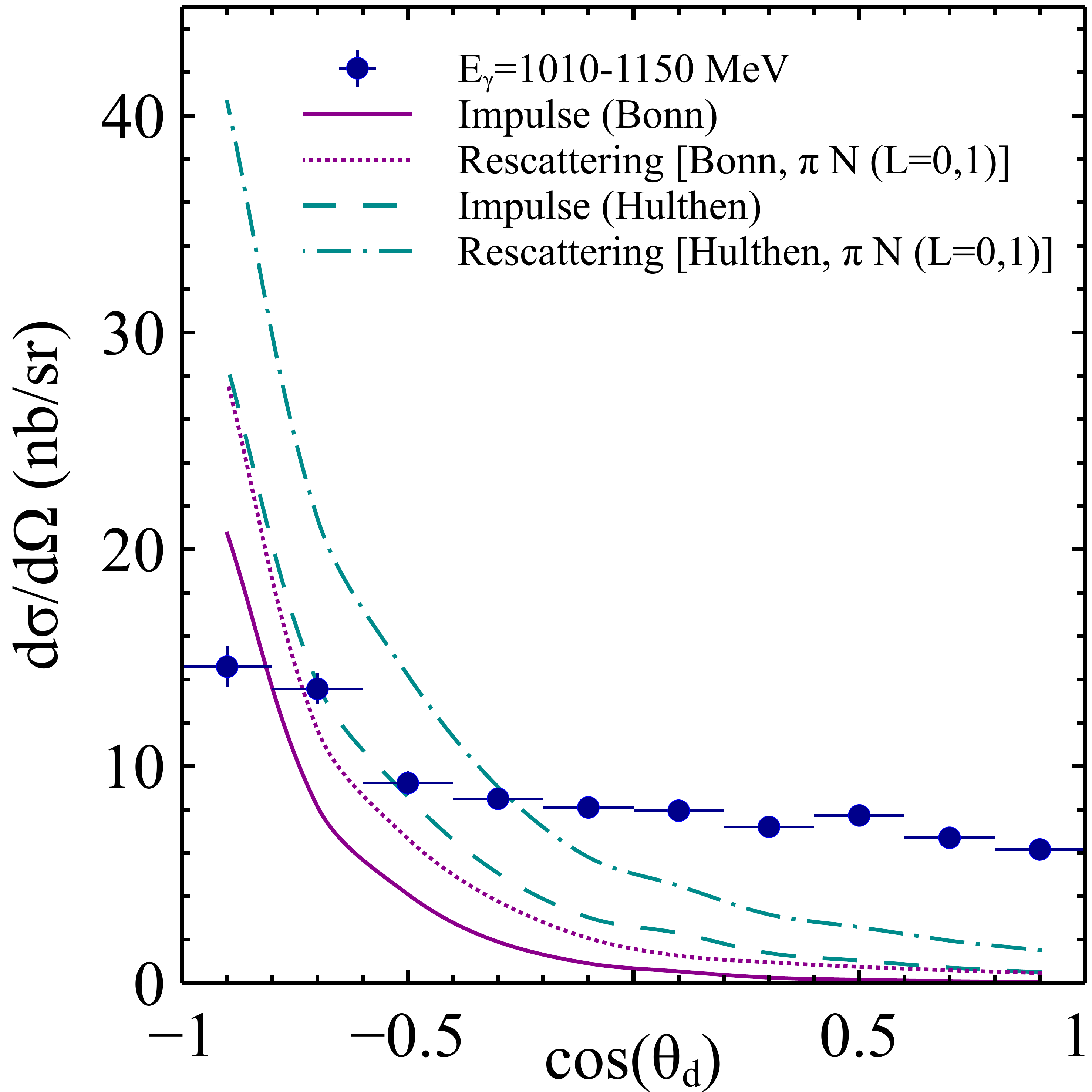}
\caption{Differential cross sections as a function of the polar angle of the outgoing deuteron. Data are taken from Ref.~\cite{Ishikawa:2022mgt}.}\label{Fig19}
\end{figure}
\begin{figure}[h!]
\includegraphics[width=0.4\textwidth]{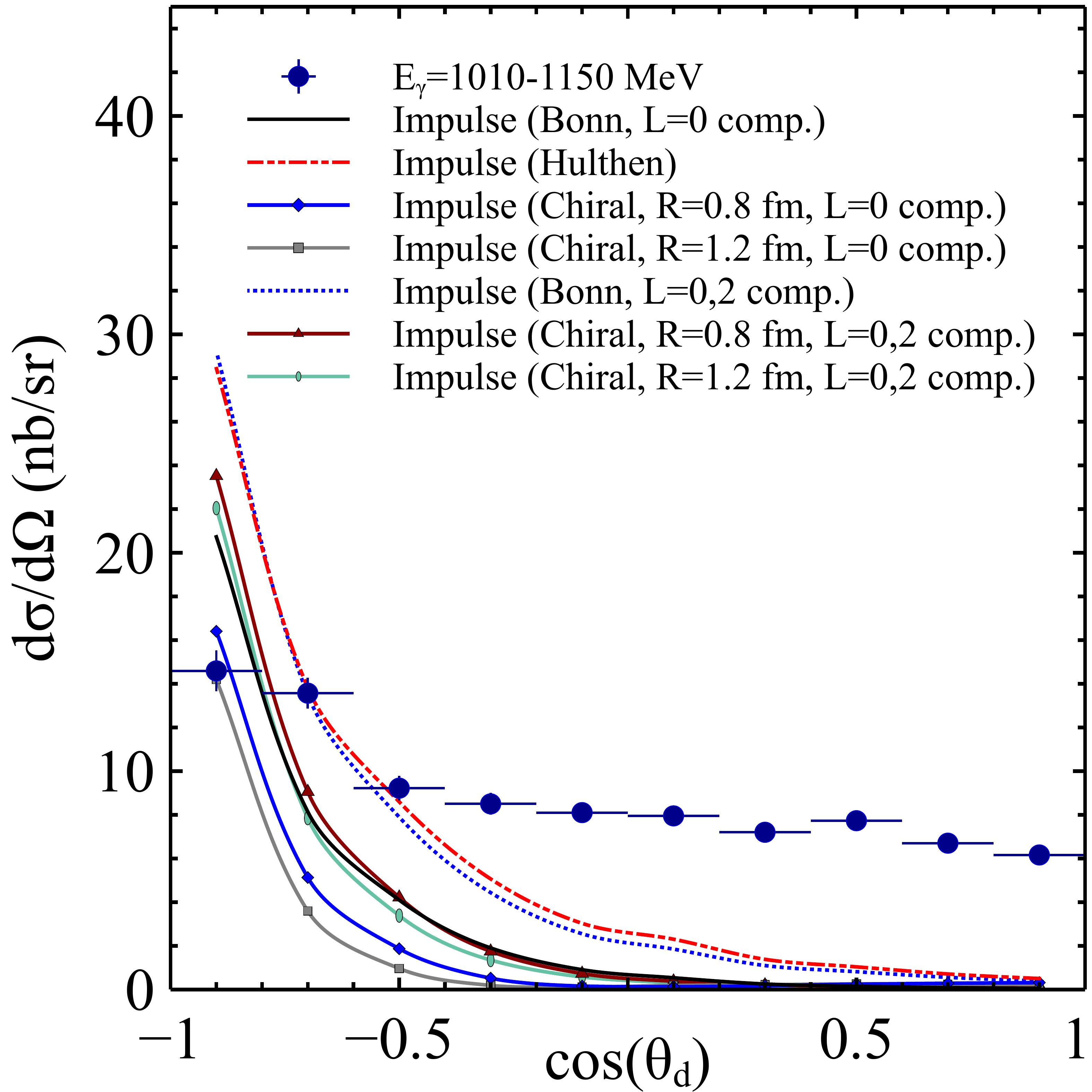}
\caption{Differential cross sections as a function of the polar angle of the outgoing deuteron including the $d$-wave component in the deuteron wave function of Refs.~\cite{Machleidt:2000ge,Epelbaum:2014efa}. Data are taken from Ref.~\cite{Ishikawa:2022mgt}.}\label{Fig20}
\end{figure}

Next, we would like to show the results on the angular distributions in Fig.~\ref{Fig19}. In this case too, we depict the results obtained with the impulse approximation and with the inclusion of the rescattering processes. Since the contribution from $s$-wave $\eta$ rescattering is not significant (see Fig.~\ref{Fig15}) we find it sufficient to consider the effects from the rescattering of a pion. The uncertainties coming from the description of the deuteron wave function (based on Bonn [$s$-wave component only] and Hulth\'en potentials) are also shown. It can be seen from the figure that the differential cross sections are  underestimated at the forward angles, while at backward angles are overestimated. One might wonder if the inclusion of the $d$-wave component of the deuteron wave function could improve the disagreement. As can be seen in Fig.~\ref{Fig20}, the $d$-wave component of the deuteron wave function increases significantly the differential cross section, producing an angular distribution using the (Bonn) wave function of Ref.~\cite{Machleidt:2000ge} which is compatible to that found with the Hulth\'en wave function of Ref.~\cite{Adler:1975ga}. In the case of the chiral wave function~\cite{Epelbaum:2014efa}, the inclusion of the $d$-wave component produces an angular distribution which is similar to the one obtained with the $L=0$ component of the wave function of Ref.~\cite{Machleidt:2000ge}. Independently of the deuteron wave function considered, the shape obtained for the angular distribution continues to differ from the data. The discrepancies shown in Figs.~\ref{Fig19},~\ref{Fig20} are striking, particularly since forward angles require large deuteron momenta. 

Similar findings have been noted in Refs.~\cite{Egorov:2013ppa,Egorov:2020xdt} too, where the $\eta NN$ and $\pi NN$ interactions are implemented with the former system giving rise to a virtual $\eta NN$ state~\cite{Fix:2000hf,Fix:2001cz}. In view of the discrepancies between the experimental data on the angular distribution and the theoretical calculations, further investigations might be necessary, including some other mechanisms which will help sharing the momentum transfer.

\begin{figure}[h!]
\begin{tabular}{cc}
\includegraphics[width=0.35\textwidth]{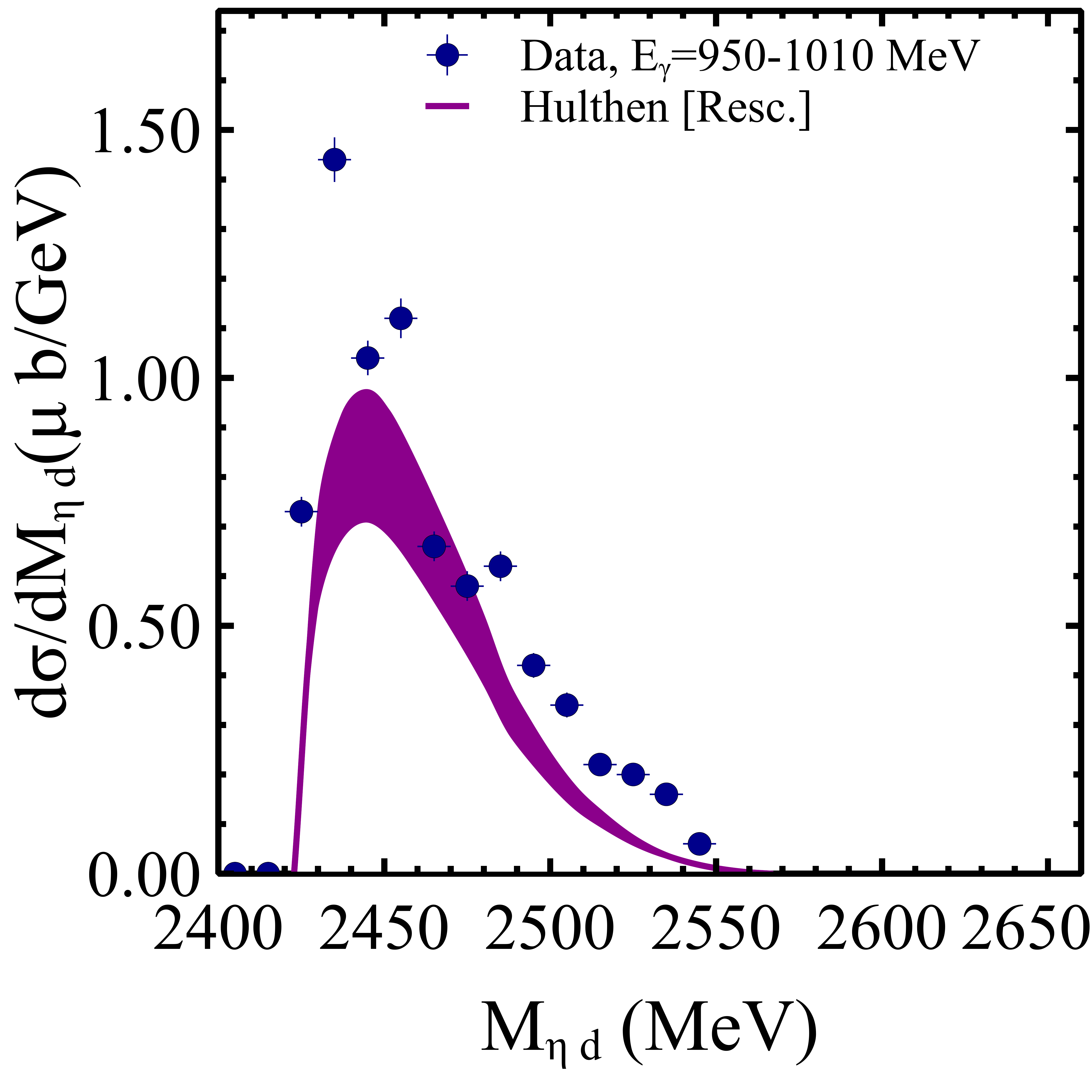}&\includegraphics[width=0.35\textwidth]{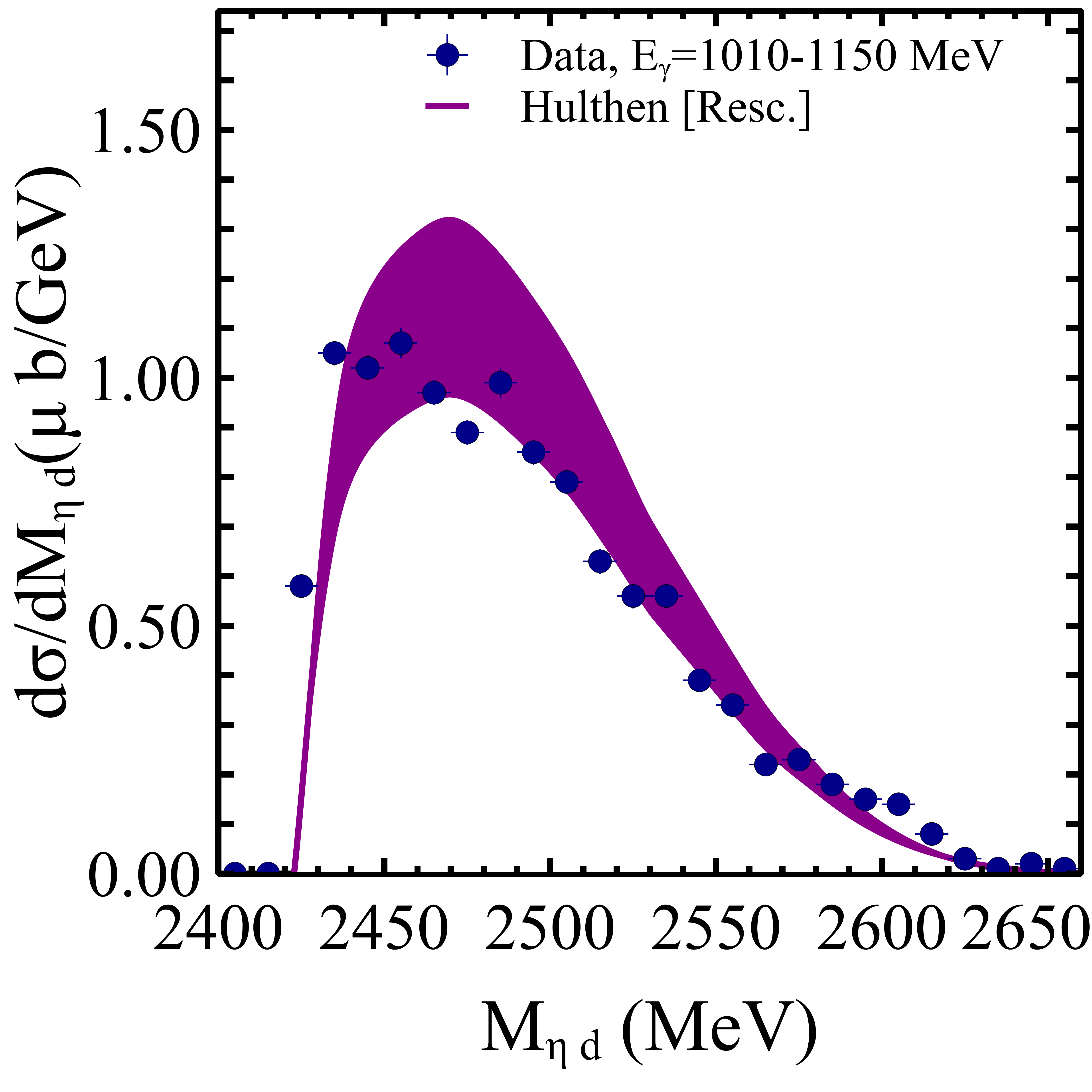}\\
\includegraphics[width=0.35\textwidth]{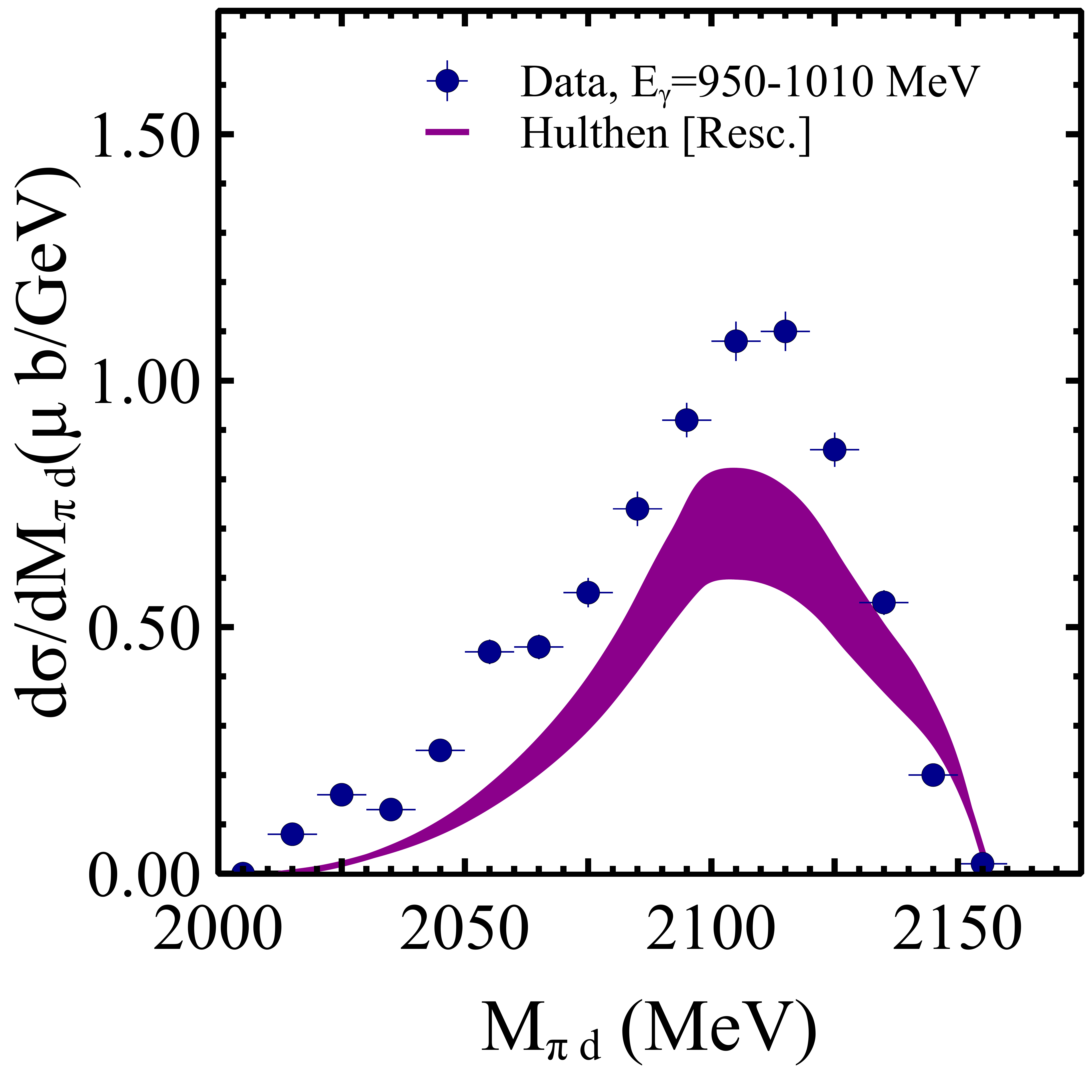}&\includegraphics[width=0.35\textwidth]{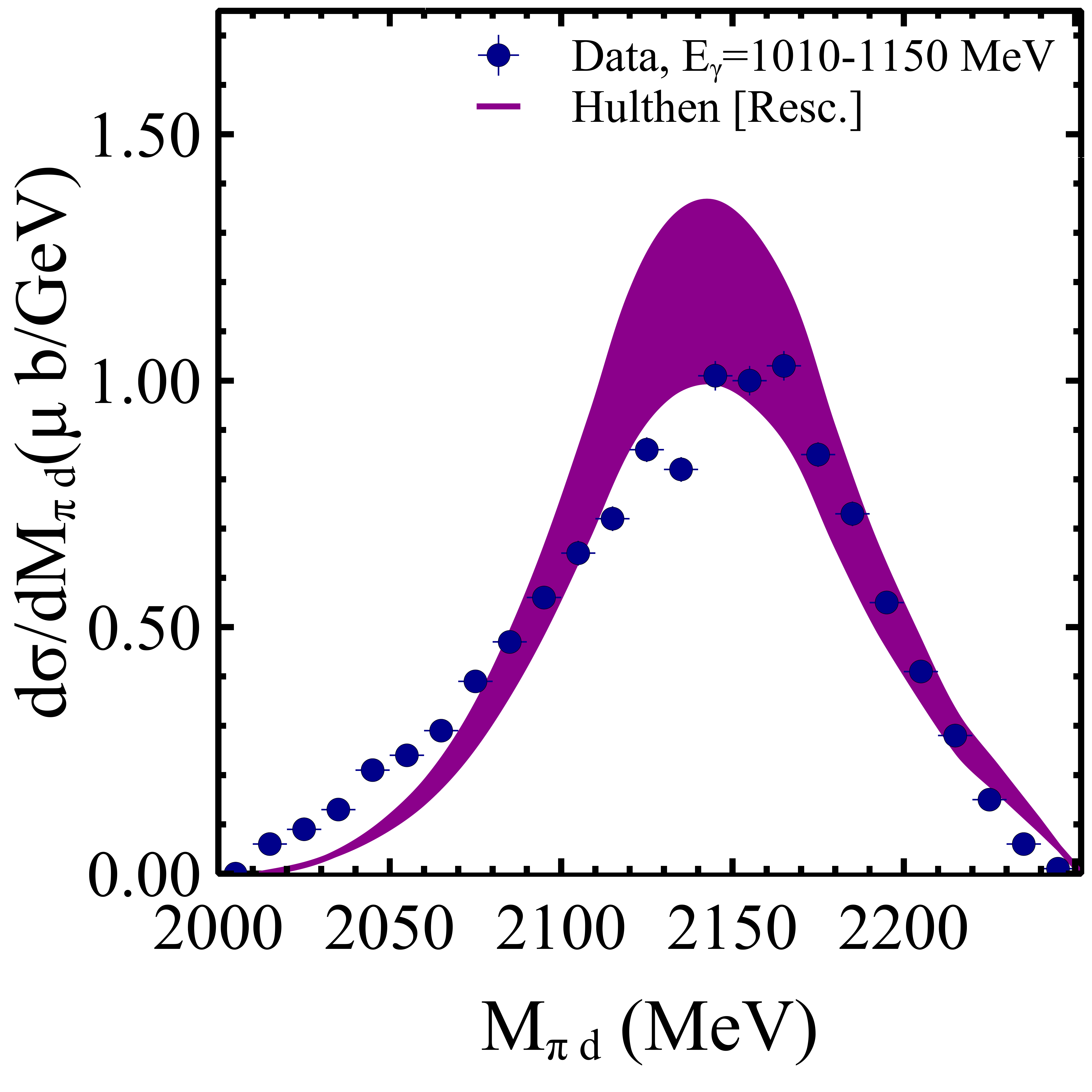}
\end{tabular}
\caption{Uncertainties produced in the invariant mass distributions when allowing variations in the couplings of $\pm 4\%$ of their values. The results shown correspond to the case of considering the Hulth\'en wave function~\cite{Adler:1975ga} and incorporate the rescattering contributions.}\label{Fig21}
\end{figure}

Finally, it is important to quantify the impact of the uncertainties present in the model. Following Ref.~\cite{Debastiani:2017dlz} we have relied on the mechanism of Fig.~\ref{Fig1}(a), but according to the results shown in Fig.~1 of Ref.~\cite{Debastiani:2017dlz}, the inclusion of the $\eta N$ rescattering through the mechanism of Fig.~\ref{Fig2} produces an increase of $\approx10\%$ in the corresponding cross section. In the strict limit of validity of the Schmid theorem~\cite{Schmid:1967ojm}, the mechanism of Fig.~\ref{Fig4} and related rescattering would incorporate the mechanism of rescattering of $\pi\eta$ of Fig.~\ref{Fig2} introducing the phase $e^{2i\delta}$, as discussed in the Introduction, which would not change the cross section. In practice one finds the small increase of $10\%$ in the $\gamma N\to\pi^0\eta N$ cross section.  Thus, the findings of Ref.~\cite{Debastiani:2017dlz} can effectively be incorporated increasing the coupling of $\Delta(1700)$ to $\gamma N$ by $\approx4\%$. One may argue that since the Kroll-Ruderman rescattering term of Fig.~\ref{Fig8} does not have this phase, including the phase would modify the interference. Yet, we proved that these Kroll-Ruderman rescattering terms are negligible and we do not worry about that, but consider the change of $4\%$ in the $\gamma N\to\Delta(1700)$ coupling when evaluating the uncertainties. Further, the couplings of $\Delta(1700)$ to $\eta N$, $N^*(1535)$ to $\pi N$ and $\eta N$ are obtained from the residues of the corresponding scattering matrices in the complex energy plane and typical uncertainties of $\approx4\%$ can also be related to them. Such an uncertainty would arise from the use of different cut-offs when regularizing the two-body loops entering in the calculation of the scattering matrix, as well as the use of physical masses instead of average masses for particles belonging to the same isospin multiplet. In Fig.~\ref{Fig21} we show the invariant mass distributions obtained for the case of the Hulth\'en wave function~\cite{Adler:1975ga} when allowing variations in the couplings and including the different rescattering mechanisms considered in this work.

\section{Conclusions}
We have made a theoretical study of the $\gamma d\to \pi^0 \eta d$ reaction based on a realistic model for the elementary $\gamma N\to \pi^0 \eta N$ reaction that has been tested before in the calculations of the cross sections and polarization observables. It is based on the dominance of the $\gamma N\to \Delta^*(1700)\to \Delta(1232) \eta\to \pi^0 \eta N$ at low energies of the photon, where $\Delta^*(1700)$ is dynamically generated from the pseudoscalar meson interaction with the decuplet of the baryons. This picture determines the $ \Delta^*(1700)\to \eta \Delta(1232)$ coupling such that a prediction without fitting to the data can be done. In fact predictions of the cross section were done prior to the measurement of the reaction and good agreement was found.

When applied to the study of the coherent $\gamma d\to \pi^0 \eta d$ reaction we find two types of mechanisms: the impulse approximation, where the amplitude comes from summing the elementary amplitudes on the $p$ and the $n$ of the deuteron, and the rescattering mechanism of both the $\pi^0$ and $\eta$. The $\pi^0$ in the $s$-wave and $p$-wave, through $\Delta(1232)$ excitation, and the $\eta$ through $N^*(1535)$ excitation in the $s$-wave. What we find is that the reaction involves large momenta of the deuteron, in a region of momenta corresponding to short distances where  the nucleons clearly overlap and it is difficult to give very precise values of the deuteron wave function. This is why we used different models which helped us quantify the uncertainties of the theoretical calculation and they were found to be sizable. With this caveat in mind it was still possible to establish that a reasonable reproduction of the mass distributions can be obtained, although at low photon energies the cross sections obtained somewhat underestimate the experimental data. One relevant feature of the experimental data, which was the shift of the mass distribution to lower invariant mass with respect to phase space for $\eta d$ is obtained and explained on the basis of the dynamical features of the model, where in $\gamma p\to \pi^0 \eta p$ the $\pi^0$ is favored to be produced at higher energies to put $\Delta(1232)$ on-shell and this makes the $\eta d$ invariant mass smaller. The same argument can be used to see that the $\pi^0 d$ mass distribution should peak at higher energies than phase space, something also observed in the experiment.

The biggest shortcoming of the model is that it predicts angular distribution clearly peaking at backward angles, something in clear conflict with experiment that gives a much flatter distribution. The disagreement persists even when considering contributions from the $d$-wave component of the deuteron wave function, which we find to be sizable. A similar discrepancy has also been reported in other theoretical models, even with the presence of an $\eta NN$ virtual state. 

Another finding of the calculations is that the rescattering of the $\pi^0$ and $\eta$ with the spectator nucleon of the impulse approximation increased the cross sections appreciable, in as much as 50$\%$. The mechanism becomes particularly relevant in this reaction because it involves large momentum transfer in the one body mechanism of the impulse approximation. Instead, when the two body mechanism of the rescattering is considered, the momentum transfer is shared between the two nucleons of the deuteron involving smaller momenta in the deuteron wave functions, enhancing the contribution of that mechanism. We also considered rescattering from the process $\gamma N\to \pi^\pm \pi^0 N^\prime$, followed by rescattering of $\pi^\pm$ to produce an $\eta$, but found the contribution of this mechanism to be extremely small.

As to using the results of the reaction to claim a possible $\eta d$ bound state, as claimed in Refs.~\cite{Ishikawa:2021yyz,Ishikawa:2022mgt}, it is a difficult task given the intrinsic uncertainties of the conventional mechanisms disclosed by our calculations. The striking experimental shape of the angular distribution will require further thoughts along other mechanisms not envisaged by us, and any other theoretical calculation so far, that help share the momentum transfer, which is extremely large for forward deuteron angles in the impulse approximation.

\section*{Acknowledgements}
The authors express gratitude towards Prof. Ishikawa for the discussions and for providing the experimental data. This work is partly supported by
the Spanish Ministerio de Econom\'ia y Competitividad and European FEDER funds under Contracts No. PID2020-112777GB-I00, and by Generalitat Valenciana under contract
PROMETEO/2020/023. This project has received funding from the European Unions 10 Horizon 2020 research and innovation programme under grant agreement No. 824093 for
the “STRONG-2020” project. K.P.K and A.M.T gratefully acknowledge the travel support from the above mentioned projects. K.P.K and A.M.T also thank the financial support provided by Funda\c c\~ao de Amparo \`a Pesquisa do Estado de S\~ao Paulo (FAPESP), processos n${}^\circ$ 2019/17149-3 and 2019/16924-3 and the Conselho Nacional de Desenvolvimento Cient\'ifico e Tecnol\'ogico (CNPq), grants n${}^\circ$ 305526/2019-7 and 303945/2019-2, which has facilitated the numerical calculations presented in this work. 
 
\appendix
\section{Spin transition elements}\label{appA}
The amplitudes for the different diagrams discussed in section~\ref{formalism} consist of different spin structures. In this section we evaluate the transitions of the spin parts of the amplitudes. Let us begin with the tree-level amplitude, given by Eq.(\ref{t3}), in which the spin structure corresponds to $\vec S\cdot\vec p_\pi \vec S^\dagger\cdot\vec \epsilon~$. Here, $\vec S$ represents the operator for spin transitions between 3/2 and 1/2. Exploiting a useful property
\begin{align}
\sum_\text{polarizations} S_i S_j^\dagger=\frac{2}{3}\delta_{ij}-\frac{i}{3}\epsilon_{ijk}\sigma_k\label{iden}
\end{align}
and considering that $\Delta$ is produced at the vertex $\Delta^*\to \Delta \eta$, which implies that the spin projections of $\Delta^*$ and  $\Delta$ always coincide, i.e., $m_{\Delta^*}=m_\Delta$, we can write
\begin{align}\nonumber
\vec S\cdot\vec p_\pi\vec S^\dagger\cdot\vec \epsilon&= \sum_{m_\Delta}~p_{\pi_i}~\epsilon_j ~S_i\mid m_\Delta\rangle\langle m_\Delta\mid S_j^\dagger\\
&=\frac{2}{3} \vec p_\pi\cdot\vec \epsilon-\frac{i}{3}\epsilon_{ijk}~p_{\pi_i}\epsilon_j\sigma_k.~\label{wij}
\end{align}
Let us denote the matrix elements for the spin structure in Eq.~(\ref{wij}) by $W_{\mu,\mu^\prime}^\lambda$, where the indices $\mu, \mu^\prime=-1, 0, 1$ represent, respectively, the spin projections $\downarrow\downarrow$, $\uparrow\downarrow+\downarrow\uparrow$, and  $\uparrow\uparrow$ of the deuteron, and $\lambda=1,2$ denotes the two possible polarizations of the photon.  Thus, for example, we can write
\begin{align}\nonumber
W_{1,1}^\lambda=&\langle \uparrow\uparrow|\vec S\cdot\vec p_\pi\vec S^\dagger\cdot\vec \epsilon_\lambda~ | \uparrow\uparrow\rangle,\\\nonumber
W_{1,0}^\lambda=&\langle \frac{1}{\sqrt{2}}\left(\uparrow\downarrow+\downarrow\uparrow\right)|\vec S\cdot\vec p_\pi \vec S^\dagger\cdot\vec \epsilon_\lambda| \uparrow\uparrow\rangle,\\\nonumber
W_{1,-1}^\lambda=&\langle \downarrow\downarrow|\vec S\cdot\vec p_\pi \vec S^\dagger\cdot\vec \epsilon_\lambda| \uparrow\uparrow\rangle.
\end{align}
The sum over polarizations in Eqs.~(\ref{invmass1}) and (\ref{invmass2}) requires calculations of $W_{\mu,\mu^\prime}^\lambda$ for different spin projections of the deuteron in the initial and final state and for the transverse polarizations of the photon [$\vec{\epsilon}_1= \left(1,~0,~0\right)$, $\vec{\epsilon}_2=\left(0,~1,~0\right)$]. We list these elements in Table~\ref{Tabapp1}.

Next, we discuss the evaluation of the spin  part  of the rescattering amplitudes, Eqs.~(\ref{tpi_s_res2}), (\ref{tpi_p_res2}) and (\ref{teta_s_res2}). The spin transition elements for the rescattering of pion, involving $s$-wave $\pi N$ interactions, given by Eq.~(\ref{tpi_s_res2}), can be obtained by replacing $\vec p_\pi \to \vec q^{\, \prime}$ in the expressions given in Table~\ref{Tabapp1}. The elements for the rescattering of pion involving the $\gamma N\to \pi N$ Kroll Ruderman vertex, as well as the elements for the $\eta$-rescattering amplitudes, are identical to those given in Table~\ref{Tabapp1}. 

Finally, the spin part of the pion rescattering amplitudes, involving $p$-wave $\pi N$ interactions [Eqs.~(\ref{tpi_p_res2})] is $\vec S_1\cdot\vec q^{\,\prime}\vec S^\dagger_1\cdot\vec \epsilon~\vec S_2\cdot\vec p_\pi\vec S_2^{\,\dagger}\cdot\vec q^{\,\prime}$, which using Eq.~(\ref{iden}) can be written as
\begin{align}
\left(\frac{2}{3} \vec q^{\,\prime}\cdot\vec \epsilon-\frac{i}{3}\epsilon_{ijk}~q^{\,\prime}_i\epsilon_j\sigma_k \right)\left(\frac{2}{3} \vec p_\pi\cdot\vec q^{\,\prime}-\frac{i}{3}\epsilon_{ijk}~p_{\pi_i}q^\prime_j\sigma_k\right)\label{wijprime}
\end{align}
Let us denote the matrix elements related to Eq.~(\ref{wijprime}) as $\mathcal{W}_{\mu,\mu^\prime}^{\lambda}$. We list these elements for the different transitions in Table~\ref{Tabapp2}.
\begin{table}[h!]
\caption{Spin transition elements $W_{\mu,\mu^\prime}^\lambda$ for different polarizations of the deuteron in the initial ($\mu$) and final ($\mu^\prime$) state. Since $W_{\mu^\prime,\mu}^\lambda$ is the negative of the complex conjugate of $W_{\mu,\mu^\prime}^\lambda$, it suffices to list any one of them.  }\label{Tabapp1}
\begin{tabular}{ccc}
\hline\hline
$\mu$& $\mu^\prime$&$W_{\mu,\mu^\prime}^\lambda$\\\hline
1&1&$\frac{2}{3} \vec p_\pi\cdot\vec \epsilon_\lambda-\frac{i}{3}\left(p_{\pi_x}\epsilon_{\lambda_y}-p_{\pi_y}\epsilon_{\lambda_x}\right)$ \\
1&0&$-\frac{i}{3\sqrt{2}}\left(-p_{\pi_z}\epsilon_{\lambda_y}+ip_{\pi_z}\epsilon_{\lambda_x}\right)$\\
1&-1&0\\
0&0&$\frac{2}{3} \vec p_\pi\cdot\vec \epsilon_\lambda$\\
0&-1&$-\frac{i}{3\sqrt{2}}\left(-p_{\pi_z}\epsilon_{\lambda_y}+ip_{\pi_z}\epsilon_{\lambda_x}\right)$\\
-1&-1&$\frac{2}{3} \vec p_\pi\cdot\vec \epsilon_\lambda+\frac{i}{3}\left(p_{\pi_x}\epsilon_{\lambda_y}-p_{\pi_y}\epsilon_{\lambda_x}\right)$\\\hline\hline
\end{tabular}
\end{table}

\begin{table}[h!]
\caption{Spin transition elements $\mathcal{W}_{\mu,\mu^\prime}^{\lambda}$ for different polarizations of the deuteron in the initial ($\mu$) and final ($\mu^\prime$) state.   }\label{Tabapp2}
\begin{tabular}{ccc}
\hline\hline
$\mu$& $\mu^\prime$&$\mathcal{W}_{\mu,\mu^\prime}^{\lambda}$\\\hline
1&1&$\left[\frac{2}{3} \vec q^{\,\prime}\cdot\vec \epsilon_\lambda-\frac{i}{3}\left(q^\prime_x\epsilon_{\lambda_y}-q^\prime_y\epsilon_{\lambda_x}\right)\right]\left[\frac{2}{3}\vec p_\pi \cdot\vec q^{\,\prime}-\frac{i}{3}\left(p_{\pi_x} q^\prime_y-p_{\pi_y}q^\prime_x\right)\right]$ \\[3ex]
\multirow{2}{*}{1}&\multirow{2}{*}{0}&$\frac{1}{\sqrt{2}}\biggl\{\left[\frac{2}{3}\vec q^{\,\prime}\cdot\vec \epsilon_\lambda-\frac{i}{3}\left(q^\prime_x\epsilon_{\lambda_y}-q^\prime_y\epsilon_{\lambda_x}\right)\right]\left[-\frac{i}{3}\left(p_{\pi_y}q^\prime_z-q^\prime_y p_{\pi_z}-i\left[p_{\pi_x}q^\prime_z-q^\prime_xp_{\pi_z}\right]\right)\right]\biggr.$\\
&&$\biggl.+\left(-\frac{i}{3}\right)\left(-q^\prime_z\epsilon_{\lambda_y}+i q^\prime_z\epsilon_{\lambda_x}\right)\left(\frac{2}{3}\vec p_\pi \cdot\vec q^{\,\prime}-\frac{i}{3}\left[p_{\pi_x} q^\prime_y-p_{\pi_y}q^\prime_x\right]\right)\biggr\}$\\[3ex]
1&-1&$\left(-\frac{i}{3}\right)^2\left(-q^\prime_z\epsilon_{\lambda_y}+i q^\prime_z\epsilon_{\lambda_x}\right)\left(p_{\pi_y}q^\prime_z-q^\prime_y p_{\pi_z}-i\left[p_{\pi_x}q^\prime_z-q^\prime_xp_{\pi_z}\right]\right)$\\[3ex]
\multirow{2}{*}{0}&\multirow{2}{*}{1}&$\frac{1}{\sqrt{2}}\biggl\{\left[\frac{2}{3}\vec q^{\,\prime}\cdot\vec \epsilon_\lambda-\frac{i}{3}\left(q^\prime_x\epsilon_{\lambda_y}-q^\prime_y\epsilon_{\lambda_x}\right)\right]\left[-\frac{i}{3}\left(p_{\pi_y}q^\prime_z-q^\prime_y p_{\pi_z}+i\left[p_{\pi_x}q^\prime_z-q^\prime_xp_{\pi_z}\right]\right)\right]\biggr.$\\
&&$\biggl.+\left(-\frac{i}{3}\right)\left(-q^\prime_z\epsilon_{\lambda_y}-i q^\prime_z\epsilon_{\lambda_x}\right)\left(\frac{2}{3}\vec p_\pi \cdot\vec q^{\,\prime}-\frac{i}{3}\left[p_{\pi_x} q^\prime_y-p_{\pi_y}q^\prime_x\right]\right)\biggr\}$\\[3ex]
\multirow{4}{*}{0}&\multirow{4}{*}{0}&$\frac{1}{2}\biggl\{\left[\frac{2}{3}\vec q^{\,\prime}\cdot\vec \epsilon_\lambda-\frac{i}{3}\left(q^\prime_x\epsilon_{\lambda_y}-q^\prime_y\epsilon_{\lambda_x}\right)\right]\left[\frac{2}{3}\vec p_\pi \cdot\vec q^{\,\prime}+\frac{i}{3}\left(p_{\pi_x} q^\prime_y-p_{\pi_y}q^\prime_x\right)\right]\biggr.$\\
&&$+\left(-\frac{i}{3}\right)^2\left(-q^\prime_z\epsilon_{\lambda_y}+i q^\prime_z\epsilon_{\lambda_x}\right)\left[\left(p_{\pi_y}q^\prime_z-q^\prime_y p_{\pi_z}+i\left[p_{\pi_x}q^\prime_z-q^\prime_xp_{\pi_z}\right]\right)\right]$\\
&&$+\left(-\frac{i}{3}\right)^2\left(-q^\prime_z\epsilon_{\lambda_y}-i q^\prime_z\epsilon_{\lambda_x}\right)\left[\left(p_{\pi_y}q^\prime_z-q^\prime_y p_{\pi_z}-i\left[p_{\pi_x}q^\prime_z-q^\prime_xp_{\pi_z}\right]\right)\right]$\\
&&$\biggl.+\left[\frac{2}{3}\vec q^{\,\prime}\cdot\vec \epsilon_\lambda+\frac{i}{3}\left(q^\prime_x\epsilon_{\lambda_y}-q^\prime_y\epsilon_{\lambda_x}\right)\right]\left[\frac{2}{3}\vec p_\pi \cdot\vec q^{\,\prime}-\frac{i}{3}\left(p_{\pi_x} q^\prime_y-p_{\pi_y}q^\prime_x\right)\right]\biggr\}$\\[3ex]
\multirow{2}{*}{0}&\multirow{2}{*}{-1}&$\frac{1}{\sqrt{2}}\biggl\{\left(-\frac{i}{3}\right)\left[-q^\prime_z\epsilon_{\lambda_y}+i q^\prime_z\epsilon_{\lambda_x}\right] \left[\frac{2}{3}\vec p_\pi \cdot\vec q^{\,\prime}+\frac{i}{3}\left(p_{\pi_x} q^\prime_y-p_{\pi_y}q^\prime_x\right)\right]\biggr.$\\
&&$\biggl.+\left(-\frac{i}{3}\right)\left(p_{\pi_y}q^\prime_z-q^\prime_y p_{\pi_z}-i\left[p_{\pi_x}q^\prime_z-q^\prime_xp_{\pi_z}\right]\right)\left[\frac{2}{3} \vec q^{\,\prime}\cdot\vec \epsilon_\lambda+\frac{i}{3}\left(q^\prime_x\epsilon_{\lambda_y}-q^\prime_y\epsilon_{\lambda_x}\right)\right]\biggr\}$\\[3ex]
-1&1&$\left(-\frac{i}{3}\right)^2\left(-q^\prime_z\epsilon_{\lambda_y}-i q^\prime_z\epsilon_{\lambda_x}\right)\left(p_{\pi_y}q^\prime_z-q^\prime_y p_{\pi_z}+i\left[p_{\pi_x}q^\prime_z-q^\prime_xp_{\pi_z}\right]\right)$\\[3ex]
\multirow{2}{*}{-1}&\multirow{2}{*}{0}&$\frac{1}{\sqrt{2}}\left(-\frac{i}{3}\right)\biggl\{\left(-q^\prime_z\epsilon_{\lambda_y}-i q^\prime_z\epsilon_{\lambda_x}\right)\left(\frac{2}{3}\vec p_\pi \cdot\vec q^{\,\prime}+\frac{i}{3}\left[p_{\pi_x} q^\prime_y-p_{\pi_y}q^\prime_x\right]\right)\biggr.$\\
&&$+\left[\frac{2}{3} \vec q^{\,\prime}\cdot\vec \epsilon_\lambda+\frac{i}{3}\left(q^\prime_x\epsilon_{\lambda_y}-q^\prime_y\epsilon_{\lambda_x}\right)\right]\left[p_{\pi_y}q^\prime_z-q^\prime_y p_{\pi_z}+i\left(p_{\pi_x}q^\prime_z-q^\prime_xp_{\pi_z}\right)\right]$\\[3ex]
-1&-1&$\left[\frac{2}{3} \vec q^{\,\prime}\cdot\vec \epsilon_\lambda+\frac{i}{3}\left(q^\prime_x\epsilon_{\lambda_y}-q^\prime_y\epsilon_{\lambda_x}\right)\right]\left[\frac{2}{3}\vec p_\pi \cdot\vec q^{\,\prime}+\frac{i}{3}\left(p_{\pi_x} q^\prime_y-p_{\pi_y}q^\prime_x\right)\right]$
\\\hline\hline
\end{tabular}
\end{table}
\clearpage
\section{Contribution of the $d$-wave component of the deuteron wave function}\label{appB}
To determine the contribution from the $d$-wave component of the deuteron wave function, we follow Ref.~\cite{Machleidt:2000ge} and consider the following wave function for the deuteron in momentum space
\begin{align}
\Psi^M_d(\vec{k})=\sqrt{4\pi}[\psi_0(k)\mathcal{Y}^{1M}_{01}(\hat k)+\psi_2(k)\mathcal{Y}^{1M}_{21}(\hat k)],\label{psiM}
\end{align}
where $\vec{k}$ is the linear momentum of the deuteron, $\hat k$ are the spherical angles associated with $\vec{k}$, $k=|\vec{k}|$, $\psi_L(k)$ is the component of the deuteron wave function associated with the two nucleon orbital angular momentum $L$, and $\mathcal{Y}^{JM}_{LS}(\hat k)$ represent the normalized eigenfunctions of the two nucleon orbital angular momentum $L$, spin $S$, and total angular momentum $J$ with projection $M$. The latter can be written in terms of spherical harmonics $Y_{Lm}$ as
\begin{align}
\mathcal{Y}^{JM}_{LS}(\hat k)=\sum\limits_m C(L,S,J;m,M-m)Y_{Lm}(\hat k)|S,M-m\rangle,
\end{align}
where $m$ is the projection of $L$, $C(L,S,J;m,M-m)$ are Clebsch-Gordan coefficients for the combination $L\otimes S=J$, and $|S,M-m\rangle$ are the corresponding spin states related to the composition of $L\otimes S$ to give $J$. The spherical harmonics are normalized as
\begin{align}
\int d\Omega\, |Y_{Lm}(\hat k)|^2=1,\quad\int d\Omega\, Y_{lm}(\hat k)Y^*_{L^\prime m^\prime}(\hat k)=\delta_{LL^\prime}\delta_{mm^\prime}.
\end{align}
The factor $\sqrt{4\pi}$ in Eq.~(\ref{psiM}), which is not included in the parametrization of Ref.~\cite{Machleidt:2000ge}, makes the wave function in Eq.~(\ref{psiM}) to be normalized as
\begin{align}
\int d^3 k|\Psi^M_d(\vec{k})|^2=1,
\end{align}
with
\begin{align}
\int d^3 k[\psi^2_0(\vec{k})+\psi^2_2(k)]=1,
\end{align}
which is compatible with the normalization considered in this work. This makes that, with this normalization, following Ref.~\cite{Machleidt:2000ge},  the s- and $d$-wave components of the deuteron wave function can be written as
\begin{align}
\psi_0(k)=\frac{1}{\sqrt{2}\pi}\sum\limits_{j=1}^n\frac{C_j}{k^2+m^2_j}=\frac{\tilde{\psi}_0(k)}{\sqrt{4\pi}},\quad \psi_2(k)=\frac{1}{\sqrt{2}\pi}\sum\limits_{j=1}^n\frac{D_j}{k^2+m^2_j}\frac{k^2}{m^2_j}=\frac{\tilde{\psi}_2(k)}{\sqrt{4\pi}},\label{psi02}
\end{align}
where $n=1,2,\dots,11$, $\tilde{\psi}_0(k)$ and $\tilde{\psi}_2(k)$ are the s- and $d$-wave components of the deuteron wave function\footnote{Note that the expression for $\tilde{\psi}_2$, which we obtain directly from the Fourier transform of the $d$-wave component of the deuteron wave function in coordinate space, $w(r)$ in Ref.~\cite{Machleidt:2000ge} [see Eq. (C20) of Ref.~\cite{Machleidt:2000ge}], is not the same as that given in Eq. (C22) of Ref.~\cite{Machleidt:2000ge}. Curiously, it can be checked that Eq.~(C22) of Ref.~\cite{Machleidt:2000ge} and the $\tilde{\psi}_2$ of Eq.~(\ref{psi02}) differ by a global minus sign. One can show that the Fourier transform of Eq.~(C22) of Ref.~\cite{Machleidt:2000ge} produces $-w(r)$ instead of $w(r)$, as it should, with $w(r)$ being given by Eq. (C20) of Ref.~\cite{Machleidt:2000ge}, which actually coincides with the results of the Table XIX of Ref.~\cite{Machleidt:2000ge}.} with the normalization followed in Ref.~\cite{Machleidt:2000ge}, and the expressions for $C_j$, $D_j$ and $m_j$ can be found in Ref.~\cite{Machleidt:2000ge}. Note that when only the $s$-wave component of the deuteron wave function is considered in the calculations, the normalization is changed such that
\begin{align}
\int d^3 k\psi^\text{only}_0(\vec{k})^2=1.\label{psionlynor}
\end{align}
 In this situation, the parametrization of $\psi^\text{only}_0(\vec{k})$ is given by
 \begin{align}
 \psi^\text{only}_0(\vec{k})=\frac{1}{N}\sum\limits_{j=1}^n\frac{C_j}{k^2+m^2_j},
 \end{align}
where $N\simeq 4.33225$ is the normalization constant needed to satisfy Eq.~(\ref{psionlynor}). In this way, we can write
\begin{align}
\psi_0(k)=\frac{N}{\sqrt{2}\pi}\frac{1}{N}\sum\limits_{j=1}^n\frac{C_j}{k^2+m^2_j}\equiv \omega_s \psi^\text{only}_0(\vec{k}),\label{psi0ws}
\end{align}
with $\omega_s\equiv N/(\sqrt{2}\pi)\simeq 0.9751$.

Let us consider, for example, the case, $J=M=1$. We have then,
\begin{align}
\mathcal{Y}^{11}_{01}(\hat k)&=Y_{00}(\hat k)|1,1\rangle=\frac{1}{\sqrt{4\pi}}|1,1\rangle,\nonumber\\
\mathcal{Y}^{11}_{21}(\hat k)&=\sqrt{\frac{3}{5}}Y_{22}|1,-1\rangle-\sqrt{\frac{3}{10}}Y_{21}|1,0\rangle+\frac{1}{\sqrt{10}}Y_{20}|1,1\rangle,\label{ycal2}
\end{align}
with
\begin{align}
Y_{22}(\hat k)&=\frac{1}{4}\sqrt{\frac{15}{2\pi}}\text{sin}^2\theta[\text{cos}\phi+i\text{sin}\phi]^2,\nonumber\\
Y_{21}(\hat k)&=-\sqrt{\frac{15}{8\pi}}\text{sin}\theta\text{cos}\theta[\text{cos}\phi+i\text{sin}\phi],\nonumber\\
Y_{20}(\hat k)&=\sqrt{\frac{5}{4\pi}}\frac{1}{2}(3\text{cos}^2\theta-1),
\end{align}
where $\theta$ and $\phi$ are the polar and azimuthal angles related to $\vec{k}$.

Using Eqs.~(\ref{psiM}), (\ref{psi0ws}), (\ref{psi02}) and (\ref{ycal2}), we have the following wave function for the deuteron for $J=M=1$,
\begin{align}
\Psi^1_d(\vec{k})&=\omega_s\psi^\text{only}_0(\vec{k})|1,1\rangle+\tilde{\psi}_2(k)\left[\sqrt{\frac{3}{5}}Y_{22}|1,-1\rangle-\sqrt{\frac{3}{10}}Y_{21}(\hat k)|1,0\rangle+\frac{1}{\sqrt{10}}Y_{20}|1,1\rangle\right]\nonumber\\
&=\omega_s\psi^\text{only}_0(\vec{k})|\uparrow\uparrow\rangle+\tilde{\psi}_2(k)\Bigg[\sqrt{\frac{3}{5}}Y_{22}|\downarrow\downarrow\rangle-\sqrt{\frac{3}{10}}Y_{21}(\hat k)\frac{1}{\sqrt{2}}\Big(|\uparrow\downarrow\rangle+|\downarrow\uparrow\rangle\Big)\nonumber\\
&\quad+\frac{1}{\sqrt{10}}Y_{20}|\uparrow\uparrow\rangle\Bigg],\label{psifin}
\end{align}
where in the last line we have written the two nucleon spin states $|1,1\rangle$, $|1,0\rangle\rangle$, $|1,-1\rangle$ in terms of the spin projections of each nucleon. Using Eq.~(\ref{psifin}), we can obtain the matrix elements determined in Appendix~\ref{appA}. For example, a combination like
\begin{align}
W^\lambda_{11}\psi^\text{only}_0(Q)\psi^\text{only}_0(Q^\prime),\nonumber\\
\end{align}
where $W^\lambda_{11}=\langle\uparrow\uparrow|\vec{S}\cdot\vec{p}_\pi\vec{S}^\dagger\cdot\vec{\epsilon}_\lambda|\uparrow\uparrow\rangle$, $\vec{Q}=\frac{\vec{p}_d}{2}-\vec{q}$, $\vec{Q}^\prime=-\frac{\vec{p}_\eta+\vec{p}_\pi}{2}-\vec{q}$, $Q=|\vec{Q}|$, $Q^\prime=|\vec{Q}^\prime|$, and which appears in the tree level amplitude, becomes
\begin{align}
\langle &\Psi^1_d(\vec{Q}^\prime)|\vec{S}\cdot\vec{p}_\pi\vec{S}^\dagger\vec{\epsilon}_\lambda|\Psi^1_d(\vec{Q})\rangle=\Bigg[\Big(\omega_s\psi^\text{only}_0(Q^\prime)+\tilde{\psi}_2(Q^\prime)\frac{1}{\sqrt{10}}Y^*_{20}(\hat Q^\prime)\Big)\langle\uparrow\uparrow|\nonumber\\
&\quad+\tilde{\psi}_2(Q^\prime)\sqrt{\frac{3}{5}}Y^*_{22}(\hat Q^\prime)\langle\downarrow\downarrow|-\sqrt{\frac{3}{10}}Y^*_{21}(\hat Q^\prime)\frac{1}{\sqrt{2}}\Big(\langle\uparrow\downarrow|+\langle\downarrow\uparrow|\Big)\Bigg]\Big(\vec{S}\cdot\vec{p}_\pi\vec{S}^\dagger\cdot\vec{\epsilon}_\lambda\Big)\nonumber\\
&\quad\times\Bigg[\Big(\omega_s\psi^\text{only}_0(Q)+\tilde{\psi}_2(Q)\frac{1}{\sqrt{10}}Y^*_{20}(\hat Q)\Big)|\uparrow\uparrow\rangle+\tilde{\psi}_2(Q)\sqrt{\frac{3}{5}}Y^*_{22}(\hat Q)|\downarrow\downarrow\rangle\nonumber\\
&\quad-\sqrt{\frac{3}{10}}Y^*_{21}(\hat Q)\frac{1}{\sqrt{2}}\Big(|\uparrow\downarrow\rangle+|\downarrow\uparrow\rangle\Big)\Bigg]\nonumber\\
&=\Big(\omega_s\psi^\text{only}_0(Q^\prime)+\tilde{\psi}_2(Q^\prime)\frac{1}{\sqrt{10}}Y^*_{20}(\hat Q^\prime)\Big)\Big(\omega_s\psi^\text{only}_0(Q^\prime)+\tilde{\psi}_2(Q^\prime)\frac{1}{\sqrt{10}}Y^*_{20}(\hat Q^\prime)\Big)W^\lambda_{11}\nonumber\\
&\quad-\sqrt{\frac{3}{10}}\Big(\omega_s\psi^\text{only}_0(Q^\prime)+\tilde{\psi}_2(Q^\prime)\frac{1}{\sqrt{10}}Y^*_{20}(\hat Q^\prime)\Big)Y_{21}(\hat Q)\tilde{\psi}_2(Q)W^\lambda_{21}\nonumber\\
&\quad+\frac{3}{5}\tilde{\psi}_2(Q^\prime)\tilde{\psi}_2(Q)Y^*_{22}(\hat Q^\prime)Y_{22}(\hat Q)W^\lambda_{33}-\frac{3}{5\sqrt{2}}Y^*_{22}(\hat Q^\prime)Y_{21}(\hat Q)\tilde{\psi}_2(Q^\prime)\psi_2(Q)W^\lambda_{23}\nonumber\\
&\quad-\sqrt{\frac{3}{10}}Y^*_{21}(\hat Q^\prime)\tilde{\psi}_2(Q^\prime)\Big(\omega_s\psi^\text{only}_0(Q)+\tilde{\psi}_2(Q)\frac{1}{\sqrt{10}}Y_{20}(\hat Q)\Big)W^\lambda_{12}\nonumber\\
&\quad-\frac{3}{5\sqrt{2}}Y^*_{21}(\hat Q^\prime)Y_{22}(\hat Q)\tilde{\psi}_2(Q^\prime)\tilde{\psi}_2(Q)W^\lambda_{32}+\frac{3}{10}Y^*_{21}(\hat Q^\prime)Y_{21}(\hat Q)\tilde{\psi}_2(Q^\prime)\tilde{\psi}_2(Q)W^\lambda_{12},
\end{align}
where we have used that $W^\lambda_{13}=W^\lambda_{31}=0$. Similarly, we can calculate the matrix elements 
\begin{align}
\langle &\Psi^{M^\prime}_d(\vec{Q}^\prime)|\vec{S}\cdot\vec{p}_\pi\vec{S}^\dagger\vec{\epsilon}_\lambda|\Psi^M_d(\vec{Q})\rangle,
\end{align}
with $M, M^\prime=-1,0,1$ and determine the differential cross section. A good estimation, however, can be obtained by realizing that the differential cross sections found with only the $s$-wave component of the deuteron wave function are dominated by transitions where the values $\mu$ and $\mu^\prime$ in $W^\lambda_{\mu,\mu^\prime}$ are the same, the latter being around 6 times bigger than that obtained from transitions where $\mu\neq \mu^\prime$ (whenever these ones are not zero). Thus, the main contribution to the differential cross section when including the $d$-wave component of the deuteron wave function comes from diagonal terms, i.e., $\mu=\mu^\prime$. At the same time, transitions where $\mu=\mu^\prime$ contributes equally to the differential cross section. The same is the case for those transitions involving $W^\lambda_{\mu,\mu^\prime}$ and $W^\lambda_{\mu^\prime,\mu}$. Then, to estimate the effect of including the $d$-wave component of the deuteron wave function in the determination of the differential cross section within the impulse approximation, we use the following expression:
\begin{align}
\frac{d\sigma}{dM_\text{inv}}=3\frac{d\sigma_{11}(\text{s+d waves})}{dM_\text{inv}}+2\left[\frac{d\sigma_{12}(\text{$s$-wave})}{dM_\text{inv}}+\frac{d\sigma_{13}(\text{$s$-wave})}{dM_\text{inv}}+\frac{d\sigma_{23}(\text{$s$-wave})}{dM_\text{inv}}\right],
\end{align}
where $\sigma_{ij}$ refers to the contribution of the transition element $ij$ to the cross section $\sigma$, and the text between brackets expresses whether we include, or not, the $L=0$ ($s$-wave) and $2$ components ($d$-wave) of the deuteron wave function when calculating the cross section.
\clearpage

 \bibliographystyle{apsrev4-1}
\bibliography{bibdib}
 \end{document}